\documentclass[11pt,reqno]{amsproc}
\allowdisplaybreaks
\usepackage{cite}
\usepackage{psfrag}
\usepackage{color}
\usepackage{mathrsfs}
\usepackage{fullpage}
\usepackage{graphicx}
\usepackage{subfigure}
\usepackage{enumerate}
\numberwithin{equation}{section}
\usepackage{wrapfig}
\usepackage[sort,semicolon,square,authoryear]{natbib} 
\DeclareGraphicsExtensions{.eps}
\usepackage{ucs}
\usepackage[utf8x]{inputenc}
\usepackage{epstopdf}
\usepackage[debug=false, colorlinks=true, pdfstartview=FitV, 
linkcolor=blue, citecolor=blue, urlcolor=blue]{hyperref}
\newtheorem{theorem}{Theorem}[section]

\usepackage{morefloats}

\newlength{\drop}
\definecolor{amethyst}{rgb}{0.6, 0.4, 0.8}
\definecolor{burgundy}{rgb}{0.5, 0.0, 0.13}

\title{Modeling flow in porous media with double 
  porosity/permeability:~Mathematical model,
  properties, and analytical solutions}

\author{\textbf{K.~B.~Nakshatrala}, \textbf{S.~H.~S.~Joodat},
  and \textbf{R.~Ballarini} \\
  {\small Department of Civil and Environmental
    Engineering, University of Houston. \\
    \textbf{Correspondence to:}~\textsf{knakshatrala@uh.edu}}}


\begin{document}

\date{\today}

\begin{titlepage}
  \drop=0.1\textheight
  \centering
  \vspace*{\baselineskip}
  \rule{\textwidth}{1.6pt}\vspace*{-\baselineskip}\vspace*{2pt}
  \rule{\textwidth}{0.4pt}\\[\baselineskip]
       {\Large \textbf{\color{burgundy}
           Modeling flow in porous media with double porosity/permeability: 
           \\[0.3\baselineskip]
           {\large 
           Mathematical 
           model, properties, and analytical solutions}}}\\[0.3\baselineskip]
       \rule{\textwidth}{0.4pt}\vspace*{-\baselineskip}\vspace{3.2pt}
       \rule{\textwidth}{1.6pt}\\[\baselineskip]
       \scshape
       An e-print of the paper is available on arXiv:~1605.07658. \par 
       \vspace*{1\baselineskip}
       Authored by \\[\baselineskip]
           
  {\Large K.~B.~Nakshatrala\par}
  {\itshape Department of Civil \& Environmental Engineering \\
  University of Houston, Houston, Texas 77204--4003 \\ 
  \textbf{phone:} +1-713-743-4418, \textbf{e-mail:} knakshatrala@uh.edu \\
  \textbf{website:} http://www.cive.uh.edu/faculty/nakshatrala}\\[0.75\baselineskip]
  
  {\Large S.~H.~S.~Joodat\par}
  {\itshape Graduate Student, University of Houston}\\[0.75\baselineskip]
    
  {\Large R.~Ballarini\par}
  {\itshape Thomas and Laura Hsu Professor and Chair \\
    Department of Civil \& Environmental Engineering,
    University of Houston.}
  
  \begin{figure}[h]
    \psfrag{x}{$x$}
    \psfrag{p1p2}{$p_1(x) - p_2(x)$}
    \includegraphics[clip,scale=0.276]{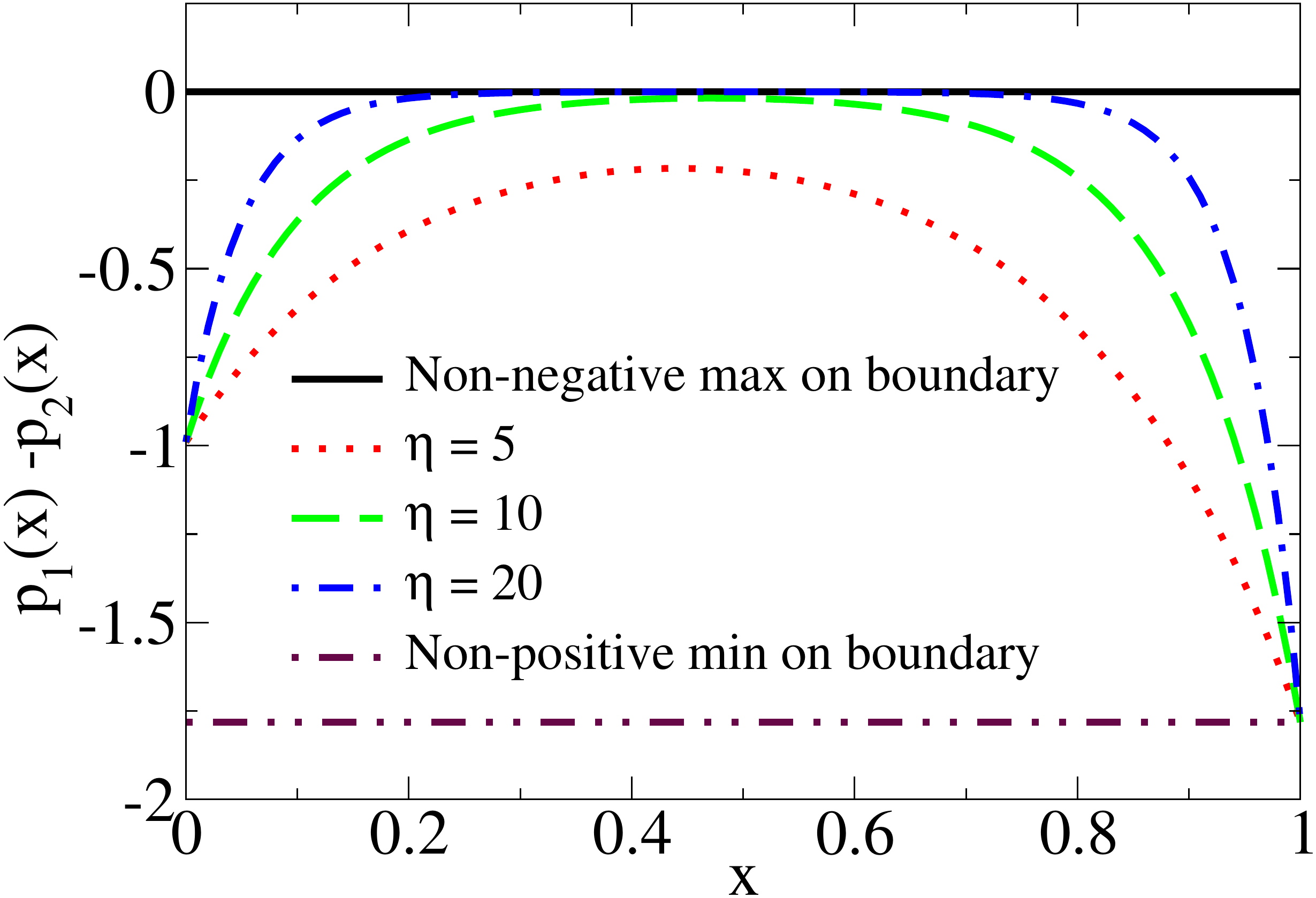}
    
    \emph{This figure numerically verifies the proposed
      maximum principle for the double porosity/permeability
      model, which implies that the difference in pressures
      in the macro-pore and micro-pore networks in the entire
      domain should lie between the non-negative maximum
      and the non-positive minimum values on the boundary.
      The parameter $\eta$ is inversely proportional to the
      square root of the harmonic mean of the permeabilities
      in the macro-pore and micro-pore networks.}
  \end{figure}
  \vfill
  {\scshape 2016} \\
  {\small Computational \& Applied Mechanics Laboratory} \par
\end{titlepage}

\begin{abstract}
  Geo-materials such as vuggy carbonates are known
  to exhibit multiple spatial scales. A common
  manifestation of spatial scales is the presence
  of (at least) two different scales of pores with
  different hydro-mechanical properties.
  Moreover, these pore-networks are connected through
  fissures and conduits. 
  Although some models are available in the literature to 
  describe flow in such porous media, they
  lack a strong theoretical basis. This paper aims to fill this
  lacuna by providing the much needed theoretical foundations of
  the flow in porous media that exhibit double porosity/permeability. 
  We first obtain a mathematical model using the 
  maximization of rate of dissipation hypothesis, 
  and thereby providing a firm thermodynamic underpinning. 
  We then present, along with mathematical proofs,
  several important mathematical properties that
  the solutions to the model satisfy.
  We also present several canonical problems and obtain
  the corresponding analytical solutions, which are used
  to gain insights into the velocity and pressure profiles,
  and the mass transfer across the two pore-networks.
  In particular, we highlight how the solutions under the
  double porosity/permeability differ from the corresponding
  ones under Darcy equations.
\end{abstract}

\maketitle

\section{INTRODUCTION AND MOTIVATION}
Most models of flow in porous media make the 
simplifying assumption that the domain consists 
of a system of similar-sized pores connected by 
a single pore-network. In reality, many geo-materials
such as aggregated soils or fissured rocks exhibit
two or more dominant pore-scales connected by multiple
pore-networks \citep{Almukhtar_1995a,Delage_1996,
	Didwania_2002, Koliji_2006,Borja_Koliji_2009,straughan2017mathematical}
that display significantly different
hydro-mechanical properties such as disparate permeabilities
and different orders of volume fractions.
As an example, let us consider a pile of soil
comprised of large pieces of clay. In such
a medium, clay pieces are considered as the
macro-pores and the existing system of fissures
and cracks form the micro-pores. It is worth
mentioning that in a system like this, the
degradation of macro-pores over the time
leads to an increase in the amount of micro-pores.
Figures \ref{Fig:Stone_wall} and \ref{Fig:Lava}
provide two examples of the double porosity structure
observed in nature; a typical wall constructed by
placing stone pieces on top of each other, and the
water-saturated lava double porosity structure formed
by putting together many pieces of lava from Mount Etna
\citep{straughan2017mathematical}. Moreover, it is possible
to obtain synthetic double porosity media as shown in
Fig. \ref{Fig:man_made}; especially due to the recent
advances in 3D printing and additive manufacturing.
For example, the pores between the spheres construct
the macro-network while the micro-network has been
generated by drilling cylindrical holes in the spheres. 

Porous materials with two dominant pore-networks
have been studied in the literature under the subject
of either the dual-porosity or dual-permeability.
(A recent work \citep{straughan2017mathematical}
even considers porous materials with multiple
pore-networks.) However, there is a subtle difference
in the phenomena the two words describe, and therefore it is necessary to clarify what we mean by dual-porosity and dual-permeability models. Note that, ``dual" and ``double" have been used equivalently in the literature, as will be done in this paper. 

The main assumption in a $\textit{dual-porosity}$ model is 
that the permeability of the macro-pores is much greater than 
the permeability of the micro-pores, while the porosity of the
former is much smaller than the porosity of the latter. In other
words, fluid is mostly trapped within the micro-pores while
macro-pores form the major fluid pathways due to their higher
permeability. Hence, the liquid phase is divided into mobile
and immobile regions with the possibility of fluid exchange
between them \citep{vanGenuchten_Wierenga_1976,vsimuunek_Jarvis_vanGenuchten_Gardenas_Journal_of_Hydrology_2003}. 
In the most general case, many rocks contain an independent system
of fractures superimposed on the porous matrix and are commonly known as an intermediate porous medium. Such media are typically idealized using the dual-porosity model which is of high interest in petroleum reservoirs. Samples of such rocks are limestones or dolomites. However, other sedimentary rocks such as cherty shale or siltstone also exhibit the same characteristics \citep{Warren_Root_1963_v3}. 
Carbonate rocks have been known to exhibit macro-pores in form of fractures and joints for a long time and the importance of such porosities in the sandstones was emphasized on by \citep{Hayes_1979,Schmidt_Mcdonald_1979}. Moreover, studies have revealed that natural soils, especially the compacted ones, have two levels of structure, leading to the appearance of two main classes of the pores (macro- and micro-pores) corresponding to the two levels of soil structure \citep{Cuisinier_2004}. Soils exhibiting such a division of pores can also be idealized using the concept of dual-porosity. 
The first dual-porosity model is commonly attributed to
\citep{Barenblatt_Zheltov_Kochina_v24_p1286_1960_ZAMM} which addressed the flow through a \emph{fractured}
porous medium. In this paper, a term has been introduced
based on dimensional analysis arguments to account for
the mass transfer across the two pore-scales (i.e., matrix
pores and fissures). \citep{Warren_Root_1963_v3} later
introduced two parameters for characterizing dual-porosity
media; one parameter measures the fluid capacitance in the
macro-pores and the other accounts for the inter-porosity
flow. \citep{Dykhuizen_v26_p351_1990_WRR} proposed a new
nonlinear coupling term for double porosity based on the
models proposed by \citep{Barenblatt_Zheltov_Kochina_v24_p1286_1960_ZAMM}
and \citep{Warren_Root_1963_v3}. This dual-porosity model, unlike the
previous ones, accounts for the diffusion across pore-networks and is
valid even for unsteady conditions.

In contrast to dual-porosity models, the term $\textit{dual-permeability}$
pertains to the case where the fluid flows through both micro-pores and macro-pores, and there can be mass transfer across the pore-networks
\citep{Vogel_Gerke_Zhang_Vangenuchten_2000,Balogun_Kazemi_Ozkan_Alkobaisi_Ramirez_2007}.
Different approaches have been used to describe flow and transport
using dual-permeability models. In some cases, the flow in both micro-pores and
macro-pores has been described using similar governing equations, while in others,
different formulations have been considered in the two pore-networks
\citep{vsimuunek_Jarvis_vanGenuchten_Gardenas_Journal_of_Hydrology_2003}. 
However, most of the works on dual-permeability have considered the
macro-network to be \emph{fractures} with much higher permeability
than the micro-network.

Herein, we generalize by assuming that there are two pore-networks with their own porosity and permeability and
there is a mass transfer across the pore-networks. The
macro-network can be a network of fractures, or can be
another pore-network. It is possible to identify the
presence of multiple pore sizes using experimental
techniques such as the Brunauer-Emmett-Teller (BET) method 
\citep{lowell_Shields_Thomas_Thommes_2012_BET_poresize}.
Moreover, the multiple pore-networks can be characterized
using modern techniques like $\mu$-CT
\citep{stock2008microcomputed}. We shall refer to the
aforementioned general treatment as the ``$\emph{double
	porosity/permeability model}$''.
Figure~\ref{Dual_porosity_Permeability_schematic} represents the fractured
porous medium idealized by dual-porosity model as well as a porous medium
with two pore-networks idealized by a double porosity/permeability model. 
The vertical and horizontal arrows represent the fluid pathways and the mass transfer within the domain. In
the fractured porous medium idealized by dual-porosity model, the mass
transfer can occur between matrix pores and the fractures, and the
fluid mostly passes through the fissures due to their higher permeability.
In the porous medium with two pore-networks, mass is transferred across the two pore-networks but in this case, both the micro-pores and
the macro-pores provide the pathways for pore fluid.

Although various models have been developed for double
porosity/permeability over the years, many of them are
applicable to simple settings and are valid only under
stringent conditions. For example, in many previously
developed models the spatial variation of pressure
within the pore-scales is neglected or the mass transfer
term is not an accurate representation of reality. 
Some mathematically-oriented works derived dual-porosity 
models using the theory of mathematical homogenization
(e.g., see \citep{Arbogast_Douglas_Hornung_1990,
	Amaziane_Pankratov_2015, Boutin_Royer_2015}).  
However, these papers did not address the relevant thermomechanical 
underpinning, and did not provide a coherent framework that makes
it possible to obtain generalizations of those models
in such a way that the thermomechanics principles
are satisfied.
Homogenization is a mathematical tool for \textit{up-scaling} differential equations. In homogenization theory a complex, rapidly-varying medium is represented by a slowly-varying medium in which the fine-scale structure is averaged out properly and a ``homogenized'' or ``effective'' system of equations is obtained at the macroscopic level \citep{Amaziane_2010}. In other words, the problem at hand is embedded in a set of problems which are parameterized by a scaling parameter \citep{Hornung}.
Most importantly, the presentations of prior works on double
porosity/permeability seem rather ad hoc, especially with respect
to the treatment of mass transfer across the pore-networks. This
is one of the main hurdles researchers are faced with while
generalizing the mathematical model to more complicated situations
like multi-phase flows and considering the effect of deformation
of the porous solid along with flow in multiple pore-networks.
Herein, we put the double porosity/permeability model under
a firm footing with strong thermodynamic and mathematical
underpinnings. In particular, we give a firm basis for the
mass transfer across the pore-networks, and a mathematical
framework amenable to further generalizations of the model. It should be emphasized that we're not proposing a new model for flow in porous media exhibiting double porosity/permeability. Rather, we're providing a thermomechanical basis for the existing models which makes further generalizations of them possible.

The basic philosophy in our modeling approach can be stated as
follows: (a) there exist (at least) two different pore-networks;
(b) each pore-network is assumed to be a continuum, and transport
of mass and chemical species can occur within each pore-network;
and (c) mass can be transferred between the pore-networks. The
parameters and quantities in the model represent values that are 
averaged over a representative volume element (RVE) whose 
existence is either tacitly or explicitly assumed in most of
the double porosity/permeability models. For simplicity, we will
model the flow in both networks using similar governing equations
(i.e., Darcy-type equations), but one can use different descriptions
of flows in the different pore-networks.

The rest of this paper is organized as follows. Section
\ref{Sec:S2_Double_Model} outlines the governing equations
for a double porosity/permeability model. Section
\ref{Sec:S3_Dual_Basis} presents a mathematical
framework for deriving porous media models using the
maximization of rate of dissipation and volume fractions
approach, and obtains the double porosity/permeability model 
as a special case. 
Several mathematical properties of this model are derived
in Section \ref{Sec:S4_Dual_Mathematical}. An analytical
solution procedure is presented in Section
\ref{Sec:S5_Double_Analytical}. Several canonical problems
along with their analytical solutions are given in
Section \ref{Sec:S6_Double_BVP}. Finally, conclusions
are drawn in Section \ref{Sec:S7_Double_CR}.

Throughout this paper, repeated indices do not imply summation.
\section{MATHEMATICAL MODEL}
\label{Sec:S2_Double_Model}
Consider a bounded domain $\Omega \subset \mathbb{R}^{nd}$, 
where ``$nd$'' denotes the number of spatial dimensions. 
The boundary $\partial \Omega$ is assumed to 
be piecewise smooth. Mathematically, $\partial 
\Omega := \mathrm{cl}(\Omega) - \Omega$, where $\mathrm{cl}(\cdot)$ denotes the set closure \citep{Evans_PDE}. A spatial
point in $\Omega$ is denoted by $\mathbf{x}$. The gradient
and divergence operators with respect to $\mathbf{x}$ are,
respectively, denoted by $\mathrm{grad}[\cdot]$ and
$\mathrm{div}[\cdot]$. The unit outward normal to the
boundary is denoted by $\widehat{\mathbf{n}}(\mathbf{x})$.  

We are interested in studying the flow of an incompressible
fluid in a rigid porous medium that consists of two distinct 
pore-networks. These pore-networks are connected by conduits 
and/or fissures, and hence there can be mass transfer across 
the pore-networks. We shall refer to these two pore-networks 
as macro-pore and micro-pore networks, and identify them 
using subscripts $1$ and $2$, respectively. The
permeability tensors for these pore-networks are denoted
by $\mathbf{K}_{1}(\mathbf{x})$ and $\mathbf{K}_{2}(\mathbf{x})$,
which are assumed to be anisotropic and spatially inhomogeneous 
second-order tensors. The porosities in these pore-networks are
denoted by $\phi_{1}(\mathbf{x})$ and $\phi_2(\mathbf{x})$.
\emph{Strictly speaking, these two parameters should
	be referred to as volume fractions.}
The true density and the coefficient of viscosity 
of the fluid are denoted by $\gamma$ and $\mu$,
respectively. The bulk densities in the macro-pores 
and micro-pores are, respectively, denoted by 
$\rho_1(\mathbf{x})$ and $\rho_2(\mathbf{x})$. 
That is,
\begin{align}
	\rho_1(\mathbf{x}) = \phi_{1}(\mathbf{x}) \gamma
	\quad \mathrm{and} \quad 
	\rho_2(\mathbf{x}) = \phi_{2}(\mathbf{x}) \gamma
\end{align}

The pressure scalar fields in the macro-pore and 
micro-pore networks are, respectively, denoted by 
$p_{1}(\mathbf{x})$ and $p_{2}(\mathbf{x})$. The 
true (or seepage) velocity vector fields in the
two pore-networks are denoted by $\mathbf{v}_{1}
(\mathbf{x})$ and $\mathbf{v}_{2}(\mathbf{x})$. 
The discharge (or Darcy) velocities, 
$\mathbf{u}_{1}(\mathbf{x})$ and $\mathbf{u}_{2}(\mathbf{x})$, 
are related to the true velocities as follows:
\begin{align}
	\mathbf{u}_{1}(\mathbf{x}) = \phi_1(\mathbf{x})
	\mathbf{v}_1(\mathbf{x}) 
	\quad \mathrm{and} \quad 
	\mathbf{u}_{2}(\mathbf{x}) = \phi_2(\mathbf{x})
	\mathbf{v}_2(\mathbf{x})
\end{align}

For the macro-pore network, we shall decompose
the boundary into two parts: $\Gamma_{1}^{v}$ and
$\Gamma_{1}^{p}$. $\Gamma_{1}^{v}$ denotes the part
of the boundary on which the normal component
of the velocity in the macro-pore network is 
prescribed. $\Gamma_{1}^{p}$ is that part of 
the boundary on which the pressure in the 
macro-pore network is prescribed. Likewise, 
for the micro-pore network, the boundary is 
decomposed into two parts:~$\Gamma_{2}^{v}$ 
and $\Gamma_{2}^{p}$. 
For mathematical well-posedness, we assume that
\begin{align}
	&\Gamma_{1}^{v} \cup \Gamma_{1}^{p} = \partial \Omega \quad \mathrm{and} \quad 
	\Gamma_{1}^{v} \cap \Gamma_{1}^{p} = \emptyset \nonumber \\
	&\Gamma_{2}^{v} \cup \Gamma_{2}^{p} = \partial \Omega
	\quad \mathrm{and} \quad 
	\Gamma_{2}^{v} \cap \Gamma_{2}^{p} = \emptyset
\end{align}
The governing equations in terms of the true 
velocities can be written as follows: 
\begin{subequations}
	\begin{alignat}{2}
		\label{Eqn:Dual_GE_BLM_1}
		&\mu \phi_{1}^{2} \mathbf{K}_{1}^{-1} \mathbf{v}_1(\mathbf{x})
		+ \phi_{1} \mathrm{grad}[p_1] = \rho_{1} \mathbf{b}(\mathbf{x})
		&&\quad \mathrm{in} \; \Omega \\
		\label{Eqn:Dual_GE_BLM_2}
		&\mu \phi_{2}^{2} \mathbf{K}_{2}^{-1} \mathbf{v}_2(\mathbf{x})
		+ \phi_{2} \mathrm{grad}[p_2] = \rho_{2} \mathbf{b}(\mathbf{x})
		&&\quad \mathrm{in} \; \Omega \\
		\label{Eqn:Dual_GE_mass_balance_1}
		&\mathrm{div}[\phi_1 \mathbf{v}_1] = +\chi(\mathbf{x}) 
		&&\quad \mathrm{in} \; \Omega \\
		\label{Eqn:Dual_GE_mass_balance_2}
		&\mathrm{div}[\phi_2 \mathbf{v}_2] = -\chi(\mathbf{x}) 
		&&\quad \mathrm{in} \; \Omega \\
		\label{Eqn:Dual_GE_vBC_1}
		&\mathbf{v}_1(\mathbf{x}) \cdot \widehat{\mathbf{n}}(\mathbf{x})
		= v_{n1}(\mathbf{x})
		&&\quad \mathrm{on} \; \Gamma^{v}_{1} \\
		\label{Eqn:Dual_GE_vBC_2}
		&\mathbf{v}_2(\mathbf{x}) \cdot \widehat{\mathbf{n}}(\mathbf{x})
		= v_{n2}(\mathbf{x})
		&&\quad \mathrm{on} \; \Gamma^{v}_{2} \\
		\label{Eqn:Dual_GE_pBC_1}
		&p_1(\mathbf{x}) = p_{01} (\mathbf{x})
		&&\quad \mathrm{on} \; \Gamma^{p}_{1} \\
		\label{Eqn:Dual_GE_pBC_2}
		&p_2(\mathbf{x}) = p_{02} (\mathbf{x})
		&&\quad \mathrm{on} \; \Gamma^{p}_{2} 
	\end{alignat}
\end{subequations}
where $\mathbf{b}(\mathbf{x})$ is the specific body
force. $v_{n1}(\mathbf{x})$ is the prescribed
normal component of the velocity on the boundary in the
macro-pores, and $v_{n2}(\mathbf{x})$ is the prescribed normal
component of the velocity on the boundary in the micro-pores. 
$p_{01}(\mathbf{x})$ is the prescribed pressure on the boundary
in the macro-pores, and $p_{02}(\mathbf{x})$ is the prescribed
pressure on the boundary in the micro-pores.
$\chi(\mathbf{x})$ is the rate of volume of the fluid 
that is exchanged between the two pore-networks per 
unit volume of the porous medium. \textit{In the rest of the 
	paper, $\chi(\mathbf{x})$ is simply referred to as 
	the mass transfer.} 
Herein, the mass transfer is modeled as follows:
\begin{align}
	\label{Eqn:Dual_GE_mass_transfer}
	\chi(\mathbf{x}) = - \frac{\beta}{\mu}
	(p_1(\mathbf{x}) - p_2(\mathbf{x}))
\end{align}
where $\beta$ is a dimensionless characteristic 
of the porous medium. The above expression for 
the mass transfer can be traced back to 
\citep{Barenblatt_Zheltov_Kochina_v24_p1286_1960_ZAMM}, 
which was derived based on a dimensional analysis argument. 
Some works in the literature refer to such an expression
for the interpososity flow as the ``Barenblatt-Zheltov"
model, for example \citep{chen1989transient}. Under this model, it is assumed that the fluid can be exchanged between the two pore-networks if there exists a sufficiently smooth change of pressure between the networks. Although such an interporosity flow equation seems simple, it has been proven to maintain the essential features of flow through the naturally fractured reservoirs\citep{chen1989transient}.
To provide a physical insight into $\beta$, consider
that the two pore-networks are connected by conduits
with radius $R$ and length $L$. Then $\beta = R^2/(8 L^2)$. 
If the two pore-networks are connected by 
fissures, which can be idealized as parallel 
plates with length $L$ and separated by a 
width of $h$, then $\beta = h^2/(12 L^2)$.
These expressions are obtained by assuming
Poiseuille flow in conduits and Couette flow in
fissures. In reality, the two pore-networks can
be connected by both conduits and fissures, and
these connectors can even be tortuous.

An alternate form of the governing equations, which is
particularly convenient for numerical formulations,
is written as follows in terms of discharge velocities:
\begin{subequations}
	\begin{alignat}{2}
		\label{Eqn:Dual_GE_Darcy_BLM_1}
		&\mu \mathbf{K}_{1}^{-1} \mathbf{u}_1(\mathbf{x})
		+ \mathrm{grad}[p_1] = \gamma \mathbf{b}(\mathbf{x})
		&&\quad \mathrm{in} \; \Omega \\
		\label{Eqn:Dual_GE_Darcy_BLM_2}
		&\mu \mathbf{K}_{2}^{-1} \mathbf{u}_2(\mathbf{x})
		+ \mathrm{grad}[p_2] = \gamma \mathbf{b}(\mathbf{x})
		&&\quad \mathrm{in} \; \Omega \\
		\label{Eqn:Dual_GE_Darcy_mass_balance_1}
		&\mathrm{div}[\mathbf{u}_1] = +\chi(\mathbf{x})
		&&\quad \mathrm{in} \; \Omega \\
		\label{Eqn:Dual_GE_Darcy_mass_balance_2}
		&\mathrm{div}[\mathbf{u}_2] = -\chi(\mathbf{x})
		&&\quad \mathrm{in} \; \Omega \\
		&\mathbf{u}_1(\mathbf{x}) \cdot
		\widehat{\mathbf{n}}(\mathbf{x})
		= \phi_{1} v_{n1}(\mathbf{x}) =: u_{n1}(\mathbf{x})
		&&\quad \mathrm{on} \; \Gamma^{v}_{1} \\
		&\mathbf{u}_2(\mathbf{x}) \cdot
		\widehat{\mathbf{n}}(\mathbf{x})
		= \phi_{2} v_{n2}(\mathbf{x}) =: u_{n2}(\mathbf{x}) 
		&&\quad \mathrm{on} \; \Gamma^{v}_{2} \\
		&p_1(\mathbf{x}) = p_{01} (\mathbf{x})
		&&\quad \mathrm{on} \; \Gamma^{p}_{1} \\
		\label{Eqn:Dual_GE_Darcy_pBC_2}
		&p_2(\mathbf{x}) = p_{02} (\mathbf{x})
		&&\quad \mathrm{on} \; \Gamma^{p}_{2} 
	\end{alignat}
\end{subequations}

\section{PROPOSED APPROACH TO DEVELOP DOUBLE POROSITY/PERMEABILITY MODELS}
\label{Sec:S3_Dual_Basis}
Several porous media models have been developed using 
the theory of interacting continua for flow, reactive-transport 
and/or deformation of multiple constituents in a single 
pore-network by treating each component to be  either 
a fluid, a solid or a chemical species. These works 
include \citep{Bowen,pekavr2014thermodynamics,
	de2012theory,Atkin_Craine_QJMAM_1976_v29_p209}, 
just to name a few. However, to the best of our knowledge, 
the theory of interacting continua has not been used to 
obtain models when the porous media exhibit multiple 
pore-networks.

The maximization of rate of dissipation hypothesis,
which is also referred to as the orthogonality principle
and is similar in spirit to the maximization of entropy
production, has been first proposed by Ziegler to derive
the constitutive relations \citep{Ziegler}. 
An attractive feature of this hypothesis is that 
prescription of two physically meaningful 
functionals (Helmholtz potential and dissipation 
functional) provides the constitutive relations even 
for a phenomenon which involves a multitude 
of interacting processes \citep{ziegler1987derivation}.
Subsequently, this hypothesis has been successfully 
employed to develop constitutive models for a wide 
variety of physical phenomena, which include inelasticity 
\citep{srinivasa2009inelasticity}, anisotropic fluids 
\citep{rajagopal2001modeling}, degradation of materials 
\citep{xu2016material}, and diffusion in viscoelastic polymers 
\citep{karra2013modeling}. However, this hypothesis has not
been utilized to derive constitutive relations for porous
media with multiple pore-networks. 

\emph{Thus, one of the goals of this paper is to 
	combine the theory of interacting continua and 
	the maximization of rate of dissipation hypothesis
	for obtaining a coherent framework to derive models
	of flow in porous media with multiple pore-networks.} 

\subsection{Theory of interacting continua:~A general setting} 
The porous medium is treated as a mixture of $\mathcal{N}$ 
constituents. We use the word ``constituent'' to refer to
the porous solid or a pore-network. This usage is slightly
different from the usual mixture theory models. In a typical
mixture theory model, a constituent refers to a different
physical/chemical component or a different phase. 

We denote the bulk density, specific body force, partial 
Cauchy stress, specific internal energy, specific Helmholtz 
potential, temperature, heat flux vector and specific entropy 
of the $i$-th constituent by $\rho_{i}$, $\mathbf{b}_{i}$, 
$\mathbf{T}_{i}$, $U_{i}$, $A_{i}$, $\theta_{i}$, 
$\mathbf{q}_{i}$ and $\eta_{i}$; respectively.

\subsubsection{Kinematics}
We denote the time by $t$. Under the theory of interacting 
continua, a mixture is treated as a superposition of multiple 
continua each following its own motion.
At a given instance of time, each spatial point $\mathbf{x}$ 
in the mixture is occupied simultaneously by $\mathcal{N}$ 
different particles $\mathbf{p}_{i}$ ($i = 1, \cdots, \mathcal{N}$), 
one from each constituent. The motion of the constituents can 
be written as:
\begin{align}
	\mathbf{x} = \boldsymbol{\varphi}_{i}(\mathbf{p}_i,t) 
	\quad i = 1, \cdots, \mathcal{N} 
\end{align}
with the corresponding velocities defined as follows: 
\begin{align}
	\mathbf{v}_{i} = \frac{\partial \boldsymbol{\varphi}_{i}(\mathbf{p}_i,t)}{\partial t} 
\end{align}
The gradient of motion of the $i$-th constituent
is denoted by $\mathbf{F}_{i}$. That is,
\begin{align}
	\mathbf{F}_i = \frac{\partial \mathbf{x}}{\partial \mathbf{p}_i}
\end{align}
Let 
\begin{align}
	\mathbf{L}_i := \mathrm{grad}[\mathbf{v}_i]
\end{align}
and
\begin{align}
	\mathbf{D}_i := \mathrm{sym}[\mathbf{L}_{i}] =
	\frac{1}{2} \left(\mathrm{grad}[\mathbf{v}_i]
	+ \mathrm{grad}[\mathbf{v}_i]^{\mathrm{T}} \right)
\end{align}
We introduce the following material time
derivative defined on arbitrary scalar
field $\psi$ and vector field $\mathbf{w}$: 
\begin{align}
	\frac{D^{(i)} \psi}{Dt} = \frac{\partial \psi}{\partial t} 
	+ \mathbf{v}_{i} \cdot \mathrm{grad}[\psi] 
\end{align}
and
\begin{align}
	\frac{D^{(i)} \mathbf{w}}{Dt} = \frac{\partial \mathbf{w}}{\partial t} 
	+ \mathrm{grad}[\mathbf{w}] \mathbf{v}_{i}
\end{align}
It is important to note that the material
derivative $D^{(i)}(\cdot)/Dt$ follows the
motion of the $i$-th constituent. 

\subsubsection{Balance laws}
The local form of the \emph{balance of mass} of
the $i$-th constituent can be written as follows:
\begin{align}
	\label{Eqn:BoM_pore}
	\frac{\partial \rho_{i}}{\partial t}
	+ \mathrm{div}[\rho_{i} \mathbf{v}_{i}]
	= m_{i}
\end{align}
where $m_{i}$ is the rate of mass transfer
into the $i$-th pore-network per unit volume
of the porous medium. The local form of the 
overall balance of mass for the porous 
medium takes the following form:
\begin{align}
	\label{Eqn:BoM_mixture}
	\sum_{i=1}^{\mathcal{N}} m_{i} = 0
\end{align}

Under the theory of interacting continua, the mechanical 
interaction between constituents is modeled using interaction 
terms \citep{Atkin_Craine_QJMAM_1976_v29_p209}. Herein, 
we denote the interaction term for the $i$-th constituent due to 
the presence of other constituents by $\mathbf{i}_{i}$.
The \emph{balance of linear momentum} of the $i$-th 
constituent, by taking into account the balance of mass (i.e.,
equation \eqref{Eqn:BoM_pore}), takes the following
form:
\begin{align}
	\label{Eqn:BoLM_constituent}
	\rho_{i} \frac{D^{(i)} \mathbf{v}_{i}}{Dt}
	= \mathrm{div}[\mathbf{T}_{i}] 
	+ \rho_{i} \mathbf{b}_{i} + \mathbf{i}_{i}
\end{align}
The local form of the overall balance of linear 
momentum for the porous medium takes the 
following form:
\begin{align}
	\label{Eqn:BoLM_mixture}
	\sum_{i=1}^{\mathcal{N}} \mathbf{i}_{i} = \mathbf{0}
\end{align}
We assume a stronger version of the \emph{balance of 
	angular momentum} for each constituent by asserting 
that 
\begin{align}
	\label{Eqn:BoAM_constituent}
	\mathbf{T}_{i} = \mathbf{T}_{i}^{\mathrm{T}} 
	\quad \forall i = 1, \cdots, \mathcal{N}
\end{align}

The \emph{balance of energy} of the $i$-th constituent,
by taking into account the balance of mass (i.e., equation 
\eqref{Eqn:BoM_pore}) and the balance of 
the linear momentum (i.e., equation \eqref{Eqn:BoLM_constituent}), 
takes the following form:
\begin{align}
	\label{Eqn:BoE_constituent}
	\rho_{i} \frac{D^{(i)} U_{i}}{Dt}
	= \mathbf{T}_{i} \cdot \mathbf{L}_{i}
	- \mathrm{div}\left[\mathbf{q}_{i}\right] 
	+ \rho_{i} r_{i} + \varepsilon^{s}_{i}
\end{align}
where $\varepsilon^{s}_{i}$ is energy supply to the 
$i$-th constituent due to the interaction with other 
constituents, and $r_{i}$ is the (external) specific 
heat supply to the $i$-th constituent. 
The local form of the overall balance of energy 
for the porous media takes the following form:
\begin{align}
	\label{Eqn:BoE_mixture}
	\sum_{i=1}^{\mathcal{N}} \left(\varepsilon^{s}_{i} + 
	\mathbf{i}_{i} \cdot \mathbf{v}_{i} \right) = 0
\end{align}

The \emph{second law of thermodynamics}, 
which is a global law, is written as follows: 
\begin{align}
	\frac{\partial}{\partial t} \sum_{i=1}^{\mathcal{N}} 
	\int_{\Omega} \rho_{i} \eta_{i} \mathrm{d} \Omega 
	&+ \sum_{i=1}^{\mathcal{N}}  \int_{\partial \Omega} 
	\rho_{i} \eta_{i} \mathbf{v}_{i} \cdot \widehat{\mathbf{n}} 
	\; \mathrm{d} \Gamma 
	\geq
	- \sum_{i=1}^{\mathcal{N}} \int_{\partial \Omega} 
	\frac{\mathbf{q}_{i} \cdot \widehat{\mathbf{n}} }{\theta_{i}}
	\; \mathrm{d} \Gamma \nonumber \\
	&+ \sum_{i=1}^{\mathcal{N}} \int_{\Omega} \frac{\rho_i r_{i}}{\theta_{i}} 
	\; \mathrm{d} \Omega
	+ \sum_{i=1}^{\mathcal{N}} \int_{\Omega} m_{i} \eta_{i}  
	\mathrm{d} \Omega
\end{align} 

Recall that $\widehat{\mathbf{n}}$ denotes 
the outward normal to the boundary. The 
above inequality can be considered as an 
extension of the Clausius-Duhem inequality  
to multi-constituent media. 
We assume the local form to hold, which is stronger 
than the second law of thermodynamics. The local form 
corresponding to the above inequality reads: 
\begin{align}
	\frac{\partial}{\partial t}  \sum_{i=1}^{\mathcal{N}}
	\rho_{i} \eta_{i}  
	+ \sum_{i=1}^{\mathcal{N}}  \mathrm{div}\left[
	\rho_{i} \eta_{i} \mathbf{v}_{i}  
	\right]
	\geq
	\sum_{i=1}^{\mathcal{N}} \left(-
	\mathrm{div}\left[\frac{\mathbf{q}_{i}}{\theta_{i}}
	\right]
	+ \frac{\rho_i r_{i}}{\theta_{i}} 
	+ m_{i} \eta_{i}  \right)
\end{align} 
Using the balance of mass (i.e., equation \eqref{Eqn:BoM_pore}), 
the above inequality can be simplified as follows: 
\begin{align}
	\label{Eqn:second_law_local}
	\sum_{i=1}^{\mathcal{N}}
	\left(\rho_{i} \frac{D^{(i)} \eta_{i}}{Dt}
	+ \mathrm{div}\left[\frac{\mathbf{q}_{i}}{\theta_{i}}
	\right]
	- \frac{\rho_{i}  r_{i}}{\theta_{i}} \right) \geq 0
\end{align} 
By diving both sides of equation \eqref{Eqn:BoE_constituent} 
by $\theta_i$, summing over the number of constituents, 
and subtracting the result from the inequality 
\eqref{Eqn:second_law_local}, we obtain the following inequality: 
\begin{align}
	\label{Eqn:second_law_simplified}
	\sum_{i=1}^{\mathcal{N}}
	\rho_{i} &\left(\frac{D^{(i)} \eta_{i}}{Dt} 
	- \frac{1}{\theta_{i}}\frac{D^{(i)} U_{i}}{Dt} \right)
	\geq \sum_{i=1}^{\mathcal{N}} \frac{1}{\theta_i} 
	\left(-\mathbf{T}_i \cdot \mathbf{L}_i 
	+ \frac{1}{\theta_i} \mathbf{q}_{i} \cdot \mathrm{grad}[\theta_i]
	- \varepsilon^{s}_{i}
	\right)
\end{align}
We now replace the specific internal energy with 
the specific Helmholtz potential using a Legendre 
transformation, which can be mathematically written 
as follows:
\begin{align}
	U_{i} = A_{i} + \theta_{i} \eta_{i} \quad \mathrm{with} \quad 
	\eta_{i} = -\frac{\partial A_{i}}{\partial \theta_i}
\end{align}
We assume the functional dependence of the 
specific Helmholtz potential to be $A_{i} = 
A_{i}(\mathbf{F}_i,\theta_i)$. Noting that 
\begin{align}
	\frac{D^{(i)} \mathbf{F}_{i}}{Dt} = \mathbf{L}_{i} 
	\mathbf{F}_{i}
\end{align} 
and using equation \eqref{Eqn:BoE_mixture}, 
inequality \eqref{Eqn:second_law_simplified} 
can be written as follows: 
\begin{align}
	\sum_{i=1}^{\mathcal{N}} \frac{1}{\theta_i} 
	&\left(\rho_{i} \frac{\partial A_{i}}{\partial \mathbf{F}_i}
	\mathbf{F}_{i}^{\mathrm{T}} -\mathbf{T}_i \right) \cdot \mathbf{L}_i
	+ \sum_{i=1}^{\mathcal{N}} \frac{1}{\theta_i}\left(\frac{1}{\theta_i} \mathbf{q}_{i} \cdot \mathrm{grad}[\theta_i]
	+ \mathbf{i}_i \cdot \mathbf{v}_{i}
	\right) \leq 0
\end{align}
The above inequality can be converted into a 
convenient equality by introducing a non-negative 
functional, $\Psi \geq 0$, and the resulting 
equality reads: 
\begin{align}
	\sum_{i=1}^{\mathcal{N}} \frac{1}{\theta_i} 
	&\left(\rho_{i} \frac{\partial A_{i}}{\partial \mathbf{F}_i}
	\mathbf{F}_{i}^{\mathrm{T}} -\mathbf{T}_i \right) \cdot \mathbf{L}_i 
	+ \sum_{i=1}^{\mathcal{N}} \frac{1}{\theta_i}\left(\frac{1}{\theta_i} \mathbf{q}_{i} \cdot \mathrm{grad}[\theta_i]
	+ \mathbf{i}_{i}\cdot\mathbf{v}_i
	\right) + \Psi = 0
\end{align}
where $\Psi$ is the rate of entropy production per 
unit volume. The above equation is referred to as 
the \emph{reduced energy-entropy equation}. 
If \emph{all} the constituents have the \emph{same} 
temperature, $\theta_i = \theta$, (i.e., the mixture is in thermal 
equilibrium) then the above equation can be 
written as follows: 
\begin{align}
	\label{Eqn:reduced_energy_dissipation_equation}
	\sum_{i=1}^{\mathcal{N}} 
	\left(\rho_{i} \frac{\partial A_{i}}{\partial \mathbf{F}_i}
	\mathbf{F}_{i}^{\mathrm{T}} -\mathbf{T}_i \right) \cdot \mathbf{L}_i 
	+ \sum_{i=1}^{\mathcal{N}} \left(\frac{1}{\theta} \mathbf{q}_{i} \cdot \mathrm{grad}[\theta]
	+ \mathbf{i}_{i}\cdot\mathbf{v}_i
	\right) + \zeta = 0
\end{align}
where the rate of dissipation per unit volume is 
defined as follows: 
\begin{align}
	\zeta = \theta \Psi
\end{align}
Since $\Psi \geq 0$ and $\theta > 0$, $\zeta \geq 0$. Equation 
\eqref{Eqn:reduced_energy_dissipation_equation} is referred 
to as the \emph{reduced energy-dissipation equation} for 
multi-constituent media. 

\subsection{A simplified framework for double porosity/permeability models}
The above framework is presented in a general setting. We now 
provide a simplified framework for the problem at hand, which
pertains to the modeling of isothermal flow of an incompressible
fluid in rigid porous media with two pore-networks. To this end,
the following choices are made:
\begin{enumerate}[(i)]
\item There are two pore-networks and a rigid porous medium. Strictly speaking, 
there are three constituents. Since the porous solid is rigid, its motion will be 
neglected and all the balance laws for this constituent are assumed to be 
trivially satisfied. 
Hence, one can take $\mathcal{N} = 2$.
\item All constituents have the same temperature (i.e., $\theta_{i} = \theta$). 
\item There is no heat transfer. That is, $\mathbf{q}_{i} = \mathbf{0}$ 
and $r_{i} = 0 $.
\item We assume that the porosities do not change with time. 
This is acceptable, as the porous solid is assumed to be rigid. 
That is, 
\begin{align}
	\label{Eqn:porosities_not_time}
	\frac{\partial \phi_i}{\partial t} = 0 \quad (i = 1, 2)
\end{align}
\item The fluid in each pore-network is incompressible, 
which, mathematically, translates into the following equations: 
\begin{align}
	\frac{D^{(i)} \gamma}{Dt} \equiv 
	\frac{\partial \gamma}{\partial t} + \mathbf{v}_{i} \cdot \mathrm{grad}\left[\gamma\right] = 0 \quad (i=1,~2)
\end{align}
Noting the above relation, equation \eqref{Eqn:porosities_not_time} 
and the balance of the mass 
for the mixture (i.e., equation \eqref{Eqn:BoM_mixture})
imply that the balance of mass for an incompressible 
fluid in each pore-network can be written as follows: 
\begin{align}
	\mathrm{div}[\phi_1 \mathbf{v}_1] = +\chi 
	\quad \mathrm{and} \quad
	\mathrm{div}[\phi_2 \mathbf{v}_2] = -\chi
\end{align}
where $\chi = m_{1}/\gamma = -m_{2}/\gamma$ 
accounts for the mass transfer from the macro-pore 
network to the micro-pore network. Note that these 
incompressibility constraints remain the same in 
both transient and steady-state responses. 
\item The velocity in each pore-network and its (spatial) 
gradient are assumed to be small so that the term 
``$\mathrm{grad}[\mathbf{v}_i] \mathbf{v}_i$" can be 
neglected. Then, the balance of linear momentum 
in each pore-network for a transient response reads: 
\begin{align}
	\rho_1 \frac{\partial \mathbf{v}_1}{\partial t} 
	= \mathrm{div}[\mathbf{T}_1] + \rho_1
	\mathbf{b}(\mathbf{x}) + \mathbf{i}_1  
\end{align}
and
\begin{align}  
	\rho_2 \frac{\partial \mathbf{v}_2}{\partial t} 
	= \mathrm{div}[\mathbf{T}_2] + \rho_2
	\mathbf{b}(\mathbf{x}) + \mathbf{i}_2
\end{align}
and the corresponding ones in a steady-state 
response reads: 
\begin{align}
	\mathrm{div}[\mathbf{T}_1] + \rho_1
	\mathbf{b}(\mathbf{x}) + \mathbf{i}_1 = \mathbf{0} 
\end{align}
and
\begin{align} 
	\mathrm{div}[\mathbf{T}_2] + \rho_2
	\mathbf{b}(\mathbf{x}) + \mathbf{i}_2 = \mathbf{0} 
\end{align}

Note that the balance of linear momentum for the mixture 
(i.e., equation \eqref{Eqn:BoLM_mixture}) does \emph{not} 
imply that the interaction terms of both the pore-networks 
add up to zero. One should not forget about the porous 
solid. Although we have assumed the porous solid to be rigid and
have not documented the balance laws pertaining to it, it does have an interaction term. The sum 
of all the three interaction 
terms (one for each pore-network and one for the porous 
solid) should add up to zero, which is according to the 
balance of linear momentum for the mixture.
\item We assume that the specific Helmholtz potentials 
satisfy the frame-indifference \citep{Truesdell}. This will imply 
that the tensor $\rho_i (\partial A_{i}/\partial \mathbf{F}_i)\mathbf{F}_i^{\mathrm{T}}$ is 
symmetric. The balance of angular momentum for each constituent 
implies that the partial Cauchy stress tensor, $\mathbf{T}_i$, is 
symmetric. The symmetry of these tensors imply that the reduced 
energy-dissipation can be written as follows:
\begin{align}
	&-\left(\mathbf{T}_1 - \rho_1 \frac{\partial A_1}{\partial \mathbf{F}_1}
	\mathbf{F}_{1}^{\mathrm{T}}\right)
	\cdot \mathbf{D}_1
	- \left(\mathbf{T}_2 - \rho_2 \frac{\partial A_2}{\partial \mathbf{F}_2}
	\mathbf{F}_{2}^{\mathrm{T}} \right)
	\cdot \mathbf{D}_2 
	+ \mathbf{i}_1 \cdot \mathbf{v}_1 
	+ \mathbf{i}_2 \cdot \mathbf{v}_2 + \zeta = 0 
\end{align}

\end{enumerate}

We now obtain the constitutive relations for the 
Cauchy stresses, interaction terms, and the 
mass transfer across the pore-networks using the
maximization of rate of dissipation hypothesis.
\subsubsection{Obtaining constitutive relations using maximization of rate of dissipation}
We handle the mass transfer across the pore-networks
using an internal variable, which will be taken as
follows:
\begin{align}
	\int_{0}^{t} \chi(\mathbf{x},\tau) \mathrm{d} \tau 
\end{align}
Then the rate of the chosen internal
variable will be 
\begin{align}
	\frac{d}{dt} \int_{0}^{t} \chi(\mathbf{x},\tau) \mathrm{d} \tau
	= \chi(\mathbf{x},t)
\end{align}
The mathematical statement of the maximization of rate
of dissipation hypothesis for multi-constituent media
can be written as follows: 
{\small
	\begin{subequations}
		\begin{align}
			\mathop{\mathrm{maximize}}_{\mathbf{D}_1,\mathbf{D}_2,
				\mathbf{v}_1,\mathbf{v}_2,\chi} 
			\quad & \zeta = \widehat{\zeta}(\mathbf{D}_1,
			\mathbf{D}_2,\mathbf{v}_1,\mathbf{v}_2,\chi) \\
			\mbox{subject to} \quad
			\label{Eqn:Dual_MREP_reduced_energy_dissipation} 
			&-\left(\mathbf{T}_1 - \rho_1 \frac{\partial A_1}{\partial \mathbf{F}_1}
			\mathbf{F}_{1}^{\mathrm{T}} \right)
			\cdot \mathbf{D}_1
			- \left(\mathbf{T}_2 - \rho_2 \frac{\partial A_2}{\partial \mathbf{F}_2}
			\mathbf{F}_{2}^{\mathrm{T}} \right)
			\cdot \mathbf{D}_2 
			\nonumber \\
			& + \mathbf{i}_1 \cdot \mathbf{v}_1 
			+ \mathbf{i}_2 \cdot \mathbf{v}_2 + \zeta = 0 \\
			\label{Eqn:Dual_MREP_macro_MT} 
			& \quad \phi_{1} \mathrm{tr}[\mathbf{D}_1] 
			+ \mathbf{v}_1 \cdot \mathrm{grad}\left[\phi_1\right] = +\chi \\
			\label{Eqn:Dual_MREP_micro_MT} 
			& \quad \phi_{2} \mathrm{tr}[\mathbf{D}_2]
			+ \mathbf{v}_2 \cdot \mathrm{grad}\left[\phi_2\right] = -\chi
		\end{align}
	\end{subequations}
}
Equation \eqref{Eqn:Dual_MREP_reduced_energy_dissipation} 
is the reduced energy-dissipation equation for
a two pore-network porous medium, and equations
\eqref{Eqn:Dual_MREP_macro_MT} and 
\eqref{Eqn:Dual_MREP_micro_MT} are, respectively,
the incompressibility constraints for the macro
and micro pore-networks.
Using the Lagrange multiplier method, one can rewrite
the above constrained optimization problem as the
following unconstrained optimization problem:
{\small 
	\begin{align}
		&\mathop{\mathrm{extremize}}_{\mathbf{D}_1,\mathbf{D}_2,
			\mathbf{v}_1,\mathbf{v}_2,\chi,p_1,p_2,\lambda} 
		\quad 
		\nonumber \\    
		&\zeta
		+ p_1 \left(\phi_{1} \mathrm{tr}[\mathbf{D}_1] 
		+ \mathbf{v}_1 \cdot \mathrm{grad}\left[\phi_1\right] - \chi \right) 
		\nonumber \\
		&\; \; +p_2 \left(\phi_{2} \mathrm{tr}[\mathbf{D}_2]
		+ 
		\mathbf{v}_2 \cdot \mathrm{grad}\left[\phi_2\right] + \chi \right) \nonumber \\
		&\;\;+ \lambda \left( 
		-\left(\mathbf{T}_1 - \rho_1 \frac{\partial A_1}{\partial \mathbf{F}_1}
		\mathbf{F}_{1}^{\mathrm{T}} \right)
		\cdot \mathbf{D}_1 \mathbf{i}_1 \cdot \mathbf{v}_1\right. \nonumber \\
		&\left. \quad \quad \; \; - \left(\mathbf{T}_2 - \rho_2 \frac{\partial A_2}{\partial \mathbf{F}_2}
		\mathbf{F}_{2}^{\mathrm{T}} \right)
		\cdot \mathbf{D}_2 
		+ \mathbf{i}_2 \cdot \mathbf{v}_2 + \zeta\right)
	\end{align}
}
where $\lambda$ is the Lagrange multiplier
corresponding to the reduced energy dissipation
equation \eqref{Eqn:Dual_MREP_reduced_energy_dissipation},
and $p_1$ and $p_2$ are the Lagrange multipliers enforcing
equations \eqref{Eqn:Dual_MREP_macro_MT} and
\eqref{Eqn:Dual_MREP_micro_MT}, respectively.
The first-order optimality conditions of
the above optimization problem yield:
\begin{subequations}
	\begin{align}
		\label{Eqn:Double_T1_relation}
		& \mathbf{T}_1 = -\phi_1 p_1 \mathbf{I}
		+ \rho_1 \frac{\partial A_1}{\partial \mathbf{F}_1}
		\mathbf{F}_{1}^{\mathrm{T}} 
		+ \left(\frac{\lambda + 1}{\lambda}\right)
		\frac{\partial \zeta}{\partial \mathbf{D}_1} \\
		& \mathbf{T}_2 = -\phi_2 p_2 \mathbf{I}
		+ \rho_2 \frac{\partial A_2}{\partial \mathbf{F}_2}
		\mathbf{F}_{2}^{\mathrm{T}} 
		+ \left(\frac{\lambda + 1}{\lambda}\right)
		\frac{\partial \zeta}{\partial \mathbf{D}_2} \\
		& \mathbf{i}_1 = \mathrm{grad}\left[\phi_1\right]p_1
		- \left(\frac{\lambda + 1}{\lambda}\right)
		\frac{\partial \zeta}{\partial \mathbf{v}_1} \\
		& \mathbf{i}_2 = \mathrm{grad}\left[\phi_2\right]p_2
		- \left(\frac{\lambda + 1}{\lambda}\right)
		\frac{\partial \zeta}{\partial \mathbf{v}_2} \\
		\label{Eqn:Double_chi_relation}
		&\frac{\partial \zeta}{\partial \chi} = -(p_1 - p_2)
	\end{align}
\end{subequations}
where $\mathbf{I}$ denotes the second-order identity
tensor. Of course, one needs to augment the aforementioned
optimality conditions with the constraints given by equations
\eqref{Eqn:Dual_MREP_reduced_energy_dissipation}--\eqref{Eqn:Dual_MREP_micro_MT}.
Equations
\eqref{Eqn:Double_T1_relation}--\eqref{Eqn:Double_chi_relation}
provide general constitutive relations.
One can obtain a specific constitutive model by specifying 
$A_1$, $A_2$, and $\zeta$ functionals. Moreover, if $\zeta$ 
is a homogeneous functional of order two with respect to its 
arguments, it can be shown that $\lambda = -2$.

\subsubsection{A specific double porosity/permeability model}
One can obtain the double porosity/permeability model 
(as given in Section \ref{Sec:S2_Double_Model}) by 
making the following choices and assumptions:
\begin{enumerate}[(i)]
	\item The specific Helmholtz potentials for the
	two fluid constituents are taken as follows:
	\begin{align}
		A_{1} = 0 \quad \mathrm{and} \quad A_{2} = 0
	\end{align}
	\item The rate of dissipation production is taken as follows:
	\begin{align}
		\label{Eqn:Double_choice_for_zeta}
		\zeta = \mu \phi_{1}^2 \mathbf{v}_{1} \cdot 
		\mathbf{K}_1^{-1} \mathbf{v}_1 + \mu \phi_{2}^2
		\mathbf{v}_{2} \cdot \mathbf{K}_2^{-1} \mathbf{v}_2 
		+ \zeta_{\mbox{MT}}(\chi)
	\end{align}
	where the first and second terms on the right-hand side of
	the equation, respectively, represent the rate of
	dissipation in the macro-pore and micro-pore
	networks, and $\zeta_{\mbox{MT}}$ accounts for the
	dissipation due to mass transfer across the two
	pore-networks.
	\item Assuming that the connectors are conduits 
	or fissures, $\zeta_{\mbox{MT}}$ can be taken as
	follows:
	\begin{align}
		\zeta_{\mbox{MT}}(\chi) = \frac{\mu}{\beta} \chi^2
	\end{align}
	where $\beta$ is a dimensionless characteristic parameter 
	of the porous medium, as mentioned in Section 
	\ref{Sec:S2_Double_Model}.
	Noting the above choice for $\zeta_{\mathrm{MT}}$,
	it is easy to verify that the functional $\zeta$
	given by equation \eqref{Eqn:Double_choice_for_zeta}
	is a homogeneous functional of order two of its
	arguments.
\end{enumerate}
By substituting the above constitutive specifications into equations
\eqref{Eqn:Double_T1_relation}--\eqref{Eqn:Double_chi_relation}, one
obtains the following constitutive relations:
%
\begin{align}
	&\mathbf{T}_1 = - \phi_1 p_1 \mathbf{I}, \quad 
	\mathbf{T}_2 = - \phi_2 p_2 \mathbf{I}, \quad 
	\mathbf{i}_1 = \mu \phi_1^2 \mathbf{K}_1^{-1} \mathbf{v}_1, \nonumber \\
	&\mathbf{i}_2 = \mu \phi_2^2 \mathbf{K}_2^{-1} \mathbf{v}_2 , \quad
	\chi = -(p_1 - p_2)
\end{align}
The above constitutive relations along with
the balance of mass and the balance of linear
momentum give rise to the following equations 
in a steady-state setting:
\begin{subequations}
	\begin{align}
		\label{Eqn:Final_BoLM_1_SS}
		& \mu \phi_1^2 \mathbf{K}_1^{-1} \mathbf{v}_1
		+ \phi_1 \mathrm{grad}\left[p_1\right] =
		\rho_1 \mathbf{b}(\mathbf{x}) \\
		\label{Eqn:Final_BoLM_2_SS}
		& \mu \phi_2^2 \mathbf{K}_2^{-1} \mathbf{v}_2
		+ \phi_2 \mathrm{grad}\left[p_2\right]
		= \rho_2 \mathbf{b}(\mathbf{x}) \\
		\label{Eqn:Double_final_BoM_macro}
		& \mathrm{div}[\phi_1 \mathbf{v}_1] = +\chi \\
		\label{Eqn:Double_final_BoM_micro}
		& \mathrm{div}[\phi_2 \mathbf{v}_2] = -\chi \\
		\label{Eqn:Double_final_MT}
		& \chi = -\frac{\beta}{\mu}(p_1 - p_2)
	\end{align}
\end{subequations}
which are the governing equations under the 
double porosity/permeability model presented 
in Section \ref{Sec:S2_Double_Model}. 

In a transient setting, equations \eqref{Eqn:Final_BoLM_1_SS} 
and \eqref{Eqn:Final_BoLM_1_SS}, will be replaced by the following:
\begin{subequations}
	\begin{align}
		\label{Eqn:Final_BoLM_1_transient}
		& \rho_1 \frac{\partial \mathbf{v}_1}{\partial t} + 
		\mu \phi_1^2 \mathbf{K}_1^{-1} \mathbf{v}_1
		+ \phi_1 \mathrm{grad}\left[p_1\right] =
		\rho_1 \mathbf{b}(\mathbf{x},t) \\
		\label{Eqn:Final_BoLM_1_transient}
		& \rho_2 \frac{\partial \mathbf{v}_2}{\partial t} 
		+ \mu \phi_2^2 \mathbf{K}_2^{-1} \mathbf{v}_2
		+ \phi_2 \mathrm{grad}\left[p_2\right]
		= \rho_2 \mathbf{b}(\mathbf{x},t)
	\end{align}
\end{subequations}
Equations \eqref{Eqn:Double_final_BoM_macro}--\eqref{Eqn:Double_final_MT} 
remain the same even in a transient setting.
However, we need to prescribe the initial
conditions for the velocity in each
pore-network for the transient case.

\subsection{An illustrative generalization:~An extension of the Brinkman model}
The above framework offers an attractive setting for
deriving porous media models in a consistent manner.
In particular, it is possible to obtain generalizations
of the double porosity/permeability model and include
other physical processes. We now illustrate how to 
generalize the Brinkman model 
\citep{Brinkman_ASR_1947_vA1_p81} to
incorporate double pore-networks and
the mass transfer across the pore-networks.
To this end, we make the following choices
for the specific Helmholtz potentials and the
dissipation functional:
\begin{subequations}
	\begin{align}
		&A_1 = A_2 = 0 \\
		&\zeta = \mu \phi_{1}^{2} \mathbf{v}_{1} \cdot
		\mathbf{K}_{1}^{-1} \mathbf{v}_{1} 
		+\mu \phi_{1} \mathbf{D}_{1} \cdot \mathbf{D}_{1}
		+\mu \phi_{2}^{2} \mathbf{v}_{2} \cdot
		\mathbf{K}_{2}^{-1} \mathbf{v}_{2} 
		\nonumber \\
		&\; \; \;+ \mu \phi_{2} \mathbf{D}_{2} \cdot \mathbf{D}_{2}
		+\zeta_{\mathrm{MT}} \\
		&\zeta_{\mathrm{MT}} = \frac{\mu}{\beta} \chi^2
	\end{align}
\end{subequations}
A physical justification of the above choice for $\zeta$ is
as follows: the first term models the dissipation due to
friction at the interface of the porous solid 
and the fluid in
the macro-pore network \citep{Rajagopal_2007}. The second term corresponds to
the dissipation due to friction in the internal layers
of the fluid in the macro-pore network. The third
and fourth terms model the corresponding phenomena
in the micro-pore network. The fifth term models
the dissipation due to mass transfer in the connectors.
The above choices give rise to the following
constitutive relations:
\begin{align}
	&\mathbf{T}_{1} = -\phi_{1} p_1 \mathbf{I} + 2 \mu \phi_{1} \mathbf{D}_1, \quad 
	\mathbf{i}_{1} =  \mu \phi_{1}^{2} \mathbf{K}_{1}^{-1} \mathbf{v}_1, \quad
	\nonumber \\
	& \mathbf{T}_{2} = -\phi_{2} p_2 \mathbf{I} + 2 \mu \phi_{2} \mathbf{D}_2, \quad 
	\mathbf{i}_{2} =  \mu \phi_{2}^{2} \mathbf{K}_{2}^{-1} \mathbf{v}_2 
\end{align}
The balance of linear momentum for the two pore-networks becomes: 
\begin{subequations}
	\begin{align}
		&\mu \phi_{1}^{2} \mathbf{K}_{1}^{-1} \mathbf{v}_1
		+ \mathrm{grad}[\phi_{1} p_1]
		- \mathrm{div}[2 \mu \phi_1 \mathbf{D}_1]
		= \rho_{1} \mathbf{b}(\mathbf{x}) \\
		&\mu \phi_{2}^{2} \mathbf{K}_{2}^{-1} \mathbf{v}_{2}
		+ \mathrm{grad}[\phi_{2} p_2]
		- \mathrm{div}[2 \mu \phi_2 \mathbf{D}_2]
		= \rho_{2} \mathbf{b}(\mathbf{x}) 
	\end{align}
\end{subequations}
The equations for the balance of mass for the two
pore-networks and the rate of mass transfer across
the pore-networks (i.e., equations \eqref{Eqn:Double_final_BoM_macro},
\eqref{Eqn:Double_final_BoM_micro} and \eqref{Eqn:Double_final_MT})
remain the same. These governing equations provide a consistent
generalization of the classical Brinkman model and the double
porosity/permeability model.

\section{MATHEMATICAL PROPERTIES}
\label{Sec:S4_Dual_Mathematical}
In this section, we shall establish various
mathematical properties that are 
satisfied by the solutions to the double porosity/permeability 
model. The mathematical proofs to these properties are provided	
in the supplementary material. These results are of very high
theoretical significance. In addition,
they can serve as valuable mechanics-based
\emph{a posteriori} measures of the accuracy 
of numerical solutions of the governing equations. 
The latter aspect is illustrated in a subsequent
paper \citep{Nakshatrala_Joodat_Ballarini_P2}.
We now introduce the required mathematical
machinery.

The body force is said to be a conservative 
vector field if there exists a scalar field 
$\psi$ such that
\begin{equation*}
	\label{Eqn:Dual_Conservative_field}
	\gamma \mathbf{b}(\mathbf{x}) = -\mathrm{grad}\left[\psi\right]
\end{equation*}
We shall assume a pair of vector
fields $\left(\widetilde{\mathbf{v}}_1,
\widetilde{\mathbf{v}}_2\right)$ to be
\emph{kinematically admissible} if the
following conditions are met:
\begin{subequations}
	\begin{align}
		&\mathrm{div}\left[\phi_{1}\widetilde{\mathbf{v}}_{1}\right] 
		+ \mathrm{div}\left[\phi_{2}\widetilde{\mathbf{v}}_{2}\right] 
		= 0 \\
		&\widetilde{\mathbf{v}}_{1}(\mathbf{x}) \cdot 
		\widehat{\mathbf{n}}(\mathbf{x}) = v_{n1}(\mathbf{x}) \\
		&\widetilde{\mathbf{v}}_{2}(\mathbf{x}) \cdot 
		\widehat{\mathbf{n}}(\mathbf{x}) = v_{n2}(\mathbf{x})
	\end{align}
\end{subequations}
Note that a kinematically admissible pair need not satisfy
the governing equations for the balance of linear momentum for
each pore-network (i.e., equations \eqref{Eqn:Dual_GE_BLM_1}
and \eqref{Eqn:Dual_GE_BLM_2}), or the pressure boundary
conditions (i.e., equations \eqref{Eqn:Dual_GE_pBC_1} and
\eqref{Eqn:Dual_GE_pBC_2}). Moreover, it is important to
note that the kinematically admissible pair need not satisfy
the mass balance equations \emph{individually} (i.e., equations
\eqref{Eqn:Dual_GE_mass_balance_1} and
\eqref{Eqn:Dual_GE_mass_balance_2}).
We shall assume $(\mathbf{v}_1(\mathbf{x}),\mathbf{v}_2
(\mathbf{x}))$ to be the pair of true velocity fields
if they satisfy all the governing equations under the
double porosity/permeability model (i.e., equations
\eqref{Eqn:Dual_GE_BLM_1}--\eqref{Eqn:Dual_GE_pBC_2}).
For convenience, we shall denote
\begin{align}
	\alpha_1 = \mu \phi_1^2 \mathbf{K}_{1}^{-1} 
	\quad \mathrm{and} \quad 
	\alpha_2 = \mu \phi_2^2 \mathbf{K}_{2}^{-1} 
\end{align}

Recently, it has been shown that the
solutions to the classical Darcy equations satisfy
a minimum principle with respect to the mechanical
dissipation \citep{2016_Shabouei_Nakshatrala_CiCP}.
Herein, we shall extend this result to the double
porosity/permeability model. 

\begin{theorem}{[Minimum dissipation theorem]}
	\label{Thm:Dual_Minimum_dissipation_Thm}
	Assume that velocity boundary conditions are enforced
	on the entire boundary (i.e., $\Gamma_{1}^{v} = \Gamma_{2}^{v}
	= \partial \Omega$). Moreover, $\gamma \mathbf{b}(\mathbf{x})$
	is assumed to be a conservative vector field.
	The dissipation functional is defined as follows: 
	{\small{
			\begin{equation}
			\label{Total_dissipation}
			\boldsymbol{\Phi}\left[\mathbf{v}_{1},\mathbf{v}_{2}\right] 
			:= \sum_{i=1}^{2} \left(\int_{\Omega} \alpha_{i} \mathbf{v}_{i}
			\cdot \mathbf{v}_{i} \mathrm{d} \Omega 
			+ \dfrac{1}{2}\int_{\Omega}\dfrac{\mu}{\beta}
			\mathrm{div}\left[\phi_{i}\mathbf{v}_{i}\right]\mathrm{div}
			\left[\phi_{i}\mathbf{v}_{i}\right] \mathrm{d} \Omega \right)
			\end{equation}
		}
	}
	Then every kinematically admissible pair
	$(\widetilde{\mathbf{v}}_1(\mathbf{x}), 
	\widetilde{\mathbf{v}}_2(\mathbf{x}))$ satisfies 
	\begin{equation}
	\boldsymbol{\Phi}\left[\mathbf{v}_{1},\mathbf{v}_{2}\right] 
	\le \boldsymbol{\Phi}\left[\widetilde{\mathbf{v}}_{1},
	\widetilde{\mathbf{v}}_{2}\right]
	\end{equation}
	where $(\mathbf{v}_{1}(\mathbf{x}),\mathbf{v}_{2}
	(\mathbf{x}))$ denotes the pair of true velocity
	vector fields. 
	To put it differently, the pair of true velocity
	vector fields admits the minimum total dissipation among
	all the possible pairs of kinematically admissible
	vector fields. 
\end{theorem}

It should be noted that the minimum dissipation theorem
is not at odds with the maximization of the rate of
dissipation hypothesis, which we discussed in the
previous section. The minimum dissipation theorem
seeks the minimum among the set of kinematically
admissible vector fields. On the other hand, the
maximization of rate of dissipation hypothesis
maximizes the rate of dissipation among the set
of all the fields that satisfy the first and
second laws of thermodynamics.

\begin{theorem}{[Uniqueness]}
	\label{Thm:Dual_Uniqueness_theorem}
	The solution under the double porosity/permeability model is unique. 
\end{theorem}

Unlike the minimum dissipation theorem, the uniqueness
theorem does not require the velocity boundary
conditions to be prescribed on the entire boundary.
That is, the uniqueness has been established under
the general boundary conditions provided by
equations \eqref{Eqn:Dual_GE_vBC_1}--\eqref{Eqn:Dual_GE_pBC_2}. 

Next, we prove a reciprocal relation for double porosity/permeability 
model. Reciprocal relations are popular in several branches of 
mechanics. For example, Betti's reciprocal relation is 
a classical result in elasticity \citep{Love_elasticity}. 
Its utility to solve a class of seemingly difficult boundary 
value problems in linear elasticity is well-documented in the 
literature; for example, see \citep{Love_elasticity,Sadd}. 
Recently, reciprocal relations have been obtained for Darcy and 
Darcy-Brinkman equations in \citep{2016_Shabouei_Nakshatrala_CiCP}. 
It should, however, be noted that the kinematically admissible 
fields in the aforementioned cases (i.e., elasticity, Darcy 
and Darcy-Brinkman equations) are different from that of the 
double porosity/permeability model.

\begin{theorem}{[Reciprocal relation]}
\label{Thm:Dual_Reciprocal_Thm}
Let $(\mathbf{v}_1^{'},p_1^{'},\mathbf{v}_2^{'},p_2^{'})$ 
and $(\mathbf{v}_1^{*},p_1^{*},\mathbf{v}_2^{*},p_2^{*})$ 
be, respectively, the solutions under the prescribed 
data-sets $(\mathbf{b}^{'},v_{n1}^{'},p_{01}^{'},v_{n2}^{'},
p_{02}^{'})$ and $(\mathbf{b}^{*},v_{n1}^{*},p_{01}^{*},
v_{n2}^{*},p_{02}^{*})$. 
The domain, $\Omega$, and the boundaries, $\Gamma_1^{v}$, 
$\Gamma_1^{p}$, $\Gamma_2^{v}$, and $\Gamma_2^{p}$, are the 
same for both  prescribed data-sets. 
The pair of solutions and the pair of prescribed 
data-sets satisfy the following reciprocal relation:
{\small{
\begin{align}
	\label{Eqn:Dual_Reciprocal_term}
	&\int_{\Omega} \phi_{1}(\mathbf{x}) \gamma 
	\mathbf{b}^{'}(\mathbf{x}) \cdot 
	\mathbf{v}_{1}^{*}(\mathbf{x}) \; \mathrm{d} \Omega
	- \int_{\Gamma^{p}_{1}} \phi_{1}(\mathbf{x}) p_{01}^{'}(\mathbf{x}) 
	\mathbf{v}_{1}^{*}(\mathbf{x}) \cdot \widehat{\mathbf{n}}(\mathbf{x}) 
	\; \mathrm{d} \Gamma 
	- \int_{\Gamma^{v}_{1}} \phi_{1}(\mathbf{x}) p_{1}^{'}(\mathbf{x}) 
	v_{n1}^{*}(\mathbf{x}) \; \mathrm{d} \Gamma \nonumber \\
	& + \int_{\Omega} \phi_{2}(\mathbf{x}) \gamma \mathbf{b}^{'}(\mathbf{x}) 
	\cdot \mathbf{v}_{2}^{*}(\mathbf{x}) \; \mathrm{d} \Omega 
	- \int_{\Gamma^{p}_2} \phi_{2}(\mathbf{x}) p_{02}^{'}(\mathbf{x}) 
	\mathbf{v}_{2}^{*}(\mathbf{x}) \cdot \widehat{\mathbf{n}}(\mathbf{x}) 
	\; \mathrm{d} \Gamma 
	- \int_{\Gamma^{v}_{2}} \phi_{2}(\mathbf{x}) p_{2}^{'}(\mathbf{x}) 
	v_{n2}^{*}(\mathbf{x}) \; \mathrm{d} \Gamma \nonumber \\
	&= \int_{\Omega} \phi_{1}(\mathbf{x}) \gamma 
	\mathbf{b}^{*}(\mathbf{x}) \cdot 
	\mathbf{v}_{1}^{'}(\mathbf{x}) \; \mathrm{d} \Omega
	- \int_{\Gamma^{p}_{1}} \phi_{1}(\mathbf{x}) p_{01}^{*}(\mathbf{x}) 
	\mathbf{v}_{1}^{'}(\mathbf{x}) \cdot \widehat{\mathbf{n}}(\mathbf{x}) 
	\; \mathrm{d} \Gamma 
	- \int_{\Gamma^{v}_{1}} \phi_{1}(\mathbf{x}) p_{1}^{*}(\mathbf{x}) 
	v_{n1}^{'}(\mathbf{x}) \; \mathrm{d} \Gamma \nonumber \\
	& + \int_{\Omega} \phi_{2}(\mathbf{x}) \gamma \mathbf{b}^{*}(\mathbf{x}) 
	\cdot \mathbf{v}_{2}^{'}(\mathbf{x}) \; \mathrm{d} \Omega 
	- \int_{\Gamma^{p}_2} \phi_{2}(\mathbf{x}) p_{02}^{*}(\mathbf{x}) 
	\mathbf{v}_{2}^{'}(\mathbf{x}) \cdot \widehat{\mathbf{n}}(\mathbf{x}) 
	\; \mathrm{d} \Gamma 
	- \int_{\Gamma^{v}_{2}} \phi_{2}(\mathbf{x}) p_{2}^{*}(\mathbf{x}) 
	v_{n2}^{'}(\mathbf{x}) \; \mathrm{d} \Gamma 
\end{align}
}
}
\end{theorem}

\subsection{Maximum principle}
Maximum principle is one of the basic qualitative
properties of second-order elliptic partial differential 
equations. It can be shown that the pressure under Darcy 
equations satisfies a maximum principle, which is valid 
even for heterogeneous and anisotropic permeabilities. 
To wit, assuming that the pressure boundary conditions
are prescribed on the entire boundary, Darcy equations
can be rewritten as follows:
\begin{subequations}
	\begin{alignat}{2}
		-&\mathrm{div}\left[\frac{1}{\mu} \mathbf{K}(\mathbf{x})
		\mathrm{grad}[p + \psi]\right] = 0
		&&\quad \mathrm{in} \; \Omega \\
		&p(\mathbf{x}) = p_0(\mathbf{x})
		&&\quad \mathrm{on} \; \partial \Omega 
	\end{alignat}
\end{subequations}
The above boundary value problem is a second-order
elliptic partial differential equation with Dirichlet
boundary conditions prescribed on the entire boundary.
From the theory of partial differential equations
\citep{Gilbarg_Trudinger}, the pressure satisfies:
\begin{align}
	\min_{\mathbf{x} \in \partial \Omega}[p_0(\mathbf{x})]
	\leq p(\mathbf{x}) \leq
	\max_{\mathbf{x} \in \partial \Omega}[p_0(\mathbf{x}))]
	\quad \forall \mathbf{x} \in \overline{\Omega}
\end{align}
That is, the maximum and minimum pressures occur on the boundary. 

On the contrary, the macro- and micro-pressures under the 
double porosity/permeability model do not individually enjoy 
such a maximum principle. One can, however, establish 
a maximum principle for the difference in pressures in the macro-pore and micro-pore networks under some 
restrictions on the nature of permeabilities and boundary 
conditions. 
%
\begin{theorem}{[Maximum principle]}
\label{Thm:Dual_Maximum_Principle}
Assume that the permeabilities are isotropic and homogeneous.
That is, $\mathbf{K}_1(\mathbf{x}) = k_1 \mathbf{I}$ and
$\mathbf{K}_{2}(\mathbf{x}) = k_{2} \mathbf{I}$, where
$\mathbf{I}$ is the second-order identity tensor. The entire
boundary is prescribed with pressure boundary conditions.
That is, $\Gamma^p_1 = \Gamma^p_2 = \partial \Omega$. The
domain $\Omega$ is bounded and the boundary is smooth. 
Then the pressure difference in the macro-pore
and micro-pore networks, $p_{1}(\mathbf{x}) -
p_{2}(\mathbf{x})$, everywhere satisfies: 
{\small{
		\begin{align}
			\label{Eqn:Dual_Maximum_principle}
			\mathrm{min}\left[0,\mathop{\mathrm{min}}_{\mathbf{x} \in
				\partial \Omega} \left[p_{01}(\mathbf{x}) - p_{02}(\mathbf{x})
			\right]\right]
			 \le p_1(\mathbf{x}) - p_2(\mathbf{x}) 
			 \le
			\mathrm{max}\left[0,\mathop{\mathrm{max}}_{\mathbf{x} \in
				\partial \Omega} \left[p_{01}(\mathbf{x}) - p_{02}(\mathbf{x})
			\right] \right] 
		\end{align}
	}
}
	
The maximum principle for the double porosity/permeability
model basically implies that the pressure difference in the
micro-pore and macro-pore networks everywhere in the
domain lies between the corresponding non-negative
maximum and the non-positive minimum values on the
boundary on which pressures are prescribed. 
\end{theorem}

The main differences between the maximum principles
of Darcy equations and the double porosity/permeability
model can be summarized as follows:
\begin{enumerate}[(i)]
\item The maximum principle for the double porosity/permeability
model holds for isotropic and homogeneous permeabilities.
There are no such restrictions for Darcy equations. 
\item The body force is assumed to be conservative under
the maximum principle for Darcy equations. Such a
restriction is not needed for the maximum principle
for the double porosity/permeability model. 
\item The maximum principle for Darcy equations is in terms
of the pressure. On the other hand, the maximum principle
for the double porosity/permeability model is with respect
to the difference in pressures in the macro-pore and
micro-pore networks.
\item In the case of Darcy equations, the maximum and
minimum occur on the boundary. In the case of double 
porosity/permeability model, the non-negative maximum
and the non-positive minimum occur on the boundary.
\end{enumerate}

\subsection{Recovery of the classical Darcy equations}
The solutions (i.e., the pressure and velocity profiles) 
under the double porosity/permeability model are, in general, more 
complicated, and qualitatively and quantitatively different 
from the corresponding ones under the classical 
Darcy equations. 
However, there are three scenarios under which the solutions 
under the double porosity/permeability model can be described 
using the Darcy equations. That is, we need to show 
that there is no mass transfer across the two pore-networks 
under these scenarios. We now discuss these three scenarios, 
of which two are trivial. 

The \emph{first} scenario is when $\phi_2(\mathbf{x}) = 0$. 
Physically, this scenario corresponds to the case where 
there is no micro-pore network in the porous medium. To 
see mathematically that equations 
\eqref{Eqn:Dual_GE_BLM_1}--\eqref{Eqn:Dual_GE_mass_balance_2} 
reduce to the classical Darcy equations, one can appeal to 
equation \eqref{Eqn:Dual_GE_mass_balance_2} and conclude 
that there is no mass transfer across the pore-networks 
(i.e., $\chi(\mathbf{x}) = 0$) in the entire domain. 
Under this condition, equations for the macro-pore network 
(i.e., equations \eqref{Eqn:Dual_GE_BLM_1} and 
\eqref{Eqn:Dual_GE_mass_balance_1}) will reduce 
to the classical Darcy equations.

The \emph{second} scenario is when $\mathbf{K}_2(\mathbf{x}) 
= \mathbf{0}$. Physically, this scenario corresponds to the 
case in which the micro-pores are not inter-connected. To 
show mathematically that one recovers the classical Darcy 
equations under $\mathbf{K}_{2}(\mathbf{x}) = \mathbf{0}$, 
one can start with equation \eqref{Eqn:Dual_GE_Darcy_BLM_2} and 
conclude that $\mathbf{u}_{2} = 0$. Equation 
\eqref{Eqn:Dual_GE_Darcy_mass_balance_2} will then imply that 
$\chi(\mathbf{x}) = 0$ in the entire domain. Similar to the 
first scenario, the governing equations for the macro-pore 
network will reduce to the Darcy equations.   

The \emph{third} scenario pertains to the case wherein 
$\mathbf{K}_1(\mathbf{x}) = \mathbf{K}_2(\mathbf{x})$, 
and the boundary conditions for the macro-pore and
micro-pore networks are the same. 
That is, $\Gamma_{1}^{p} = \Gamma_{2}^{p}$, $\Gamma_{1}^{v} 
= \Gamma_{2}^{v}$, $p_{01}(\mathbf{x}) = p_{02}(\mathbf{x})$ 
and $v_{n1}(\mathbf{x}) = v_{n2}(\mathbf{x})$. Note that it is
\emph{not} necessary for $\phi_1(\mathbf{x})$ to be equal to $\phi_2(\mathbf{x})$. 
Under these conditions, flow in 
the porous medium can be modeled using 
the classical Darcy equations with porosity equal to 
$\phi(\mathbf{x}) = \phi_{1}(\mathbf{x}) +\phi_{2}(\mathbf{x})$. 
To see that the mass transfer across the pore
networks is zero, one can proceed as follows.
For convenience, assume $\mathbf{K}_1(\mathbf{x}) 
= \mathbf{K}_2(\mathbf{x}) = \mathbf{K}(\mathbf{x})$.
Then the aforementioned conditions give rise to the
following boundary value problem:
{\small
	\begin{subequations}
		\begin{alignat}{2}
			\label{Eqn:Dual_scenario_3_LM}
			&\mu\mathbf{K}^{-1} \left(\mathbf{u}_{1} - \mathbf{u}_{2}\right) 
			+ \mathrm{grad}[p_1 - p_2] = \mathbf{0}
			&&\quad \mathrm{in} \; \Omega \\
			&\mathrm{div}[\mathbf{u}_1 - \mathbf{u}_2] = -\frac{2 \beta}{\mu}(p_1 - p_2)
			&&\quad \mathrm{in} \; \Omega  \\
			&\left(\mathbf{u}_{1}(\mathbf{x}) - \mathbf{u}_2(\mathbf{x})\right)
			\cdot \widehat{\mathbf{n}}(\mathbf{x}) = 0
			&&\quad \mathrm{on} \; \Gamma^{v} = \Gamma_1^{v} = \Gamma_2^{v} \\
			\label{Eqn:Dual_scenario_3_p_BC}
			&p_1(\mathbf{x}) - p_2(\mathbf{x}) = 0
			&&\quad \mathrm{on} \; \Gamma^{p} =
			\Gamma_1^{p} = \Gamma_2^{p} 
		\end{alignat}
	\end{subequations}
}
Clearly, the pair $p_1(\mathbf{x}) - p_2(\mathbf{x})= 0$
and $\mathbf{u}_1(\mathbf{x}) - \mathbf{u}_2(\mathbf{x})
= \mathbf{0}$ is a solution to the above boundary value
problem
\eqref{Eqn:Dual_scenario_3_LM}--\eqref{Eqn:Dual_scenario_3_p_BC}.
By the uniqueness theorem \ref{Thm:Dual_Uniqueness_theorem},
this is the only solution to the above boundary value
problem. Since $p_1(\mathbf{x}) = p_2(\mathbf{x})$, the
mass transfer across the pore-networks is zero.  

In all the above three scenarios, it is important
to note that there will be no contribution to the
dissipation from the connectors (i.e., conduits/fissures),
as there is no flow in the connectors. 

\section{ANALYTICAL SOLUTION BASED ON GREEN'S FUNCTION APPROACH}
\label{Sec:S5_Double_Analytical}
In this section, we present an analytical solution
procedure for a general boundary value problem arising
from the double porosity/permeability model. We provide
a formal mathematical derivation based on the Green's
function approach.

We start by rewriting the governing equations
\eqref{Eqn:Dual_GE_Darcy_BLM_1}--\eqref{Eqn:Dual_GE_Darcy_pBC_2}.
By eliminating $\mathbf{u}_1(\mathbf{x})$ from these
equations, we obtain the following boundary value
problem for the macro-pore network:
{\small{
\begin{subequations}
	\begin{alignat}{2}
		\label{Eqn:Dual_Green_BLM_1}
		&\mathrm{div}\left[\frac{1}{\mu}\mathbf{K}_{1}(\mathbf{x})
		\left(\gamma \mathbf{b}(\mathbf{x}) -
		\mathrm{grad}[p_1]\right)\right]
		= \chi(\mathbf{x}) 
		&&\quad \mathrm{in} \; \Omega \\
		\label{Eqn:Dual_Green_vBC_1}
		&\frac{1}{\mu}\widehat{\mathbf{n}}(\mathbf{x})
		\cdot \mathbf{K}_{1}(\mathbf{x})
		\left(\gamma \mathbf{b}(\mathbf{x}) -
		\mathrm{grad}[p_1]\right)
		= u_{n1}(\mathbf{x})
		&&\quad \mathrm{on} \; \Gamma_{1}^{v} \\
		\label{Eqn:Dual_Green_pBC_1}
		&p_1(\mathbf{x}) = p_{01}(\mathbf{x})
		&&\quad \mathrm{on} \; \Gamma_{1}^{p}
	\end{alignat}
\end{subequations}
	}
}
By multiplying equation \eqref{Eqn:Dual_Green_BLM_1}
with $G_1$, integrating over the domain, employing the 
Green's identity, and noting the boundary conditions
(i.e., equations \eqref{Eqn:Dual_Green_vBC_1} and
\eqref{Eqn:Dual_Green_pBC_1}), we obtain:
{\small
\begin{align}
	& -\int_{\Omega} \mathrm{div}\left[\frac{1}{\mu} \mathbf{K}_{1}
	\mathrm{grad}[G_1]\right] p_1 \mathrm{d} \Omega
	+\int_{\Gamma_{1}^{p}} \frac{1}{\mu} G_1 \widehat{\mathbf{n}}
	\cdot \mathbf{K}_1 \left(\gamma \mathbf{b} - \mathrm{grad}[p_1]
	\right) \mathrm{d} \Gamma 
	\nonumber \\
	& + \int_{\Gamma_{1}^{v}} \frac{1}{\mu} \widehat{\mathbf{n}}
	\cdot \mathbf{K}_1 \mathrm{grad}[G_1] p_1 \mathrm{d} \Gamma
	= \int_{\Omega} G_{1} \chi \mathrm{d} \Omega
	+ \int_{\Omega} \frac{1}{\mu} \mathrm{grad}[G_1] \cdot
	\mathbf{K}_1 \gamma \mathbf{b} \; \mathrm{d} \Omega 
	\nonumber \\
	&- \int_{\Gamma_{1}^{v}} G_{1} u_{n1} \mathrm{d} \Gamma 
	- \int_{\Gamma_{1}^{p}} \frac{1}{\mu} \widehat{\mathbf{n}} \cdot
	\mathbf{K}_1 \mathrm{grad}[G_{1}] p_{01} \mathrm{d} \Gamma 
\end{align}
}
This suggests to construct the Green's function $G_{1}
(\mathbf{x},\mathbf{y})$ to be the solution of the 
following boundary value problem: 
\begin{subequations}
	\label{Eqn:Dual_G1_Green_function}
	\begin{alignat}{2}
		-&\mathrm{div}\left[\frac{1}{\mu}\mathbf{K}_{1}(\mathbf{x})
		\mathrm{grad}[G_1(\mathbf{x},\mathbf{y})]\right]
		= \delta(\mathbf{x}-\mathbf{y})
		&&\quad \mathrm{in} \; \Omega \\
		-&\frac{1}{\mu} \widehat{\mathbf{n}}(\mathbf{x})
		\cdot \mathbf{K}_{1}(\mathbf{x}) \mathrm{grad}[G_{1}
		(\mathbf{x},\mathbf{y})] = 0
		&&\quad \mathrm{on} \; \Gamma_{1}^{v} \\
		&G_{1}(\mathbf{x},\mathbf{y}) = 0
		&&\quad \mathrm{on} \; \Gamma_{1}^{p} 
	\end{alignat}
\end{subequations}
where $\delta(\mathbf{x}-\mathbf{y})$ denotes
the Dirac-delta distribution \citep{Lighthill}.
Then the macro-pressure $p_1(\mathbf{x})$ can
be written in terms of mass transfer between the
pore-networks $\chi(\mathbf{x})$ as follows: 
{\small{
	\begin{align}
		\label{Eqn:Dual_p1_solution}
		p_{1}(\mathbf{x})
		&= \int_{\Omega} G_{1}(\mathbf{x},\mathbf{y})
		\chi(\mathbf{y}) \mathrm{d} \Omega_{y}
		+ \int_{\Omega} \frac{1}{\mu} \mathrm{grad}_{y}[G_1(\mathbf{x},\mathbf{y})]
		\cdot \mathbf{K}_1(\mathbf{y}) \gamma \mathbf{b}(\mathbf{y})
		\; \mathrm{d} \Omega_{y} \nonumber \\
		&- \int_{\Gamma_{1}^{v}} G_{1}(\mathbf{x},\mathbf{y})
		u_{n1}(\mathbf{y}) \; \mathrm{d} \Gamma_{y} 
		- \int_{\Gamma_{1}^{p}} \frac{1}{\mu}
		\widehat{\mathbf{n}}(\mathbf{y}) \cdot 
		\mathbf{K}_1(\mathbf{y})
		\mathrm{grad}_{y}[G_1(\mathbf{x},\mathbf{y})] 
		p_{01}(\mathbf{y}) \mathrm{d} \Gamma_{y} 
	\end{align}
	}
}
where $\mathrm{d} \Omega_{y}$ and $\mathrm{d}\Gamma_{y}$,
respectively, denote the volume element and the surface
area element with respect to $\mathbf{y}$-coordinates,
and the gradient with respect to $\mathbf{y}$-coordinates
is denoted by $\mathrm{grad}_{y}[\cdot]$. 

By carrying out a similar procedure for the micro-pore
network, the Green's function $G_2(\mathbf{x},\mathbf{y})$
is taken to be the solution of the following boundary
value problem:
\begin{subequations}
	\label{Eqn:Dual_G2_Green_function}
	\begin{alignat}{2}
		-&\mathrm{div}\left[\frac{1}{\mu}\mathbf{K}_{2}(\mathbf{x})
		\mathrm{grad}[G_2(\mathbf{x},\mathbf{y})]\right]
		= \delta(\mathbf{x}-\mathbf{y})
		&&\quad \mathrm{in} \; \Omega \\
		-&\frac{1}{\mu} \widehat{\mathbf{n}}(\mathbf{x})
		\cdot \mathbf{K}_{2}(\mathbf{x}) \mathrm{grad}[G_{2}
		(\mathbf{x},\mathbf{y})] = 0
		&&\quad \mathrm{on} \; \Gamma_{2}^{v} \\
		&G_{2}(\mathbf{x},\mathbf{y}) = 0
		&&\quad \mathrm{on} \; \Gamma_{2}^{p} 
	\end{alignat}
\end{subequations}
The micro-pressure $p_2(\mathbf{x})$ can then
be written in terms of $\chi(\mathbf{x})$ as
follows: 
{\small{
	\begin{align}
		\label{Eqn:Dual_p2_solution}
		p_{2}(\mathbf{x})
		= &-\int_{\Omega} G_{2}(\mathbf{x},\mathbf{y})
		\chi(\mathbf{y}) \mathrm{d} \Omega_{y}
		+ \int_{\Omega} \frac{1}{\mu} \mathrm{grad}_{y}[G_2(\mathbf{x},\mathbf{y})]
		\cdot \mathbf{K}_2(\mathbf{y}) \gamma \mathbf{b}(\mathbf{y})
		\; \mathrm{d} \Omega_{y} \nonumber \\
		&- \int_{\Gamma_{2}^{v}} G_{2}(\mathbf{x},\mathbf{y})
		u_{n2}(\mathbf{y}) \; \mathrm{d} \Gamma_{y} 
		- \int_{\Gamma_{2}^{p}} \frac{1}{\mu}
		\widehat{\mathbf{n}}(\mathbf{y}) \cdot 
		\mathbf{K}_2(\mathbf{y})
		\mathrm{grad}_{y}[G_2(\mathbf{x},\mathbf{y})] 
		p_{02}(\mathbf{y}) \mathrm{d} \Gamma_{y} 
	\end{align}
	}
}
Note that $G_1(\mathbf{x},\mathbf{y})$ and
$G_2(\mathbf{x},\mathbf{y})$ are Green's
functions for scalar diffusion equations.
Since the permeabilities, $\mathbf{K}_1
(\mathbf{x})$ and $\mathbf{K}_2(\mathbf{x})$,
are symmetric tensors, it is easy to establish
that the Green's functions, $G_1(\mathbf{x},
\mathbf{y})$ and $G_2(\mathbf{x},\mathbf{y})$,
are symmetric. That is,
{\small
	\begin{align}
		G_{1}(\mathbf{x},\mathbf{y}) 
		= G_{1}(\mathbf{y},\mathbf{x})
		\quad \mathrm{and} \quad 
		G_{2}(\mathbf{x},\mathbf{y}) 
		= G_{2}(\mathbf{y},\mathbf{x})
		\quad \forall \mathbf{x}, \mathbf{y} 
	\end{align}
}
Equations \eqref{Eqn:Dual_p1_solution} and
\eqref{Eqn:Dual_p2_solution} give rise to
the following integral equation for the
mass transfer between the pore-networks: 
\begin{align}
	\label{Eqn:Dual_integral_equation}
	\frac{\mu}{\beta} \chi(\mathbf{x}) 
	+ \int_{\Omega} (G_1(\mathbf{x},\mathbf{y})
	+ G_2(\mathbf{x},\mathbf{y})) \chi(\mathbf{y})
	\; \mathrm{d} \Omega_{y}
	= h(\mathbf{x})
\end{align}
where
{\small{
	\begin{align}
		&h(\mathbf{x})
		:= \nonumber \\
		&\int_{\Omega} \frac{1}{\mu} \left(\mathbf{K}_{2}
		(\mathbf{y})\mathrm{grad}_{y}[G_2(\mathbf{x},\mathbf{y})]
		- \mathbf{K}_{1}(\mathbf{y}) \mathrm{grad}_{y}[G_1(\mathbf{x},
		\mathbf{y})]\right) \cdot \gamma \mathbf{b}(\mathbf{y})
		\mathrm{d} \Omega_{y} + \nonumber \\
		&\int_{\Gamma_1^{v}} G_{1}(\mathbf{x},\mathbf{y})
		u_{n1}(\mathbf{y}) \mathrm{d} \Gamma_{y}
		+\int_{\Gamma_{1}^{p}} \frac{1}{\mu} \widehat{\mathbf{n}}
		(\mathbf{y}) \cdot \mathbf{K}_{1}(\mathbf{y})
		\mathrm{grad}_{y}[G_1(\mathbf{x},\mathbf{y})]
		p_{01}(\mathbf{y}) \mathrm{d} \Gamma_{y} - \nonumber \\
		&\int_{\Gamma_2^{v}} G_{2}(\mathbf{x},\mathbf{y})
		u_{n2}(\mathbf{y}) \mathrm{d} \Gamma_{y}
		-\int_{\Gamma_{2}^{p}} \frac{1}{\mu} \widehat{\mathbf{n}}
		(\mathbf{y}) \cdot \mathbf{K}_{2}(\mathbf{y})
		\mathrm{grad}_{y}[G_2(\mathbf{x},\mathbf{y})]
		p_{02}(\mathbf{y}) \mathrm{d} \Gamma_{y} 
	\end{align}
	}
}
Equation \eqref{Eqn:Dual_integral_equation} is 
a non-homogeneous Fredholm integral equation of 
second type with symmetric kernel \citep{Tricomi_IE}. 
The symmetry of the kernel stems from the fact that 
the Green's functions, $G_1(\mathbf{x},\mathbf{y})$ 
and $G_2(\mathbf{x},\mathbf{y})$, are symmetric.

The overall analytical solution procedure
can be compactly written as follows:
\begin{enumerate}
	\item[(i)] Construct the Green's functions, $G_{1}
	(\mathbf{x},\mathbf{y})$ and $G_{2}(\mathbf{x},
	\mathbf{y})$, that are, respectively, the solutions
	of the boundary value problems given by equations 
	\eqref{Eqn:Dual_G1_Green_function} and 
	\eqref{Eqn:Dual_G2_Green_function}.
	\item[(ii)] Using $G_1(\mathbf{x},\mathbf{y})$ and $G_2(\mathbf{x},\mathbf{y})$,
	solve the integral equation \eqref{Eqn:Dual_integral_equation} to
	obtain the mass transfer between the pore-networks $\chi(\mathbf{x})$.
	\item[(iii)] Using the solution for $\chi(\mathbf{x})$,
	compute the pressures, $p_1(\mathbf{x})$ and $p_2(\mathbf{x})$, 
	using equations \eqref{Eqn:Dual_p1_solution} and
	\eqref{Eqn:Dual_p2_solution}, respectively.
	\item[(iv)] Once the pressures, $p_1(\mathbf{x})$
	and $p_2(\mathbf{x})$, are known, the discharge
	velocities, $\mathbf{u}_1(\mathbf{x})$ and
	$\mathbf{u}_2(\mathbf{x})$, can be computed
	using equations \eqref{Eqn:Dual_GE_Darcy_BLM_1}
	and \eqref{Eqn:Dual_GE_Darcy_BLM_2}.
\end{enumerate}

The solution procedure presented above is quite general,
as it can be applied even to those problems with anisotropic
and heterogeneous medium properties. The procedure is built
upon obtaining Green's functions for scalar diffusion equations
and solving a linear scalar Fredholm integral equation of second
type. There are numerous existing works that provide Green's
functions for scalar diffusion equations (e.g., see 
\citep{stackgoldgreen}). A good deal of work exists on Fredholm 
integral equations in terms of mathematical theory, analytical 
solutions, and numerical techniques. For instance, see 
\citep{Tricomi_IE,Atkinson_IE,Polyanin_Manzhirov_IE}. 

The boundary value problems corresponding to the Green's 
functions assumed $\Gamma_1^{p}$ and $\Gamma_{2}^{\mathrm{p}}$ 
to be non-empty. That is, it is assumed that there is a (non-empty) 
portion of the boundary on which Dirichlet boundary conditions are 
prescribed. One needs to modify the procedure if velocity boundary 
conditions are prescribed on the entire boundary for 
a pore-network, which gives rise to a boundary 
value problem with Neumann boundary conditions for 
the construction of the Green's function for that 
particular pore-network. However, one can find 
in the literature procedures to construct modified 
Green's functions for diffusion-type equations 
with Neumann boundary conditions (e.g., 
see \citep{stackgoldgreen}). This aspect will be illustrated in the next section. 

\section{CANONICAL BOUNDARY VALUE PROBLEMS}
\label{Sec:S6_Double_BVP}
We now present various boundary value problems to highlight 
the differences between the bulk response (e.g., single 
pore-network modeled using Darcy equations) and the one 
obtained by incorporating the double porosity/permeability 
model. These problems are specifically designed to be 
simple, as the primary aim is to illustrate that a number 
of features and characteristics will be lost in the bulk 
response. We believe that these findings will be 
valuable to the subsurface modeling community.

\subsection{One-dimensional problem \#1}
Consider a one-dimensional domain of length $L$. For
the macro-pore network, pressures $p_{1}^{\mathrm{L}}$
and $p_{1}^{\mathrm{R}}$ are prescribed on the left and
right ends of the domain, respectively. Similarly,
pressures of $p_{2}^{\mathrm{L}}$ and $p_{2}^{\mathrm{R}}$
are, respectively, prescribed on the left and right
ends of the domain for the micro-pore network.
The purpose of this boundary value problem is four-fold:
\begin{enumerate}[(i)]
	\item The problem will be used to illustrate the various
	steps in the analytical solution procedure that was
	presented in Section \ref{Sec:S5_Double_Analytical}. 
	\item It will be shown that the integral equation
	\eqref{Eqn:Dual_integral_equation} can provide
	the appropriate and consistent boundary conditions
	for the mass transfer across the pore-networks in
	terms of the prescribed velocity and pressure
	boundary conditions.
	\item It will be shown the maximum and minimum
	pressures need not occur on the boundary under
	the double porosity/permeability model for a
	boundary value problem with pressures prescribed
	on the \emph{entire} boundary. On the contrary,
	the maximum and minimum pressures occur on
	the boundary under Darcy equations for a
	pressure-prescribed boundary value problem.
	\item The maximum principle proposed
	in Theorem \ref{Thm:Dual_Maximum_Principle} will be
	verified for this problem.
\end{enumerate}

\subsubsection{Non-dimensionalization}
We take the length of the domain $L$ [L], $p_1^{\mathrm{L}} -
p_1^{\mathrm{R}}$ $\mathrm{[ML^{-1}T^{-2}]}$, and $\beta/\mu$
$\mathrm{[M^{-1}LT]}$ as the reference quantities. The time and the mass scales are taken as $\frac{\mu / \beta}{(p_1^{\mathrm{L}} -
	p_1^{\mathrm{R}})}$ $\mathrm{[T]}$ and $\frac{L (\mu / \beta )^2}{(p_1^{\mathrm{L}} -
	p_1^{\mathrm{R}})}$ $\mathrm{[M]}$, respectively. We
also take the datum for the pressure to be $p_1^{\mathrm{R}}$. 
These reference quantities give rise to the
following non-dimensional quantities, which
are denoted by a superposed bar: 
\begin{align}
	&\overline{x} = \frac{x}{L}, \;
	\overline{p}_{1} =
	\frac{p_1 - p_1^{\mathrm{R}}}{p_1^{\mathrm{L}} - p_1^{\mathrm{R}}}, \;
	\overline{p}_{2} =
	\frac{p_2 - p_1^{\mathrm{R}}}{p_1^{\mathrm{L}} - p_1^{\mathrm{R}}}, \;
	\overline{k}_{1} = \frac{k_1}{L^2}, 
	\overline{k}_{2} = \frac{k_2}{L^2}, \; 
	\overline{\mu} = \frac{\mu}{\mu_{\mathrm{ref}}}, \;
	\overline{u}_1 = \frac{u_1}{u_{\mathrm{ref}}}, \;
	\overline{u}_2 = \frac{u_2}{u_{\mathrm{ref}}}
\end{align}
where
\begin{align}
	\mu_{\mathrm{ref}} := \frac{\mu}{\beta}, \;
	u_{\mathrm{ref}} := \frac{\beta L}{\mu}
	\left(p_1^{\mathrm{L}} - p_1^{\mathrm{R}}\right)
	%
\end{align}
The non-dimensional form of the governing
equations can be written as follows:
\begin{subequations}
	\begin{alignat}{2}
		&\frac{\overline{\mu}}{\overline{k}_1}
		\overline{u}_1 + \frac{d\overline{p}_1}{d\overline{x}} = 0, \; 
		\frac{d\overline{u}_1}{d\overline{x}}
		= - (\overline{p}_1 - \overline{p}_2) \quad \mathrm{in} \; (0,1) \\
		&\frac{\overline{\mu}}{\overline{k}_2}
		\overline{u}_2 + \frac{d\overline{p}_2}{d\overline{x}} = 0, \; 
		\frac{d\overline{u}_2}{d\overline{x}}
		= + (\overline{p}_1 - \overline{p}_2) \quad \mathrm{in} \; (0,1) \\
		&\overline{p}_1(\overline{x} = 0) = 1, \;
		\overline{p}_1(\overline{x} = 1) = 0, \\
		&\overline{p}_{2}(\overline{x} = 0) = \overline{p}_{2}^{\mathrm{L}}, \; 
		\overline{p}_{2}(\overline{x} = 1) = \overline{p}_{2}^{\mathrm{R}} 
	\end{alignat}
\end{subequations}
The mass transfer across the two pore-networks takes
the following form:
\begin{align}
	\overline{\chi}(\overline{x}) =
	\overline{p}_{2}(\overline{x})
	- \overline{p}_{1}(\overline{x})
\end{align}
In this boundary value problem, $\overline{k}_1$
and $\overline{k}_2$ are assumed to be independent
of $\overline{x}$. For simplicity, we drop the
over-lines, as all the quantities below will
be non-dimensional. 

\subsubsection{Analytical solution}
For convenience, let us introduce the following parameter: 
\begin{align}
	\label{Eqn:Dual_1D_problem_1_eta}
	\eta
	:= \sqrt{\frac{\mu \left(k_1 + k_2\right)}{k_1 k_2}}
\end{align}
Note that $\eta$ is inversely proportional to the square root
of the harmonic average of the permeabilities in the
macro-pore and micro-pore networks.
The Green's functions $G_{1}(x,y)$ and $G_{2}(x,y)$ will be:
\begin{align}
	\frac{k_1}{\mu} G_{1}(x,y) = \frac{k_2}{\mu} G_{2}(x,y)
	= \left\{\begin{array}{ll}
		x - xy & x \leq y \\
		y - xy & x > y 
	\end{array} \right. 
\end{align}
The integral equation for the mass transfer becomes:
{\small{
		\begin{align}
			\label{Eqn:Dual_1D_problem_1_chi_integral}
			\chi(x) - \int_{0}^{1} \eta^{2} x y \chi(y) \mathrm{d}y
			+ \int_{0}^{x} \eta^2 y \chi(y) \; \mathrm{d} y
			+ \int_{x}^{1} \eta^2 x \chi(y) \; \mathrm{d} y
			= h(x) 
		\end{align}
	}
}
Using the Leibniz integral rule and noting that
$h^{''}(x) = 0$, the above equation implies that 
\begin{align}
	\label{Eqn:Dual_1D_problem_1_chi}
	\frac{d^2 \chi}{d x^2} = \eta^2 \chi
\end{align}
The boundary conditions for the mass transfer
$\chi(x)$ in terms of the prescribed pressure boundary
conditions take the following form:
\begin{subequations}
	\begin{align}
		\label{Eqn:Dual_1D_problem_1_chi_BC1}
		\chi(x = 0) = h(x = 0^{+})
		&= p_{2}^{\mathrm{L}} - 1 \\
		\label{Eqn:Dual_1D_problem_1_chi_BC2}
		\chi(x = 1) = h(x = 1^{-})
		&= p_{2}^{\mathrm{R}} 
	\end{align}
\end{subequations}
The solution of the boundary value problem given
by equations \eqref{Eqn:Dual_1D_problem_1_chi} and
\eqref{Eqn:Dual_1D_problem_1_chi_BC1}--\eqref{Eqn:Dual_1D_problem_1_chi_BC2},
which will also be the solution of the integral equation
\eqref{Eqn:Dual_1D_problem_1_chi_integral}, takes
the following form: 
\begin{align}
	\chi(x) = C_1 \exp[\eta x] + C_2 \exp[-\eta x]
\end{align}
where
{\small{
\begin{align}
	C_{1} = \frac{p_{2}^{\mathrm{R}} + (1 - p_{2}^{\mathrm{L}})\exp[-\eta]}
	{\exp[\eta] - \exp[-\eta]} 
	\quad \mathrm{and} \quad 
	C_{2} = -\frac{p_{2}^{\mathrm{R}} + (1 - p_{2}^{\mathrm{L}})\exp[\eta]}
	{\exp[\eta] - \exp[-\eta]} 
\end{align}
	}
}
The solution for the pressures in the macro-pore
and micro-pore networks can be written as follows:
{\small{
		\begin{subequations}
			\begin{align}
				p_1(x) &=
				\underbrace{1 - x}_{\mbox{{\small Darcy solution}}}
				- \underbrace{\frac{\mu}{k_1 \eta^2} \left\{
					\left(1 - p_2^{\mathrm{L}}\right) (1 - x) 
					- p_2^{\mathrm{R}} x
					+ C_1 \exp[\eta x] + C_2 \exp[-\eta x]\right\}}_{\mbox{{\small deviation due to mass transfer}}}  \\
				p_2(x) &=
				\underbrace{p_{2}^{\mathrm{L}}(1 - x) + p_{2}^{\mathrm{R}} x}_{\mbox{{\small Darcy solution}}}
			   + \underbrace{\frac{\mu}{k_2 \eta^2} \left\{ 
					\left(1 - p_{2}^{\mathrm{L}}\right) \left(1 - x\right)
					- p_{2}^{\mathrm{R}} x 
					+ C_1 \exp[\eta x] + C_2 \exp[-\eta x]\right\}}_{\mbox{{\small deviation due to mass transfer}}} 
			\end{align}
		\end{subequations}
	}
}
The solution for the discharge velocities in
the pore-networks can be written as follows:
{\small{
		\begin{subequations}
			\begin{align}
				u_1(x) &= \underbrace{\frac{k_1}{\mu}}_{\mbox{{\small Darcy solution}}}
				+ \underbrace{\frac{1}{\eta^2} \left(p_{2}^{\mathrm{L}} - p_{2}^{\mathrm{R}} - 1\right) 
					+ \frac{1}{\eta} \left(C_1 \exp[\eta x] 
					- C_2 \exp[-\eta x]\right)}_{\mbox{{\small deviation due to mass transfer}}} \\
				u_2(x) &= \underbrace{\frac{k_{2}}{\mu} \left(p_{2}^{\mathrm{L}} - p_{2}^{\mathrm{R}}\right)}_{\mbox{{\small Darcy solution}}}
				- \underbrace{\frac{1}{\eta^2} \left(p_{2}^{\mathrm{L}} - p_{2}^{\mathrm{R}} - 1 \right)
					- \frac{1}{\eta} \left(C_1 \exp[\eta x] - C_2 \exp[-\eta x]\right)}_{\mbox{{\small deviation due to mass transfer}}}
			\end{align}
		\end{subequations}
	}
}

The deviation of the solution under the double
porosity/permeability model from the
corresponding one under Darcy equations
is indicated in the above expressions for the
analytical solution. One fact that is clear
from the analytical solution is that the nature
of the solution (i.e., the pressure and velocity
profiles) depends on the parameter $\eta$. 

Figure~\ref{Fig:Double_1D_P1_pressure_analytical}
illustrates that the maximum and minimum pressures
for macro-pore and micro-pore networks need not occur
on the boundary. On the other hand, as mentioned earlier,
the maximum and minimum pressures occur on the boundary
under Darcy equations for a boundary value problem with
pressures prescribed on the entire boundary.
Figure~\ref{Fig:1D_problem_1_maximum_principle}
numerically verifies the maximum principle proposed
in Theorem \ref{Thm:Dual_Maximum_Principle} for the
double porosity/permeability model. As one can see
from this figure, the non-negative maximum and the
non-positive minimum of the pressure difference,
$p_1(x) - p_2(x)$, occur on the boundary under the
double porosity/permeability model.

\subsection{One-dimensional problem \#2}
\label{Subsec:Double_1D_P2}
In this problem, pressure boundary conditions
are applied to the macro-pore network, and no-flux 
(i.e., zero normal velocity) boundary
conditions are enforced on the micro-pore
network. This problem highlights the
following important points:
\begin{enumerate}[(i)]
	\item One can have discharge (i.e., non-zero velocity) in
	the micro-pore network even if the micro-pore network
	does not extend to the boundary (i.e., there is no
	discharge on the boundary of the micro-pore network).
	This implies that, for complicated porous media, it
	is essential to know the internal pore-structure (e.g.,
	using $\mu$-CT \citep{stock2008microcomputed}). It 
	is not sufficient to know the surface pore-structure on the boundary. 
	\item One can find the solution uniquely for all
	the fields (i.e., pressures, velocities, and mass
	transfer) even if the velocity boundary conditions
	are prescribed on the entire boundary for one of
	the pore-networks. On the other hand, one cannot
	find the pressure uniquely under Darcy equations
	if the velocity boundary conditions are prescribed
	on the entire boundary.
	\item This problem will be utilized to illustrate the 
	construction of modified Green's function for problems 
	involving velocity boundary conditions for either 
	macro-pore or micro-pore network. 
\end{enumerate}

We shall employ the same reference quantities,
as defined in problem \#1. The non-dimensional
form of the governing equations for this boundary
value problem can be written as follows:
\begin{subequations}
	\begin{align}
		&\frac{\mu}{k_1} u_1 + \frac{d p_1}{dx} = 0, \;
		\frac{d u_1}{dx} = - (p_1 - p_2)
		\quad \mathrm{in} \; (0,1) \\  
		&\frac{\mu}{k_2} u_2 + \frac{d p_2}{dx} = 0, \;
		\frac{d u_2}{dx} = + (p_1 - p_2)
		\quad \mathrm{in} \; (0,1) \\
		& p_1(x = 0) = 1, \; p_1(x = 1) = 0 \\
		& u_{2}(x = 0) = 0, \; u_{2}(x = 1) = 0 
	\end{align}
\end{subequations}
The expression for the mass transfer across
the pore-networks is the same as before.

\subsubsection{Analytical solution}
The Green's function $G_{1}(x,y)$ will be:
\begin{align}
	G_{1}(x,y) = \frac{\mu}{k_1}
	\left\{\begin{array}{ll}
		x - xy & x \leq y \\
		y - xy & x > y 
	\end{array} \right. 
\end{align}
Since the boundary value problem for the
micro-pore network has Neumann boundary
conditions on the entire boundary, one
needs to modify the procedure for finding
the Green's function. 
Following the technique provided in
\citep{stackgoldgreen}, the Green's 
function $G_{2}(x,y)$ is constructed
in such a way that it satisfies the following
boundary value problem:
\begin{align}
	-&\frac{\mu}{k_2} \frac{d^2 G_2}{d x^2} = \delta(x-y) - 1
	\quad \mathrm{in} \; (0,1) \nonumber \\
	& \frac{\mu}{k_2} \frac{d G_2}{d x} \Bigg |_{x = 0} = 0, \quad 
	\frac{\mu}{k_2} \frac{d G_2}{d x} \Bigg |_{x = 1} = 0
\end{align}
The Green's function $G_{2}(x,y)$ takes the following form: 
\begin{align}
	G_{2}(x,y) = \frac{\mu}{k_2} \left\{\begin{array}{ll}
		\frac{x^2}{2} - y + C_{0}  & \quad \mathrm{for} \; x \leq y \\
		\frac{x^2}{2} - x + C_{0}  & \quad \mathrm{for} \; x > y 
	\end{array} \right. 
\end{align}
where $C_0$ is a constant which needs to be determined.
It should be noted that $C_0$ is independent of $x$, but
could depend on $y$. The integral equation for the mass
transfer becomes: 
{\small{
		\begin{align}
			\label{Eqn:Dual_1D_P2_integral_eqn}
			&\chi(x) + \int_{0}^{1} p_2(y) dy
			- \frac{1}{k_1} \int_{0}^{1} xy \chi(y) \mathrm{d} y
			+ \frac{1}{k_2} \int_{0}^{1} (C_{0} + x^2/2) \chi(y) \mathrm{d} y
			\nonumber \\  
			&+ \int_{0}^{x} \left(\frac{y}{k_1} - \frac{x}{k_2}\right) \chi(y) \mathrm{d} y 
			+ \int_{x}^{1} \left(\frac{x}{k_1} - \frac{y}{k_2}\right) \chi(y) \; \mathrm{d} y
			= h(x)
		\end{align}
	}
}
Using the Leibniz integration rule, the integral
equation \eqref{Eqn:Dual_1D_P2_integral_eqn} can
be shown to be equivalent to the following
differential equation:
\begin{align}
	\frac{d^2 \chi}{dx^2}
	+ \frac{1}{k_2} \int_{0}^{1} \chi(y) \mathrm{d} y
	= \eta^2 \chi(x) 
\end{align}
where the parameter $\eta$ is defined in equation
\eqref{Eqn:Dual_1D_problem_1_eta}. The solution
for the above differential equation takes the
following form:
\begin{align}
	\chi(x) = D_1 \exp[\eta x] + D_2 \exp[-\eta x]
\end{align}
The boundary conditions give rise to 
{\small{
		\begin{align}
			&D_1 = \frac{k_1 + k_2}{k_1 \eta (\exp[\eta] + 1) + 2 k_2 (\exp[\eta] - 1)}, \quad 
			D_2 = - \exp[\eta] D_1 
		\end{align}
}}
The analytical solution can be compactly written as follows: 
\begin{subequations}
	\begin{align}
		p_1(x) &= 1 - x - \frac{1 - 2 x}{2 + \coth[\eta/2]\frac{k_1 \eta}{k_2}} 
		- \frac{\mu}{k_1}\frac{1}{\eta^2} \left(D_1 \exp[\eta x]
		+ D_2 \exp[-\eta x] \right) \\
		p_2(x) &= 1 - x - \frac{1 - 2 x}{2 + \coth[\eta/2]\frac{k_1 \eta}{k_2}} 
		+ \frac{\mu}{k_2}\frac{1}{\eta^2} \left(D_1 \exp[\eta x]
		+ D_2 \exp[-\eta x] \right) \\
		u_{1}(x) &= \frac{k_1}{\mu} - \frac{k_1}{\mu} \frac{2}{2 + \coth[\eta/2]\frac{k_1 \eta}{k_2}} 
		+ \frac{1}{\eta} (D_1 \exp[\eta x] - D_2 \exp[-\eta x]) \\
		u_{2}(x) &= \frac{k_2}{\mu} - \frac{k_2}{\mu} \frac{2}{2 + \coth[\eta/2]\frac{k_1 \eta}{k_2}} - \frac{1}{\eta} (D_1 \exp[\eta x] - D_2 \exp[-\eta x]) 
	\end{align}
\end{subequations}
Note that one cannot find $p_2(x)$ uniquely under
Darcy equations, as the boundary conditions for
the micro-pore network are all velocity boundary
conditions. All one can say about the solution
for $p_2(x)$ under Darcy equations is that it
is an arbitrary constant. On the other hand,
one can find uniquely the solution for $p_2(x)$
under the double porosity/permeability model. 

\subsubsection{Comparison with Darcy equations}
If only the macro-pore network is present, the boundary 
conditions of the macro-pore network imply that the 
pressure and the velocity take the following forms:
\begin{align}
	u_{1}(x) = \frac{k_1}{\mu} 
	\quad \mathrm{and} \quad  
	p_{1}(x) = 1 - x 
\end{align}
If only the micro-pore network is present, the boundary 
conditions of the micro-pore network imply that 
\begin{align}
	u(x) = 0 
	\quad \mathrm{and} \quad 
	p(x) = \mbox{an arbitrary constant}
\end{align}
If the standard permeability test is performed on 
a porous medium (which has both pore-networks)
the permeability will be:
\begin{align}
	\label{Dual_K_eff}
	k_{\mathrm{eff}} = \mu u_1(0) = \frac{k_1 + k_2}{2 \tanh[\eta/2]\frac{k_2}{k_1 \eta}+ 1}
\end{align}
Equation \eqref{Dual_K_eff} implies that one can relate 
the experimental value of the effective permeability to 
the permeabilities of macro- and micro-pore networks. 
This clearly shows the need to know the internal 
pore-structure for an accurate modeling of porous 
media. 

Figure \ref{Fig:Double_1D_P2_macro_micro_velocities} shows
the variation of velocity in micro-pore network and the mass 
transfer across the pore-networks
for various values of $\eta$ and for two different cases
$k_{1} < k_{2}$ and $k_{1} > k_{2}$.
It is observed that although there is no
supply of fluid on the boundaries of the micro-pore network, 
there will still be discharge (i.e., non-zero velocity) within
the micro-pore network. This reveals that the internal pore-structure is an important factor characterizing the flow in a complicated porous medium. One can analyze the internal pore-structure using modern techniques such as $\mu$-CT. In other words, the surface pore-structure cannot solely specify the flow within the domain. Moreover, whether the permeability of the macro-pores is larger or smaller than the permeability of the micro-pores, we will still have flow in the micro-pore network.

Figure \ref{Fig:Comparison_Velocities_Darcy_Dual} compares
the velocities under the double porosity/permeability
model and the Darcy equations. Here, the permeability 
used in the Darcy equations is the effective permeability 
introduced in equation \eqref{Dual_K_eff}. Macro- and 
micro-velocities and their summation (i.e., $u_{1} + u_{2}$) under the double porosity/permeability model as well as the velocity under the Darcy equations are displayed. As it can be seen for both cases $k_{1}> k_{2}$ and $k_{1}< k_{2}$, under the Darcy equations the velocity throughout the domain for this one-dimensional boundary value problem is a horizontal line where the constant value is equal to the summation of the macro- and micro-velocities under the double porosity/permeability model. This implies that the effective permeability $k=k_{\mathrm{eff}}$, which is obtained by the classical Darcy experiment, cannot completely capture the complex internal pore-structure of the porous medium. This is due to the fact that the experimental value obtained for $k_{\mathrm{eff}}$ does not account for the case of multiple pore-networks within the domain. It just assumes a single pore-network, and the effective permeability is calculated based on the surface pore-structure of the specimen. 

\subsection{Two-dimensional boundary value problem}
This problem pertains to the flow of water in candle
filters, which are widely used for purifying drinking
water \citep{dickenson1997filters}. Consider a circular
disc of inner radius $r_i = a$ and outer radius of
$r_o = 1$. 
The inner surface of the cylinder is subjected to
a pressure, and the outer surface of the cylinder
is exposed to the atmosphere. For the micro-pore 
network, there is no discharge from the inner and
outer surfaces of the cylinder.
Figure~\ref{Fig:Double_description_and_pressure}
provides a pictorial description of the problem. 

We shall employ cylindrical polar coordinates.
Noting the underlying symmetry in the problem,
the variables, $u_1$, $u_2$, $p_1$ and $p_2$,
are assumed to be functions of $r$ only. The
governing equations can be written as follows:
%
{\small{
		\begin{subequations}
			\begin{align}
				&\frac{\mu}{k_1} u_{1} + \frac{dp_1}{dr} = 0, \;  
				\frac{1}{r} \frac{d (r u_{1})}{dr} + (p_1 - p_2) = 0 
				\quad \forall r \in (a,1) \\ 
				&\frac{\mu}{k_2} u_{2} + \frac{dp_2}{dr} = 0, \; 
				\frac{1}{r} \frac{d (r u_{2})}{dr} - (p_1 - p_2) = 0
				\quad \forall r \in (a,1) \\
				&p_{1}(r = a) = 1, \quad p_{1}(r = 1) = 0 \\  
				&u_{2}(r = a) = 0, \quad u_{2}(r = 1) = 0
			\end{align}
		\end{subequations}
	}
}
This implies that the mass transfer $\chi(r)$
satisfies the following differential equation:
\begin{align}
	\chi^{''} + \frac{1}{r}\chi^{'} - \eta^2 \chi = 0 
\end{align}
which is a (homogeneous) modified Bessel ordinary
differential equation \citep{Bowman_Bessel}. A
general solution to the above ordinary differential
equation can be written as follows:
\begin{align}
	\chi(r) = p_2(r) - p_1(r) = C_{3} I_{0}(\eta r)
	+ C_{4} K_{0}(\eta r)
\end{align}
where $I_{0}(z)$ and $K_{0}(z)$ are, respectively, zeroth-order
modified Bessel functions of first and second kinds. Noting
that $I_{0}^{'}(z) = I_{1}(z)$ and $K_{0}^{'}(z) = -K_{1}(z)$,
the analytical solution can be written as follows:
\begin{subequations}
	\begin{align}
		p_{1}(r) &= 
		\frac{\ln[r]}{\ln[a]} C_1 + C_2 - \frac{\alpha_1}{\eta^2} \left( C_3 I_{0}(\eta r)
		+ C_4 K_{0}(\eta r) \right) \\
		p_{2}(r) &= 
		\frac{\ln[r]}{\ln[a]} C_1 + C_2 + \frac{\alpha_2}{\eta^2} \left( C_3 I_{0}(\eta r)
		+ C_4 K_{0}(\eta r) \right) \\
		u_{1}(r) &= -\frac{k_1}{\mu} \frac{C_1}{r\ln[a]}
		+ \frac{1}{\eta} \left(C_3 I_{1}(\eta r) - C_4 K_{1}(\eta r)\right)  \\
		u_{2}(r) &= -\frac{k_2}{\mu} \frac{C_1}{r\ln[a]}
		- \frac{1}{\eta} \left(C_3 I_{1}(\eta r) - C_4 K_{1}(\eta r)
		\right)
	\end{align}	
\end{subequations}
The boundary conditions give rise to the following coefficients:
\begin{subequations}
	\begin{align}
		&C_{1} = \delta \eta^{2} a \ln[a]
		\left(I_1(a \eta) K_1(\eta) - I_1(\eta) K_1(a \eta) \right)
		\alpha_2 \\
		&C_{2} = \delta \left( -1 + a \eta I_1(a \eta)
		K_0(\eta)+ a \eta I_0(\eta)K_1(a \eta) \right)\alpha_1 \\
		&C_{3} =  \delta \eta ^{3} (a K_1(a \eta) - K_1(\eta)) \\
		&C_{4} =  \delta \eta ^{3}(a I_1(a \eta) - I_1(\eta))
	\end{align}
\end{subequations}
where
\begin{align}
	\delta^{-1} = & \{ -2 + a \eta I_1(a \eta) K_0(\eta) + \eta I_1(\eta) K_0(a \eta) \nonumber \\
	&+ \eta I_0(a \eta) K_1(\eta) + a \eta I_0(\eta) K_1(a\eta) \} \alpha_1  \nonumber \\
	&+ \eta^{2} a \ln[a] \left(I_1(a \eta) K_1(\eta) - I_1(\eta) K_1(a \eta) \right)\alpha_2 
\end{align}

For comparison, the pressure and the discharge
velocity under Darcy equations with constant
permeability $k$ and with boundary conditions
$p(r = a) = 1$ and $p(r=1) = 0$ can be written
as follows:
\begin{align}
	p(r) = \frac{\ln[r]}{\ln[a]}
	\quad \mathrm{and} \quad
	u(r) = -\frac{k}{\mu} \frac{1}{r \ln[a]}
\end{align}
It is evident that the velocity and pressure profiles
under the double porosity/permeability model are much
more complicated than the corresponding profiles under
Darcy equations. Figure~\ref{Fig:Double_description_and_pressure}
illustrates the qualitative difference between the pressures under
the double porosity/permeability model and Darcy equations. The
graph of the pressures under Darcy equations is always convex
while the graph of the macro-pressure under the double
porosity/permeability model has both convex and concave
parts.
It should also be noted that, although there is no discharge
from the micro-pore network on the boundary, there is discharge
in the micro-pore network within the domain. 

\section{CONCLUDING REMARKS}
\label{Sec:S7_Double_CR}
In this paper, several contributions have been made
to the modeling of fluid flow in porous media with
double porosity/permeability.
\emph{First}, a thermodynamic basis for models 
studying flow in porous media exhibiting double 
porosity/permeability has been provided using the 
maximization of rate of dissipation hypothesis. 
This model nicely allows for further generalizations 
of the existing models. The mass transfer across the 
macro-pore and micro-pore networks has been obtained 
in a systematic manner by treating it as an internal 
variable and maximizing a prescribed (physical) 
dissipation functional.
\emph{Second}, various mathematical properties that
the solutions under the double porosity/permeability
model satisfy have been presented along with their
proofs.
\emph{Third}, a maximum principle has been established
for the double porosity/permeability model. The main 
differences between the maximum principles of Darcy 
equations and the double porosity/permeability model 
have been discussed.
\emph{Fourth}, an analytical solution procedure
based on the Green's function method has been presented
for a general boundary value problem under the double
porosity/permeability model.
\emph{Last but not least}, using the analytical
solutions of some canonical problems, the salient features
of the pressures and velocities in the macro-pore and
micro-pore networks under the double porosity/permeability
model have been highlighted.
Some of the significant findings of the paper can be summarized as follows:
\begin{enumerate}[(C1)]
\item[(C1)] In general, the pressure and velocity profiles under the double
  porosity/permeability model are qualitatively and quantitatively different
  from the corresponding ones under the classical Darcy equations. 
  Moreover, the solution using 
  the effective permeability, which is obtained by 
  the classical Darcy experiment, does not necessarily 
  match the solution under the double porosity/permeability 
  model. 
  These differences can be attributed to 
  the complex nature of a porous medium 
  that exhibits double porosity/permeability. 
  However, there are situations under 
  which the solutions under the double 
  porosity/permeability model can be 
  adequately described by Darcy 
  equations, which have been discussed 
  in this paper. 
\item[(C2)] The maximum and minimum pressures need not occur on the boundary
  under the double porosity/permeability model. This is in contrast with the
  case of Darcy equations under which the maximum and minimum pressures
  occur on the boundary for a pressure-prescribed boundary value problem.
\item[(C3)] The solution under the double porosity/permeability
  model is unique even for those boundary value problems in which
  pressure conditions are prescribed on the entire boundary for either
  macro-pore network or micro-pore network. This is not the case
  with Darcy equations, as the pressure can be found up to an
  arbitrary constant if pressure conditions are prescribed on
  the entire boundary. 
\item[(C4)] There will be discharge in the micro-pore network even if
  there is no fluid supply on the boundaries of the micro-pore network.
  Therefore, it can be concluded that the surface pore-structure is not
  the only factor in characterizing the flow through a complex porous
  medium which highlights the need to use modern techniques (e.g., 
  $\mu$-CT) for studying the internal pore-structure.
\item[(C5)] There will be mass transfer across the two
  pore-networks whether the permeability of the macro-pore
  network is greater than the permeability of the micro-pore
  network or vice-versa. This means that the path that the
  fluid takes is not necessarily through the network with higher
  permeability. 
\end{enumerate}

An interesting extension of the research presented herein
can be towards the study of the flow of multi-phase fluids
in a porous medium with double porosity/permeability. Another
research endeavor can be towards coupling the deformation
of the porous solid with the flow in a porous medium with
double porosity/permeability.

\appendix
\section{SUPPLEMENTARY MATERIAL}
\label{Sec:Dual_SM}
We provide mathematical proofs to all the
theorems that have been stated in the main
manuscript. 

\subsection{Proof of minimum dissipation theorem (i.e.,
  Theorem \ref{Thm:Dual_Minimum_dissipation_Thm})}
\begin{proof}
  Let $\delta \mathbf{v}_{i}(\mathbf{x})
    := \widetilde{\mathbf{v}}_{i}(\mathbf{x})
    - \mathbf{v}_{i}(\mathbf{x})$. 
 Clearly, $\delta \mathbf{v}_1(\mathbf{x})$ and
$\delta \mathbf{v}_2(\mathbf{x})$ satisfy 
\begin{align}
\label{Eqn:Dual_delta_v_kinematic}
&\mathrm{div}\left[\phi_{1}\delta \mathbf{v}_{1}\right] 
+ \mathrm{div}\left[\phi_{2}\delta \mathbf{v}_{2}\right] 
= 0, \; 
\delta \mathbf{v}_{1}(\mathbf{x}) \cdot 
\widehat{\mathbf{n}}(\mathbf{x}) = 0,
\; \mathrm{and} \; \delta \mathbf{v}_{2}(\mathbf{x}) \cdot 
\widehat{\mathbf{n}}(\mathbf{x}) = 0
\end{align}
Now consider 
{\small
\begin{align}
\boldsymbol{\Phi}\left[\widetilde{\mathbf{v}}_{1},
  \widetilde{\mathbf{v}}_{2}\right]-\boldsymbol{\Phi}
\left[\mathbf{v}_{1},\mathbf{v}_{2}\right]
& = \sum_{i=1}^{2} \left(\int_{\Omega} \alpha_{i}
\delta\mathbf{v}_{i} \cdot \delta\mathbf{v}_{i}
\mathrm{d} \Omega  + \dfrac{1}{2}\int_{\Omega}
\dfrac{\mu}{\beta}\left(\mathrm{div}
\left[\phi_{i}\delta\mathbf{v}_{i}\right]\right)^{2}
\mathrm{d} \Omega \right)  \nonumber \\
& + \sum_{i=1}^{2} \left(2 \int_{\Omega} \alpha_{i} \delta
\mathbf{v}_{i} \cdot \mathbf{v}_{i} \mathrm{d} \Omega 
+ \int_{\Omega}\dfrac{\mu}{\beta} \mathrm{div}\left[\phi_{i}
  \delta\mathbf{v}_{i}\right]\mathrm{div}
\left[\phi_{i}\mathbf{v}_{i}\right] \mathrm{d} \Omega \right)
\end{align}
}
  Noting that the first two terms on the right
  side of the above equation are non-negative,
  we have 
  {\small
  \begin{align}
    \boldsymbol{\Phi}\left[\widetilde{\mathbf{v}}_{1},
      \widetilde{\mathbf{v}}_{2}\right] &- \boldsymbol{\Phi}
    \left[\mathbf{v}_{1},\mathbf{v}_{2}\right]
    \geq  \sum_{i=1}^{2} \left(2 \int_{\Omega} \alpha_{i}
    \delta\mathbf{v}_{i} \cdot \mathbf{v}_{i} \mathrm{d}
    \Omega
    + \int_{\Omega}\dfrac{\mu}{\beta} \mathrm{div}
    \left[\phi_{i}\delta\mathbf{v}_{i}\right]\mathrm{div}
    \left[\phi_{i}\mathbf{v}_{i}\right] \mathrm{d} \Omega
    \right) 
  \end{align}
}
  Using the balance of linear momentum for each pore-network
  (i.e., equations \eqref{Eqn:Dual_GE_BLM_1} and
  \eqref{Eqn:Dual_GE_BLM_2}), noting that $\gamma
  \mathbf{b}(\mathbf{x}) = -\mathrm{grad}[\psi]$,
  and employing Green's identity, we obtain 
   {\small
  \begin{align}
    \boldsymbol{\Phi}\left[\widetilde{\mathbf{v}}_{1},
      \widetilde{\mathbf{v}}_{2}\right] &- \boldsymbol{\Phi}
    \left[\mathbf{v}_{1},\mathbf{v}_{2}\right]
     -2 \sum_{i=1}^{2} \left(\int_{\partial \Omega} \phi_{i}
    \delta \mathbf{v}_{i} \cdot \widehat{\mathbf{n}}(\mathbf{x})
    \left(\psi + p_i\right) \mathrm{d} \Gamma \right) + 2 \int_{\Omega} \left(\sum_{i=1}^{2}\mathrm{div}
    \left[\phi_{i} \delta \mathbf{v}_{i}\right] \right)
    \psi \mathrm{d} \Omega \nonumber \\
    &+ \sum_{i=1}^{2} \left(2 \int_{\Omega} \mathrm{div}
    \left[\phi_{i} \delta\mathbf{v}_{i}\right] 
    p_{i} \mathrm{d} \Omega + \int_{\Omega}\dfrac{\mu}{\beta} \mathrm{div}
    \left[\phi_{i}\delta\mathbf{v}_{i}\right]\mathrm{div}
    \left[\phi_{i}\mathbf{v}_{i}\right] \mathrm{d} \Omega
    \right)
  \end{align}
}
  Using equation \eqref{Eqn:Dual_delta_v_kinematic} we have
  {\small
  \begin{align}
    &\boldsymbol{\Phi}\left[\widetilde{\mathbf{v}}_{1},
      \widetilde{\mathbf{v}}_{2}\right] - \boldsymbol{\Phi}
    \left[\mathbf{v}_{1},\mathbf{v}_{2}\right]
     \geq \sum_{i=1}^{2} \left(2 \int_{\Omega} \mathrm{div}
    \left[\phi_{i} \delta\mathbf{v}_{i}\right] 
    p_{i} \mathrm{d} \Omega 
    + \int_{\Omega}\dfrac{\mu}{\beta} \mathrm{div}
    \left[\phi_{i}\delta\mathbf{v}_{i}\right]\mathrm{div}
    \left[\phi_{i}\mathbf{v}_{i}\right] \mathrm{d} \Omega
    \right)
  \end{align}
}
  Using equations \eqref{Eqn:Dual_GE_mass_balance_1}
  and \eqref{Eqn:Dual_GE_mass_balance_2}, we obtain
  {\small 
  \begin{align}
    \boldsymbol{\Phi}\left[\widetilde{\mathbf{v}}_{1},
      \widetilde{\mathbf{v}}_{2}\right]
    - \boldsymbol{\Phi}\left[\mathbf{v}_{1},\mathbf{v}_{2}\right]
    \geq \int_{\Omega} \left(\sum_{i=1}^{2} \mathrm{div}
    \left[\phi_{i} \delta \mathbf{v}_{i} \right] \right)
    (p_1 + p_2) \mathrm{d} \Omega 
  \end{align}
}
  Invoking equation $\eqref{Eqn:Dual_delta_v_kinematic}_{1}$
  gives the desired result.
\end{proof}

\subsection{Proof of uniqueness of solutions
  (i.e., Theorem \ref{Thm:Dual_Uniqueness_theorem})}
\begin{proof}
  On the contrary, assume that $\left(\mathbf{v}_{1}^{'},
  \mathbf{v}_{2}^{'}, p_{1}^{'}, p_{2}^{'}, \chi^{'} \right)$
  and $\left(\mathbf{v}_{1}^{*}, \mathbf{v}_{2}^{*}, p_{1}^{*},
  p_{2}^{*}, \chi^{*} \right)$ are two sets of solutions.
  Now consider the following quantity, which
  is a sum of three non-negative integrals:
    {\small 
  	\begin{align}
  	\label{Eqn:Dual_Definition_of_I}
  	\mathcal{I}
  	&:= \int_{\Omega} \alpha_{1} \left(\mathbf{v}_{1}^{'} - \mathbf{v}_{1}^{*}\right)
  	\cdot \left(\mathbf{v}_{1}^{'} - \mathbf{v}_{1}^{*} \right) \mathrm{d} \Omega
  	+\int_{\Omega} \alpha_{2} \left(\mathbf{v}_{2}^{'} - \mathbf{v}_{2}^{*}\right)
  	\cdot \left(\mathbf{v}_{2}^{'} - \mathbf{v}_{2}^{*} \right) \mathrm{d} \Omega
  	+ \int_{\Omega} \dfrac{\mu}{\beta} \left(\chi^{'} - \chi^{*} \right)^{2}
  	\mathrm{d} \Omega
  	\end{align}
  }
  We start by simplifying the first integral. Noting
  that both $\mathbf{v}^{'}_{1}(\mathbf{x})$ and
  $\mathbf{v}^{*}_{1}(\mathbf{x})$ satisfy the
  balance of linear momentum for the macro-pore
  network (i.e., equation \eqref{Eqn:Dual_GE_BLM_1}),
  and using Green's identity, we obtain:
    {\small
  	\begin{align}
  	\int_{\Omega} \alpha_{1} \left(\mathbf{v}^{'}_{1}
  	- \mathbf{v}^{*}_{1}\right) \cdot \left(\mathbf{v}^{'}_{1}
  	- \mathbf{v}^{*}_{1}\right) \mathrm{d} \Omega 
  	&= \int_{\Omega} \left(p^{'}_{1} - p^{*}_{1} \right)
  	\mathrm{div} \left[\phi_{1} \left(\mathbf{v}^{'}_{1} -
  	\mathbf{v}^{*}_{1}\right) \right] \mathrm{d}\Omega
  	\nonumber \\
  	&- \int_{\Gamma_{1}^{p}} \left(p^{'}_{1} - p^{*}_{1} \right)
  	\phi_{1} \left(\mathbf{v}^{'}_{1} - \mathbf{v}^{*}_{1}\right)
  	\cdot \widehat{\mathbf{n}}(\mathbf{x}) \mathrm{d} \Gamma \nonumber \\
  	&- \int_{\Gamma_{1}^{v}} \left(p^{'}_{1} - p^{*}_{1} \right)
  	\phi_{1} \left(\mathbf{v}^{'}_{1} - \mathbf{v}^{*}_{1}\right)
  	\cdot \widehat{\mathbf{n}}(\mathbf{x}) \mathrm{d} \Gamma 
  	\end{align}
  }
  Noting that $p^{'}_1(\mathbf{x}) = p^{*}_1(\mathbf{x})
  = p_{01}(\mathbf{x})$ on $\Gamma_{1}^{p}$,
  $\mathbf{v}^{'}_1(\mathbf{x}) \cdot \widehat{\mathbf{n}}
  (\mathbf{x}) = \mathbf{v}^{*}_1(\mathbf{x}) \cdot
  \widehat{\mathbf{n}}(\mathbf{x}) = v_{n1}(\mathbf{x})$
  on $\Gamma_{1}^{v}$, and using equation
  \eqref{Eqn:Dual_GE_mass_balance_1}, we obtain 
  {\small
  \begin{align}
    \label{Eqn:Dual_Thm_term1}
    \int_{\Omega} \alpha_{1} \left(\mathbf{v}^{'}_{1}
    - \mathbf{v}^{*}_{1}\right) \cdot \left(\mathbf{v}^{'}_{1}
    - \mathbf{v}^{*}_{1}\right) \mathrm{d} \Omega
    = \int_{\Omega} \left(p^{'}_{1} - p^{*}_{1} \right)
    \left(\chi^{'} - \chi^{*} \right) \mathrm{d} \Omega 
  \end{align}
}
  A similar simplification of the second integral gives:
  {\small
  \begin{align}
    \label{Eqn:Dual_Thm_term2}
    \int_{\Omega} \alpha_{2} \left(\mathbf{v}^{'}_{2}
    - \mathbf{v}^{*}_{2}\right) \cdot \left(\mathbf{v}^{'}_{2}
    - \mathbf{v}^{*}_{2}\right) \mathrm{d} \Omega
    = -\int_{\Omega} \left(p^{'}_{2} - p^{*}_{2} \right)
    \left(\chi^{'} - \chi^{*} \right) \mathrm{d} \Omega 
  \end{align}
}
  Note that
  \begin{align}
    \label{Eqn:Dual_Thm_pressure_chi}
   & p_1^{'}(\mathbf{x}) - p_2^{'}(\mathbf{x})
    = -\frac{\mu}{\beta} \chi^{'}(\mathbf{x}), \quad p_1^{*}(\mathbf{x}) - p_2^{*}(\mathbf{x})
    = -\frac{\mu}{\beta} \chi^{*}(\mathbf{x})
  \end{align}
  Equations \eqref{Eqn:Dual_Thm_term1}, \eqref{Eqn:Dual_Thm_term2}
  and \eqref{Eqn:Dual_Thm_pressure_chi} imply that $\mathcal{I} = 0$.
  Since each integral and each integrand in $\mathcal{I}$
  is non-negative, we conclude that
  \begin{align*}
    \mathbf{v}^{'}_{1}(\mathbf{x})
    = \mathbf{v}^{*}_{1}(\mathbf{x}), \; 
    \mathbf{v}^{'}_{2}(\mathbf{x})
    = \mathbf{v}^{*}_{2}(\mathbf{x}), 
    \; \mathrm{and} \; 
    \chi^{'}(\mathbf{x}) = \chi^{*}(\mathbf{x})
  \end{align*}
  and thus the uniqueness of the
  solution can be established. 
\end{proof}

  The mechanics basis in the proof for the 
  uniqueness will be evident by noting that 
  the integral $\mathcal{I}$, which is defined 
  in equation \eqref{Eqn:Dual_Definition_of_I}, 
  is related to the physical dissipation 
  functional, $\boldsymbol{\Phi}$, defined in equation 
  \eqref{Total_dissipation}, for the solution 
  fields. That is, 
  \[
  \mathcal{I} = \Phi[\mathbf{v}_1^{'} - \mathbf{v}_1^{*},
    \mathbf{v}_2^{'} - \mathbf{v}_2^{*}]
  \]
  if $\left(\mathbf{v}_{1}^{'}, \mathbf{v}_{2}^{'}, p_{1}^{'},
  p_{2}^{'}, \chi^{'} \right)$ and $\left(\mathbf{v}_{1}^{*},
  \mathbf{v}_{2}^{*}, p_{1}^{*}, p_{2}^{*}, \chi^{*} \right)$
  are the solutions of the double porosity/permeability
   model. In general, $\mathcal{I}$ is not equal
  to $\Phi[\mathbf{v}_1^{'} - \mathbf{v}_1^{*},\mathbf{v}_2^{'}
  - \mathbf{v}_2^{*}]$. 
  One can alternatively employ the mathematical tools
  from functional analysis to establish uniqueness.
  Herein, uniqueness was established using mechanics-based
  arguments, and hence, it is believed that the above proof
  will have some pedagogical value in engineering education.  

\subsection{Proof of the reciprocal relation (i.e.,
  Theorem \ref{Thm:Dual_Reciprocal_Thm})}
\begin{proof}
  Using equations \eqref{Eqn:Dual_GE_BLM_1}--\eqref{Eqn:Dual_GE_pBC_2} 
  and \eqref{Eqn:Dual_GE_mass_transfer}, and 
  invoking Green's identity, one can show that each side 
  of the equality in equation \eqref{Eqn:Dual_Reciprocal_term} 
  is:
   {\small{
  		\begin{align*}
  		\int_{\Omega} \alpha_{1} \mathbf{v}_{1}^{'}(\mathbf{x}) 
  		\cdot \mathbf{v}_{1}^{*}(\mathbf{x}) \; \mathrm{d} 
  		\Omega &+ \int_{\Omega} \alpha_{2} \mathbf{v}_{2}^{'}(\mathbf{x}) 
  		\cdot \mathbf{v}_{2}^{*}(\mathbf{x}) \; \mathrm{d} 
  		\Omega 
  		+ \int_{\Omega} \frac{\beta}{\mu} (p_1^{'} - p_2^{'}) 
  		(p_1^{*} - p_2^{*}) \; \mathrm{d} \Omega 
  		\end{align*}
  	}
  }
  This completes the proof. 
\end{proof}

\subsection{Proof of maximum principle (i.e.,
  Theorem \ref{Thm:Dual_Maximum_Principle})}
\begin{proof}
  Noting that the permeabilities are isotropic
  and homogeneous, equations
  \eqref{Eqn:Dual_GE_BLM_1}--\eqref{Eqn:Dual_GE_vBC_2}
  give rise to the following boundary value problem:
  \begin{subequations}
    \begin{alignat}{2}
      \label{Eqn:Dual_Diffusion_with_decay}
      &\beta \left(\frac{1}{k_1} + \frac{1}{k_2}\right) (p_{1} - p_{2}) 
      - \Delta (p_{1} - p_{2}) = 0
      &&\quad \mathrm{in} \; \Omega \\
      \label{Eqn:Dual_Diffusion_with_decay_BC}
      &p_1(\mathbf{x}) - p_2(\mathbf{x})
      = p_{01}(\mathbf{x}) - p_{02}(\mathbf{x})
      &&\quad \mathrm{on} \; \partial \Omega 
    \end{alignat}
  \end{subequations}
  where $\Delta$ is the Laplacian operator.
  Note that
  \begin{align}
    \beta \left(\frac{1}{k_1} + \frac{1}{k_2} \right)
    \geq 0 
  \end{align}
  The boundary value problem given by equations
  \eqref{Eqn:Dual_Diffusion_with_decay}--\eqref{Eqn:Dual_Diffusion_with_decay_BC} is a diffusion
  equation with decay defined in terms of $p_1(\mathbf{x}) - p_2(\mathbf{x})$.
  Moreover, equation \eqref{Eqn:Dual_Diffusion_with_decay}
  is a homogeneous partial differential equation, and
  Dirichlet boundary conditions are prescribed on the
  entire boundary. From the theory of partial differential
  equations \cite{Gilbarg_Trudinger}, it is well-known that
  the solutions to such a boundary value problem satisfy a
  maximum principle, which implies that the non-negative
  maximum and the non-positive minimum occur on the boundary.
  Specifically, using \cite[Theorem 1]{Gilbarg_Trudinger},
  we conclude that 
    {\small{
  		\begin{align}
  		\mathrm{min}\left[0,\mathop{\mathrm{min}}_{\mathbf{x} \in
  			\partial \Omega} \left[p_{01}(\mathbf{x}) - p_{02}(\mathbf{x})
  		\right] \right]
  		\le p_1(\mathbf{x}) - p_2(\mathbf{x}) \le
  		\mathrm{max}\left[0,\mathop{\mathrm{max}}_{\mathbf{x} \in
  			\partial \Omega} \left[p_{01}(\mathbf{x}) - p_{02}(\mathbf{x})
  		\right] \right] 
  		\end{align}
  	}
  }
\end{proof}

It needs to be mentioned that it is possible to extend 
the above maximum principle to the case in which Neumann 
boundary conditions are prescribed on a part of the 
boundary. Such a discussion, however, may need a functional 
analysis treatment. It is also possible to extend it to weak 
solutions (i.e., where the difference in pressures is not 
twice differentiable). But, weak solutions and a functional 
analysis treatment of the double porosity/permeability model 
are beyond the scope of this paper. Some of these aspects 
will be addressed in a subsequent paper
\cite{Nakshatrala_Joodat_Ballarini_P2}.

\bibliographystyle{plainnat}
\bibliography{Master_References/Books,Master_References/Master_References}

\newpage
%
\begin{figure}
	\subfigure[Double porosity structure of a typical stone wall. \label{Fig:Stone_wall}]{
		\includegraphics[clip,scale=0.34]{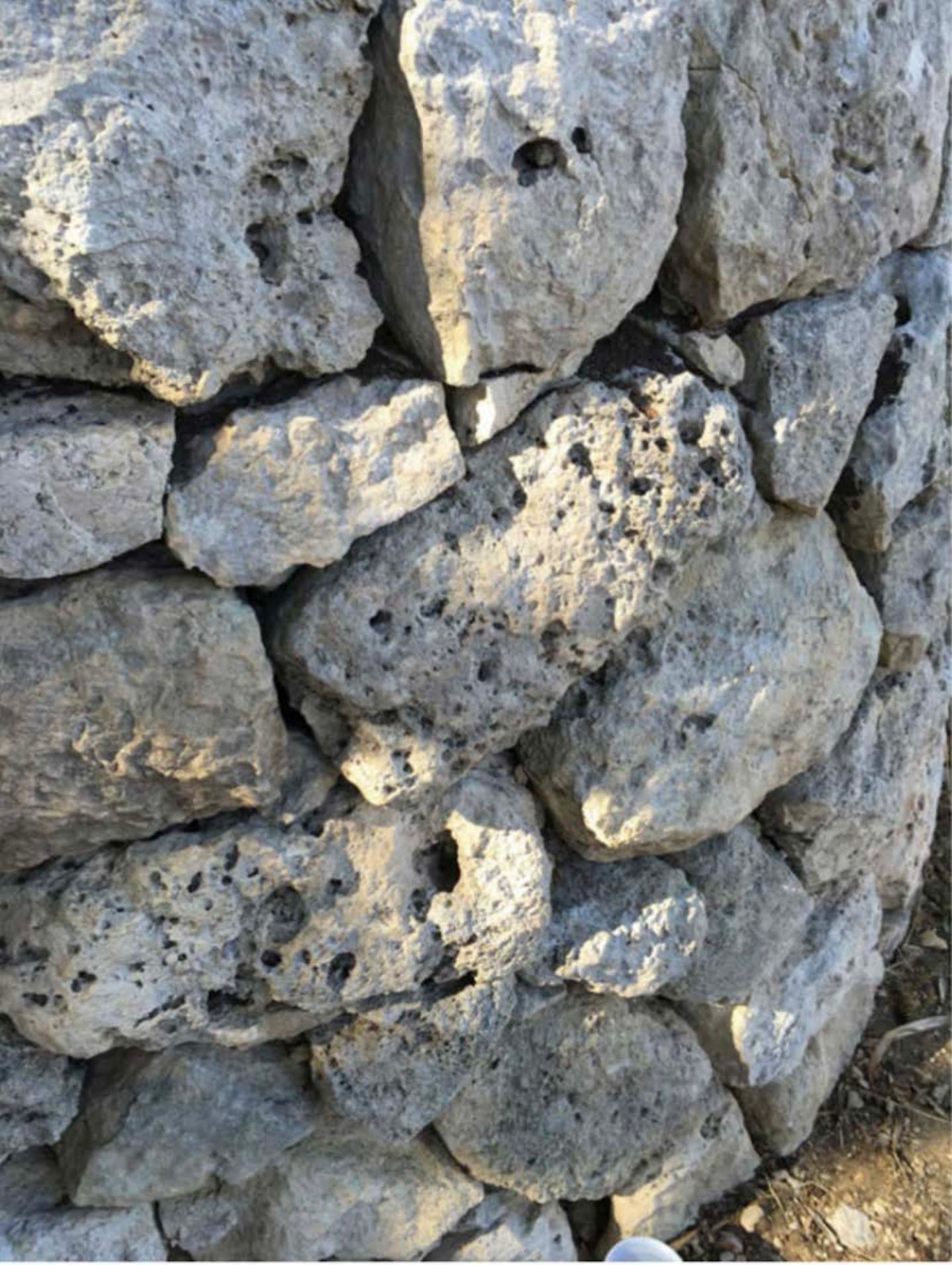}}
	\subfigure[Double porosity structure in lava. \label{Fig:Lava}]{
		\includegraphics[clip,scale=0.266]{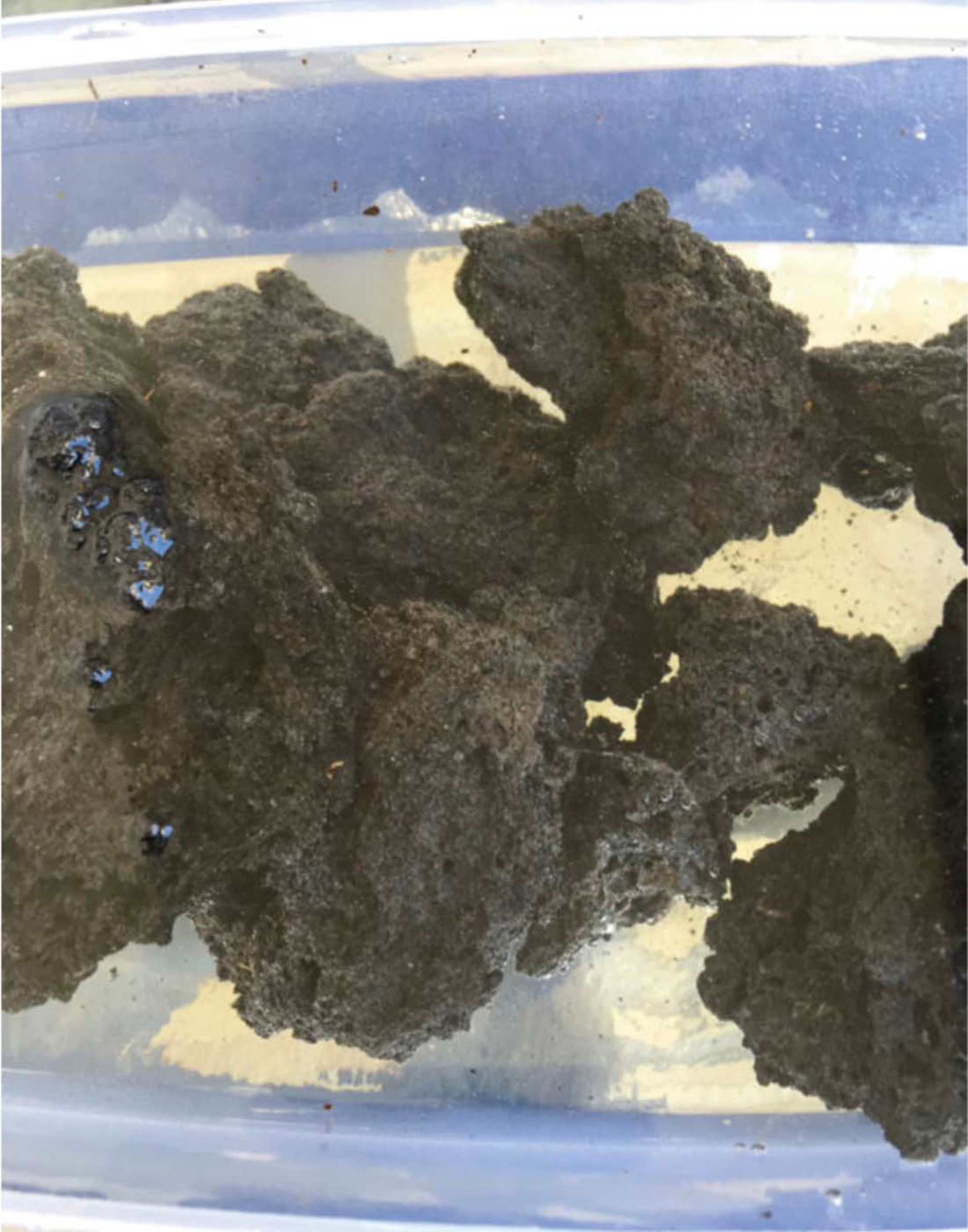}}
	\subfigure[Conceptualization of a synthetic double porosity medium. \label{Fig:man_made}]{
		\includegraphics[clip,scale=0.5]{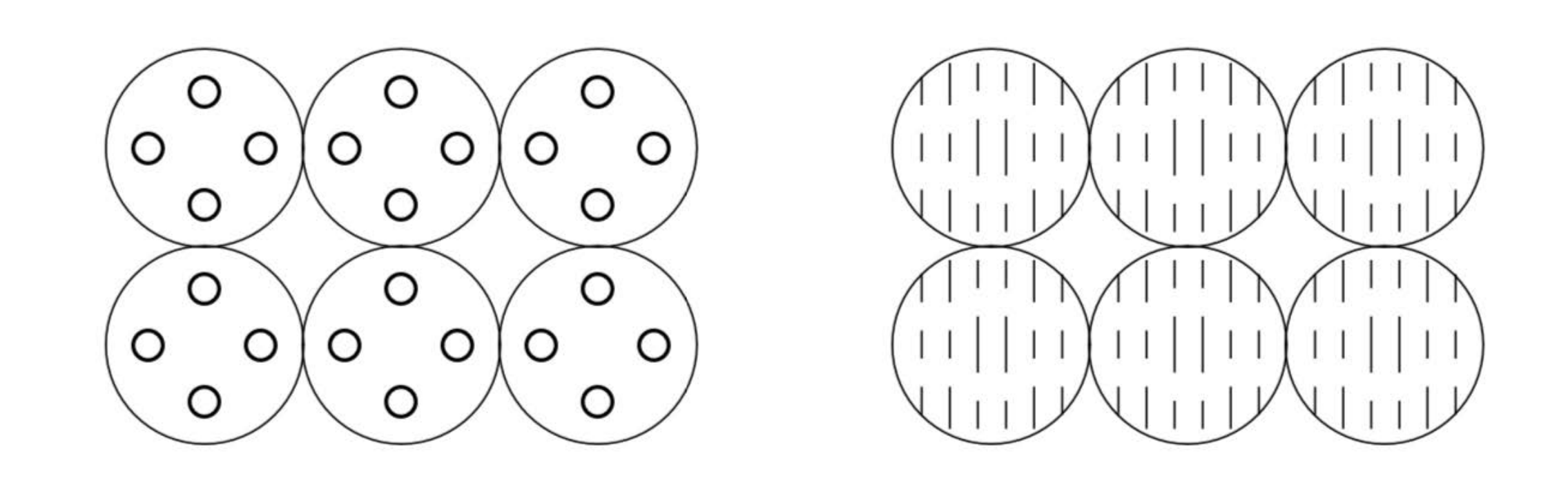}}
	\caption{Examples of natural and synthetic double
		porosity materials. Figures are taken from
		\cite{straughan2017mathematical}.
		\label{Dual_porosity_examples}}
\end{figure}

\begin{figure}
	\includegraphics[clip,scale=0.18]{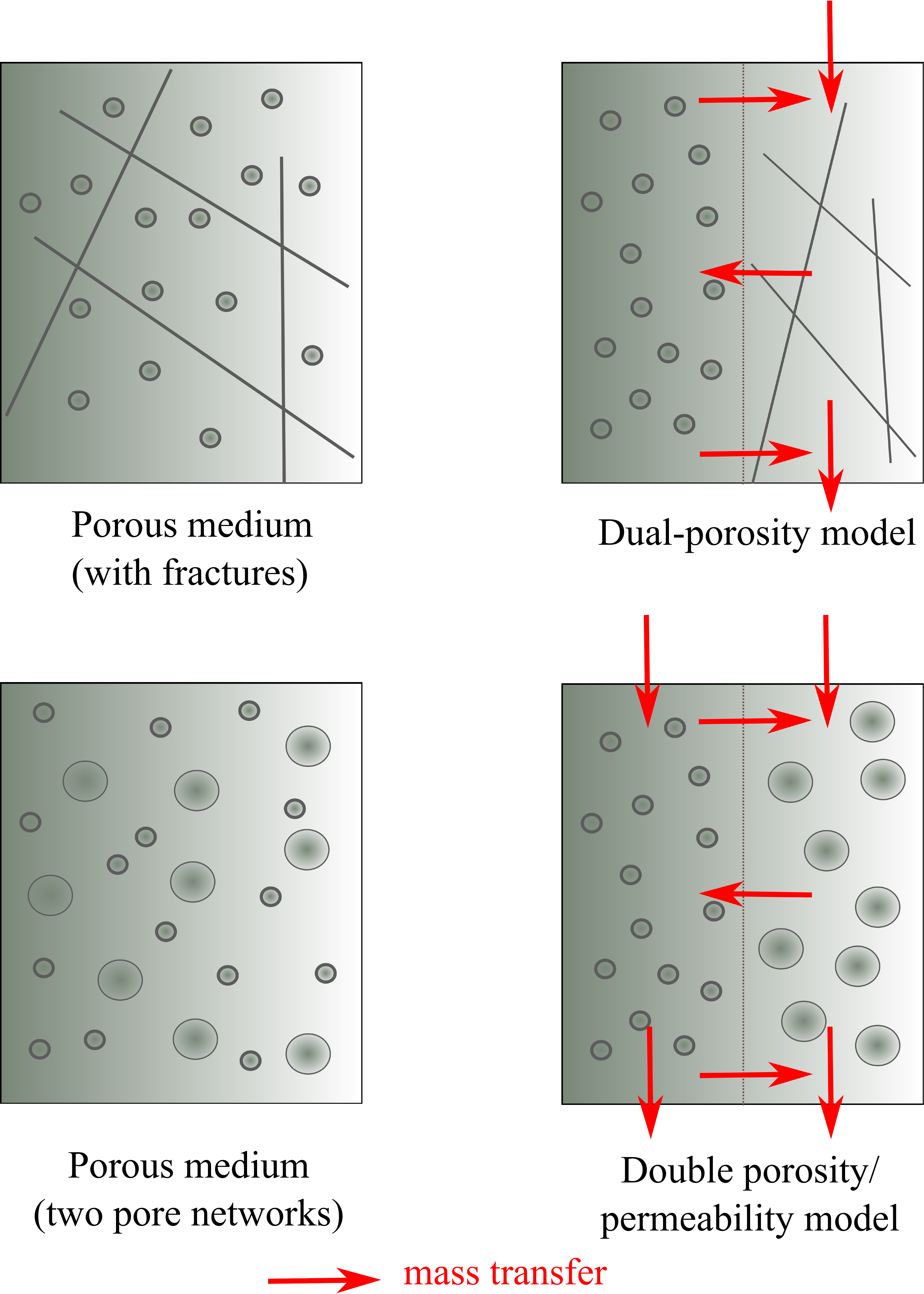}
	\caption{Porous media and their idealizations: Top
		part of the figure displays the idealization of a fractured
		porous medium using the dual-porosity model and the
		bottom part shows the idealization of a porous medium
		with two distinct pore-networks using the double
		porosity/permeability model. The arrows represent
		the fluid pathways and the mass transfer within the
		domain.
		\label{Dual_porosity_Permeability_schematic}}
\end{figure}

\begin{figure}[!h]
	\centering
	\subfigure[$p_2^{\mathrm{R}} < p_2^{\mathrm{L}} < 1$]{
		\includegraphics[clip,scale=0.26]{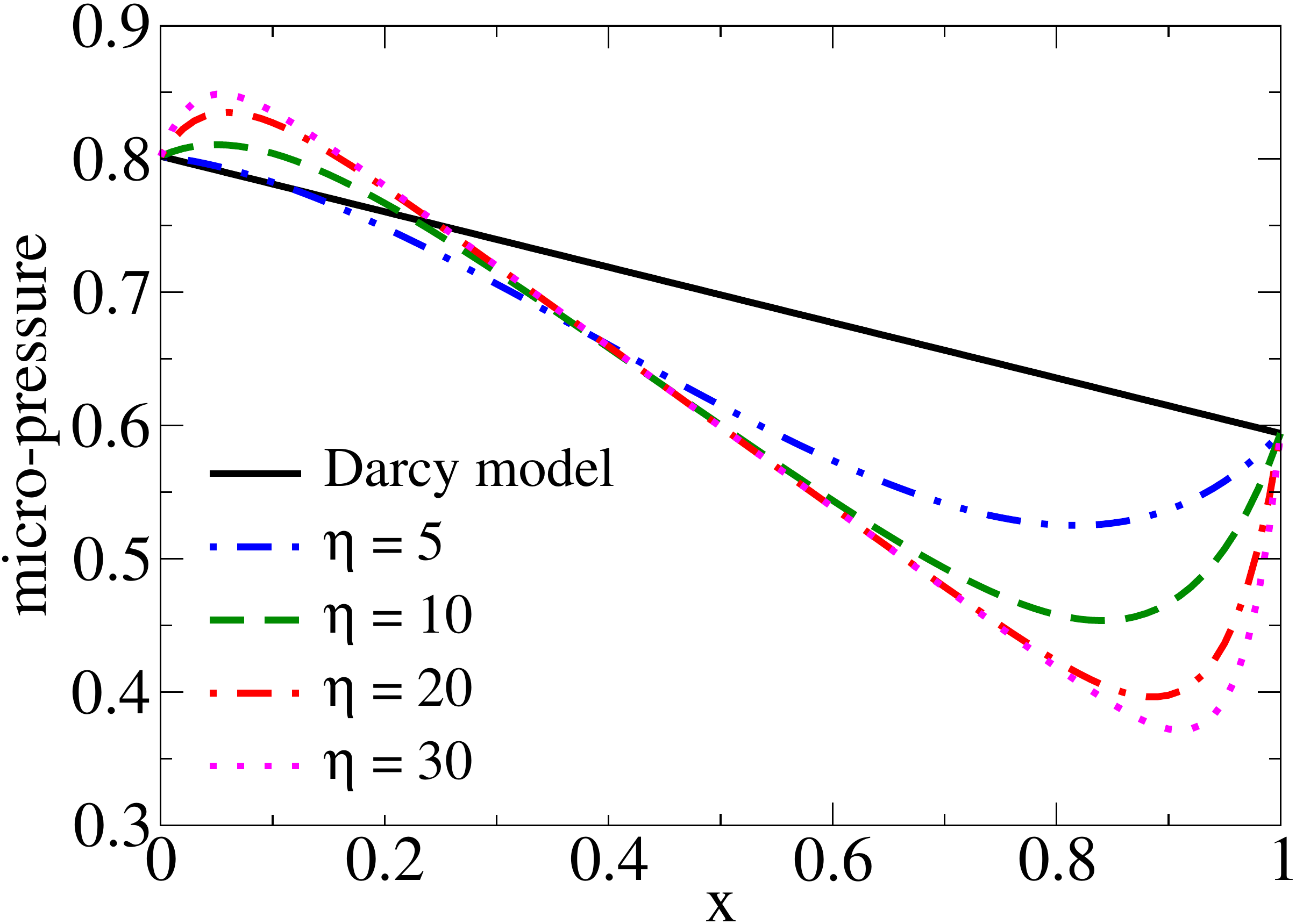}} 
	\subfigure[$1 < p_2^{\mathrm{R}} < p_2^{\mathrm{L}}$]{
		\includegraphics[clip,scale=0.26]{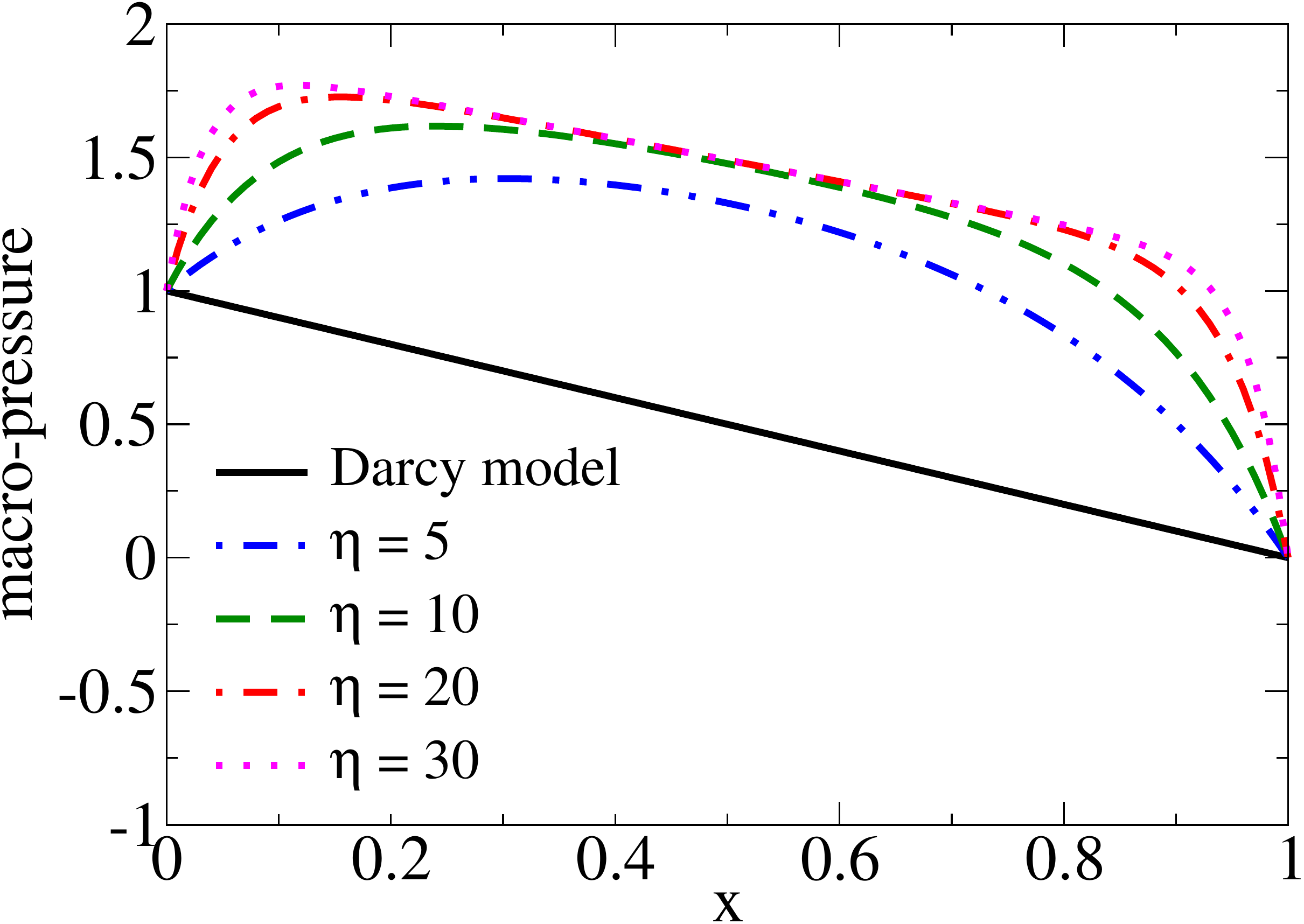}}
	\caption{\textsf{\small One-dimensional problem \#1:}~Variation
		of micro-pressure and macro-pressure in the one-dimensional
		domain. For comparison, the analytical solution under
		Darcy equations is also plotted. The maximum and minimum
		pressures in the pore-networks need not occur on
		the boundary in the case of double porosity/permeability
		model. The parameter $\eta$ is defined in equation
		\eqref{Eqn:Dual_1D_problem_1_eta}.
		\label{Fig:Double_1D_P1_pressure_analytical}}
\end{figure}

\begin{figure}[!h]
	\subfigure[$0 = p_1^{\mathrm{R}} < 0.3 = p_2^{\mathrm{R}}
	< p_2^{\mathrm{L}} = 0.9 < p_1^{\mathrm{L}} = 1$]{
		\includegraphics[clip,scale=0.26]{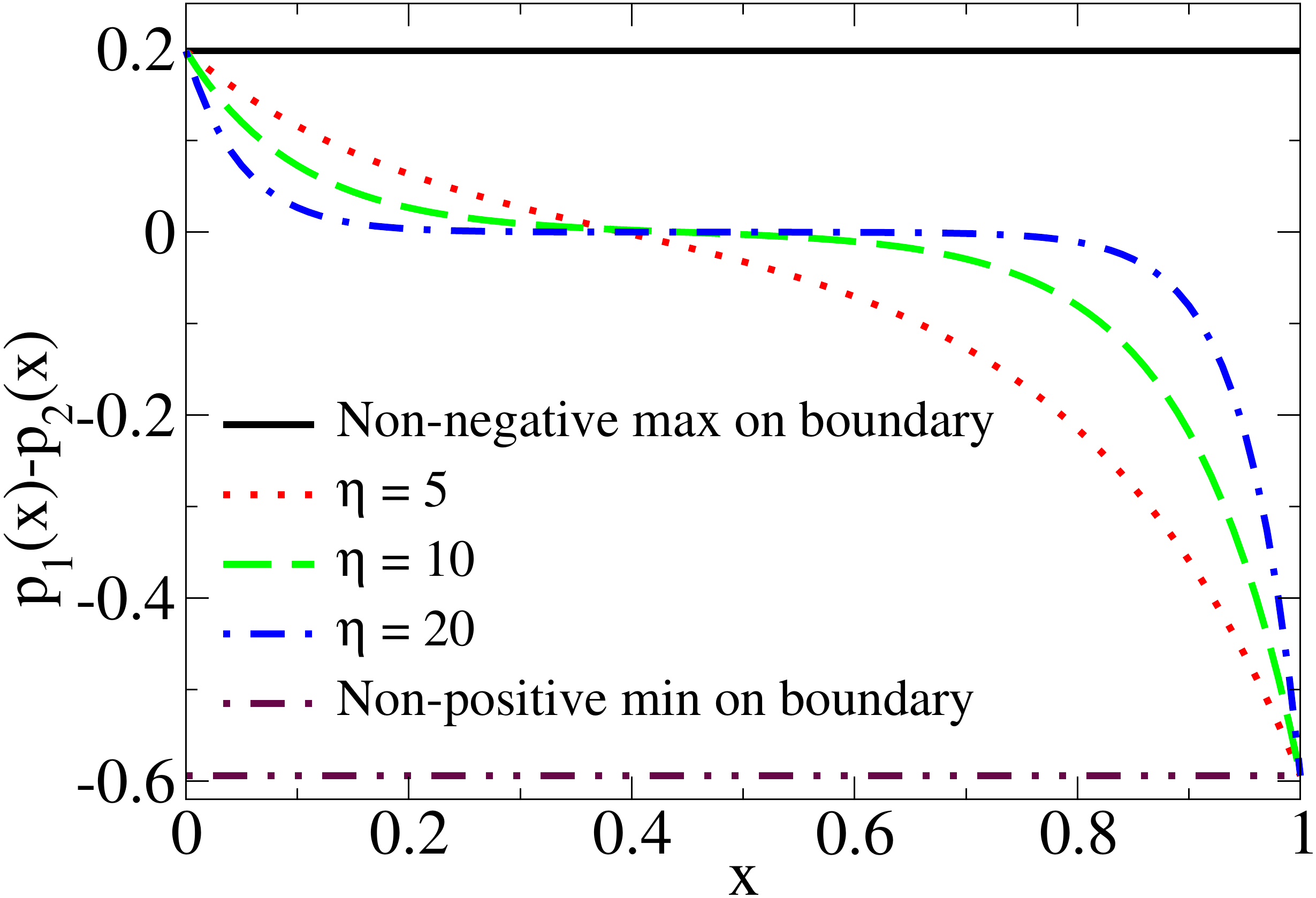}}
	\subfigure[$0 = p_1^{\mathrm{R}} < 0.9 = p_2^{\mathrm{R}}
	< p_1^{\mathrm{L}} = 1 < p_2^{\mathrm{L}} = 1.5$]{
		\includegraphics[clip,scale=0.26]{Figures/Fig_4b_maximum_principle_case2.pdf}}
	\caption{\textsf{\small One-dimensional problem \#1:}~This
		figure numerically verifies the maximum principle
		given by Theorem \ref{Thm:Dual_Maximum_Principle}.
		According to the maximum principle, $p_1(x) - p_2(x)$
		in the entire domain lies between the non-negative
		maximum and non-positive minimum values on the boundary.
		Note that the medium properties are isotropic and homogeneous.
		\label{Fig:1D_problem_1_maximum_principle}}
\end{figure}

\begin{figure}[!h]
	\centering
	\subfigure[Micro-velocity for $k_{1} < k_{2}$]{
		\includegraphics[clip,scale=0.25]{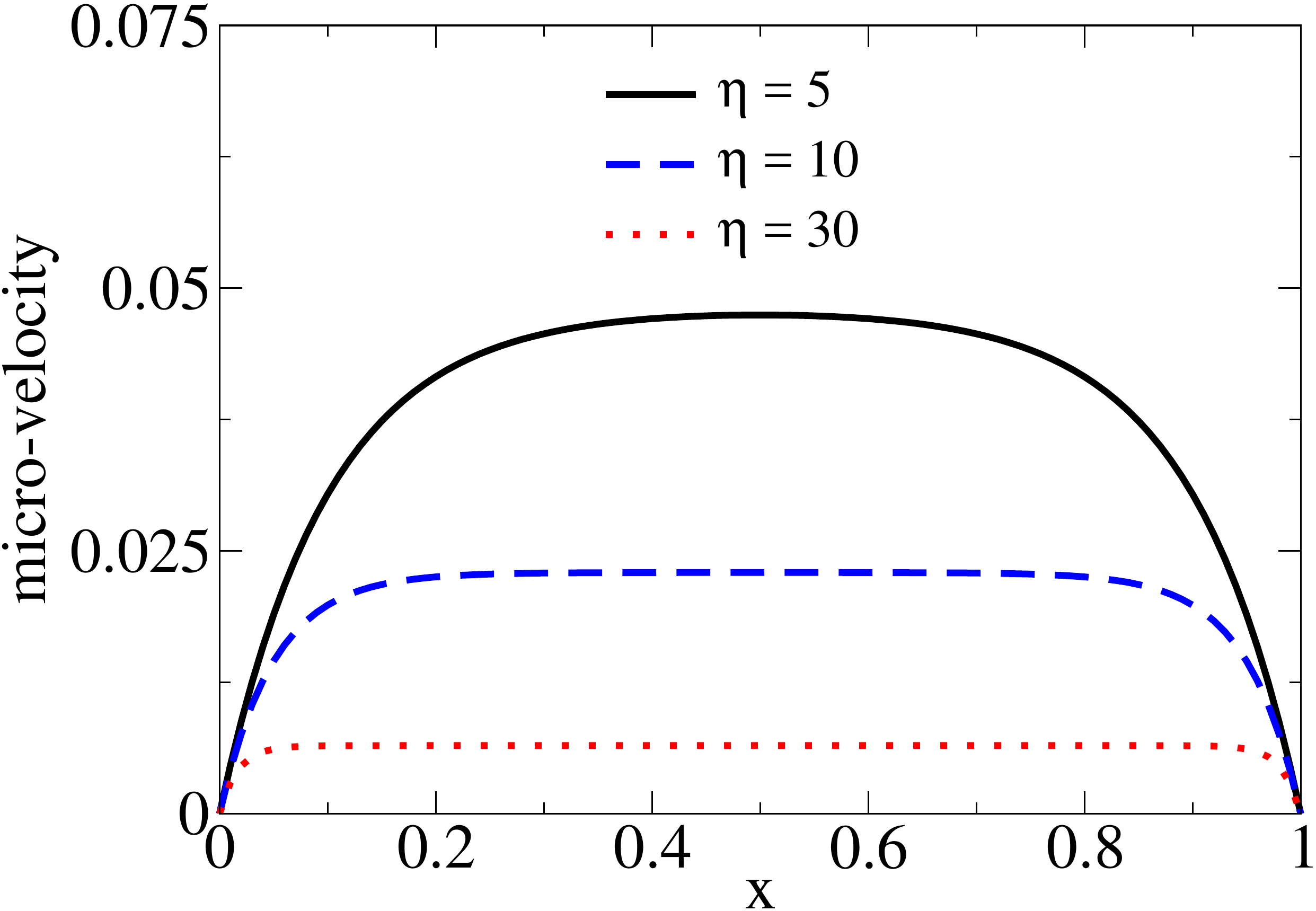}}
	\subfigure[Micro-velocity for $k_{1} > k_{2}$]{
		\includegraphics[clip,scale=0.25]{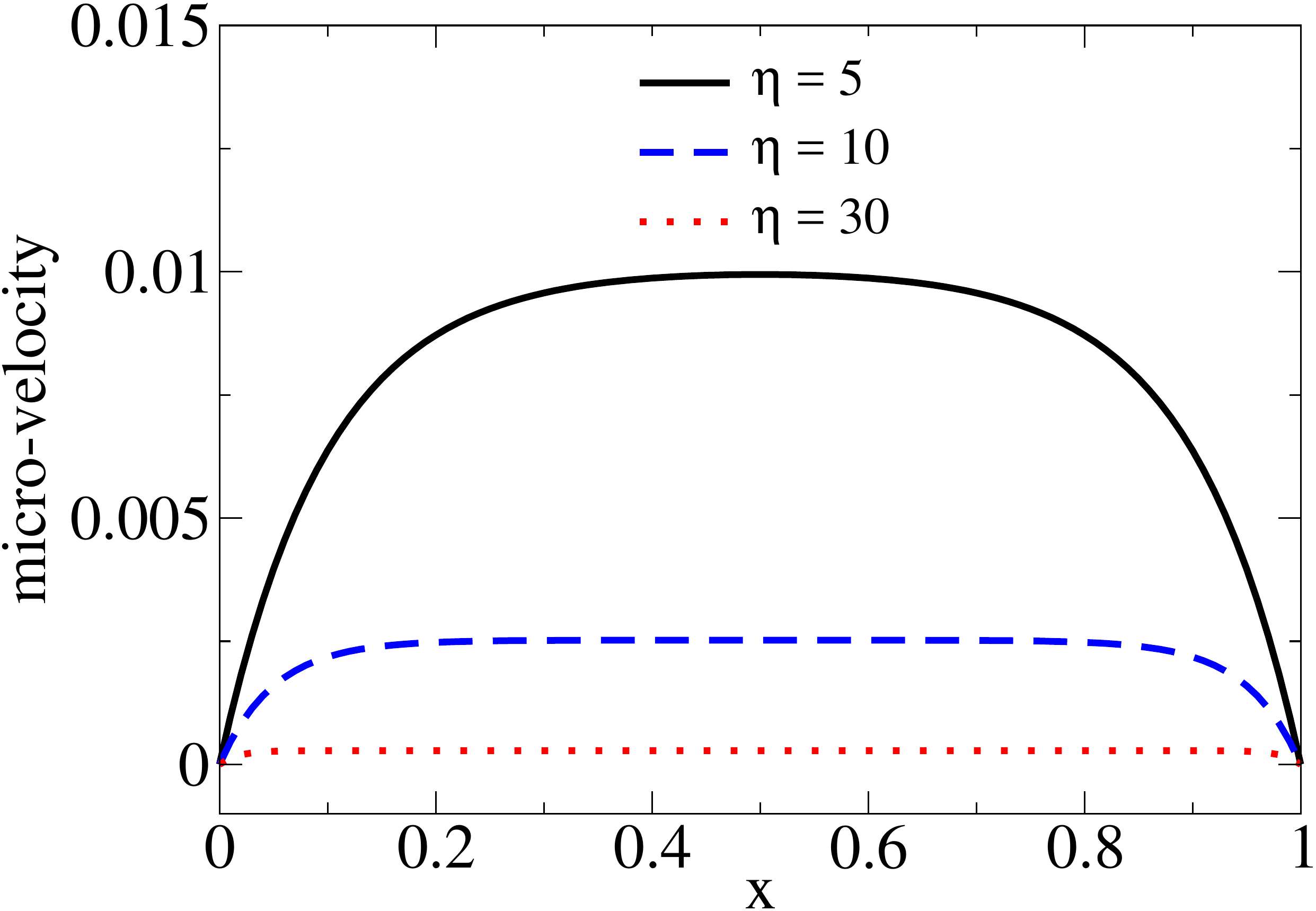}}
	\subfigure[Mass transfer for $k_1 < k_2$]{
		\includegraphics[clip,scale=0.25]{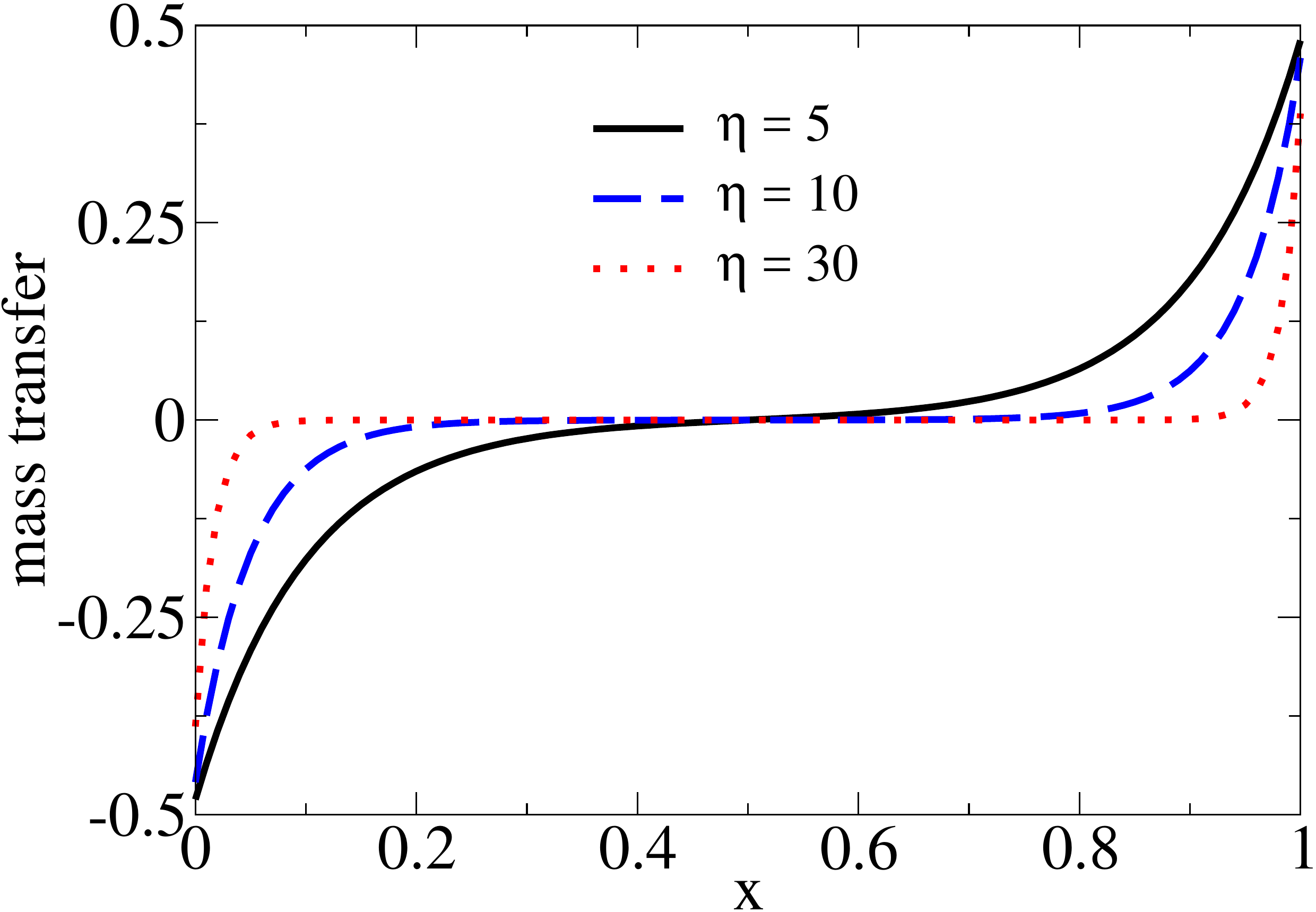}}
	\subfigure[Mass transfer for $k_1 > k_2$]{
		\includegraphics[clip,scale=0.25]{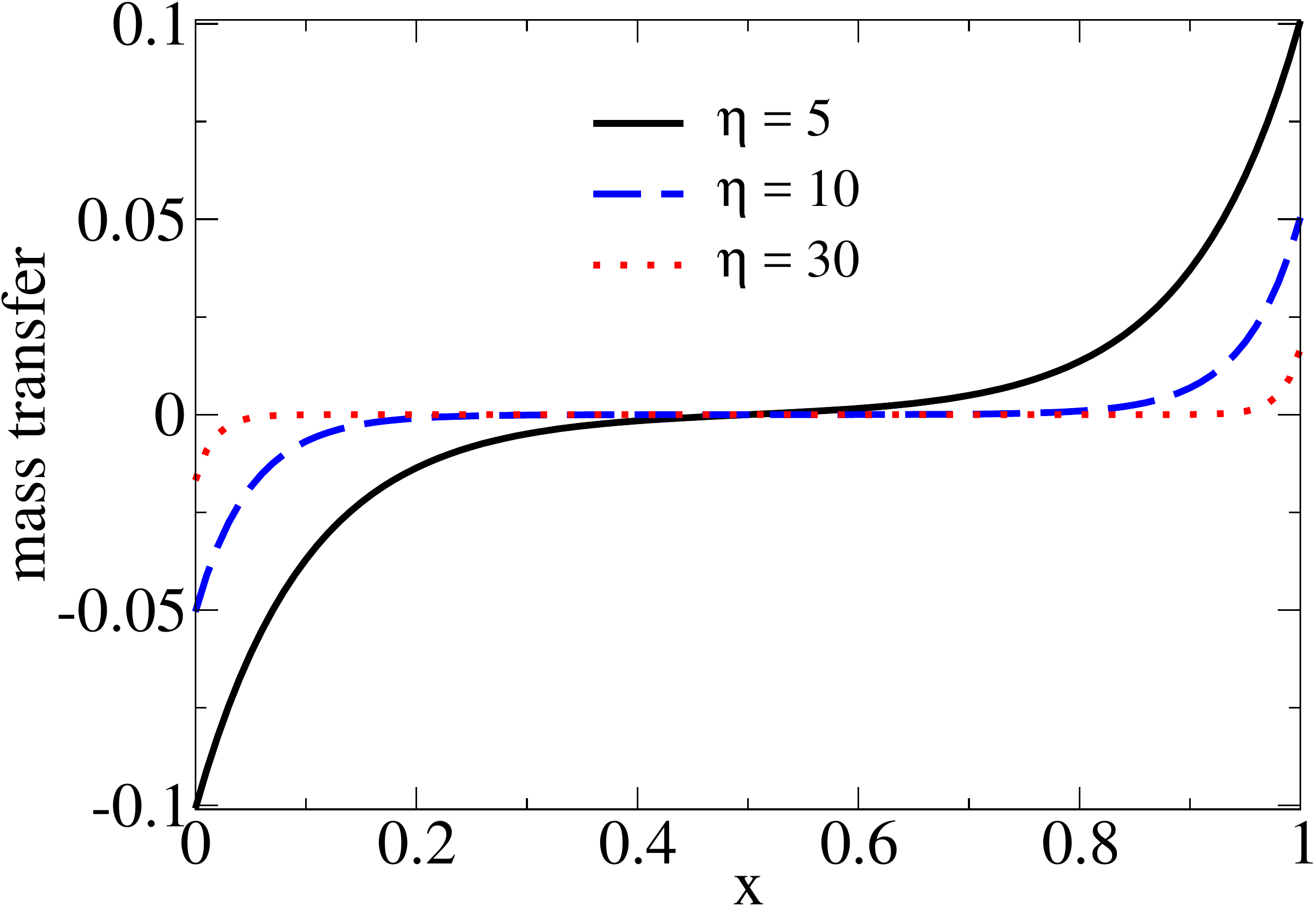}}
	\caption{\textsf{\small One-dimensional problem \#2:}~Variation
		of the micro-velocity and mass transfer
		for various $\eta$ values for the cases $k_{1} < k_{2}$
		and $k_{1} > k_{2}$.
		Although there is no supply of fluid on the boundaries of 
		the micro-pore network, there is still a discharge (i.e.,
		non-zero velocity) in the micro-pore network, and there is
		a mass transfer across the pore-networks. 
		\label{Fig:Double_1D_P2_macro_micro_velocities}}
\end{figure}
%
\begin{figure}
	\centering
	\subfigure[Case 1: $k_{1} = 1.0$ and $k_{2} = 0.1$]{
		\includegraphics[clip,scale=0.26]{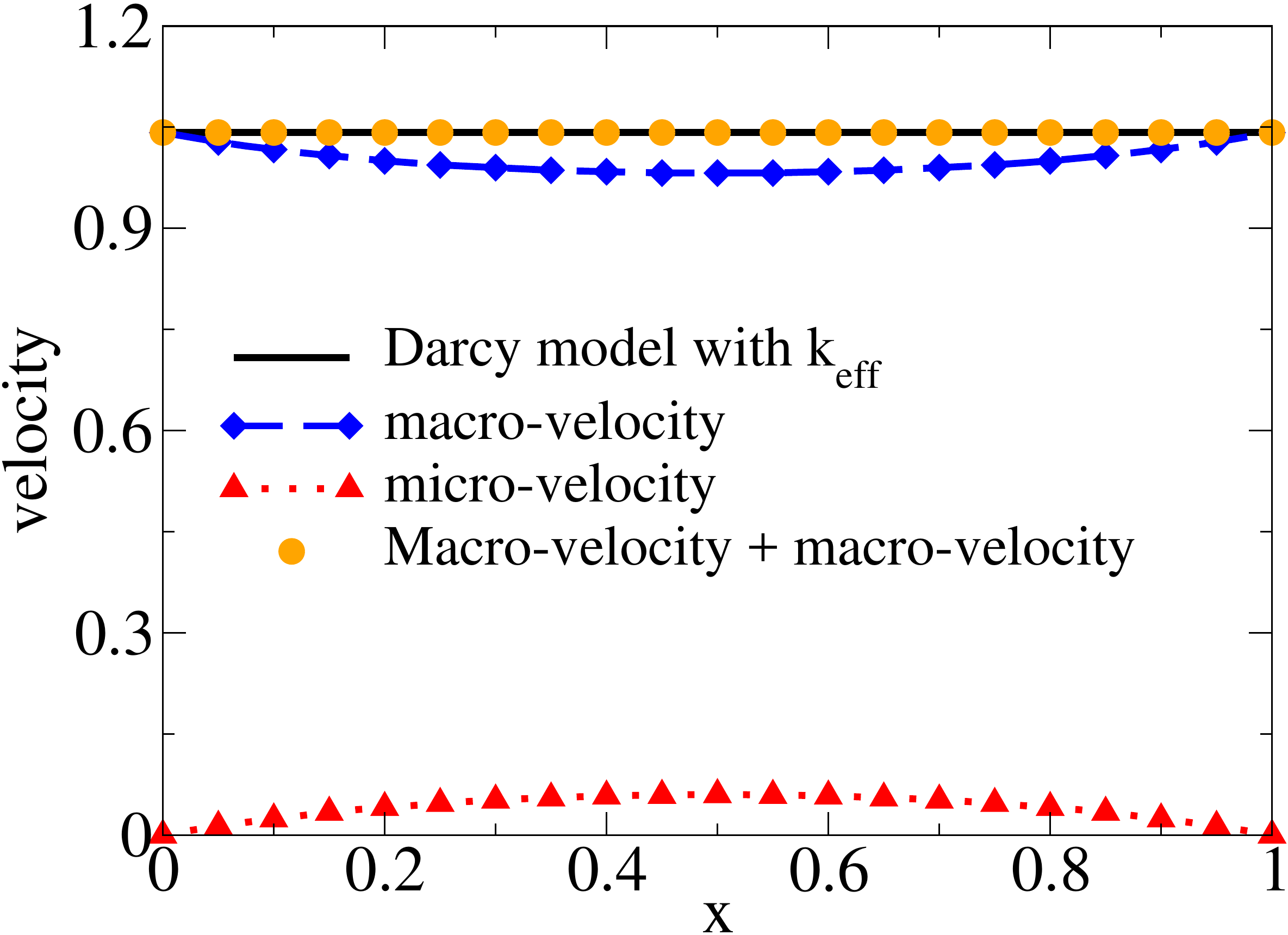}}
	\subfigure[Case 2: $k_{1} = 0.1$ and $k_{2} = 1.0$]{
		\includegraphics[clip,scale=0.26]{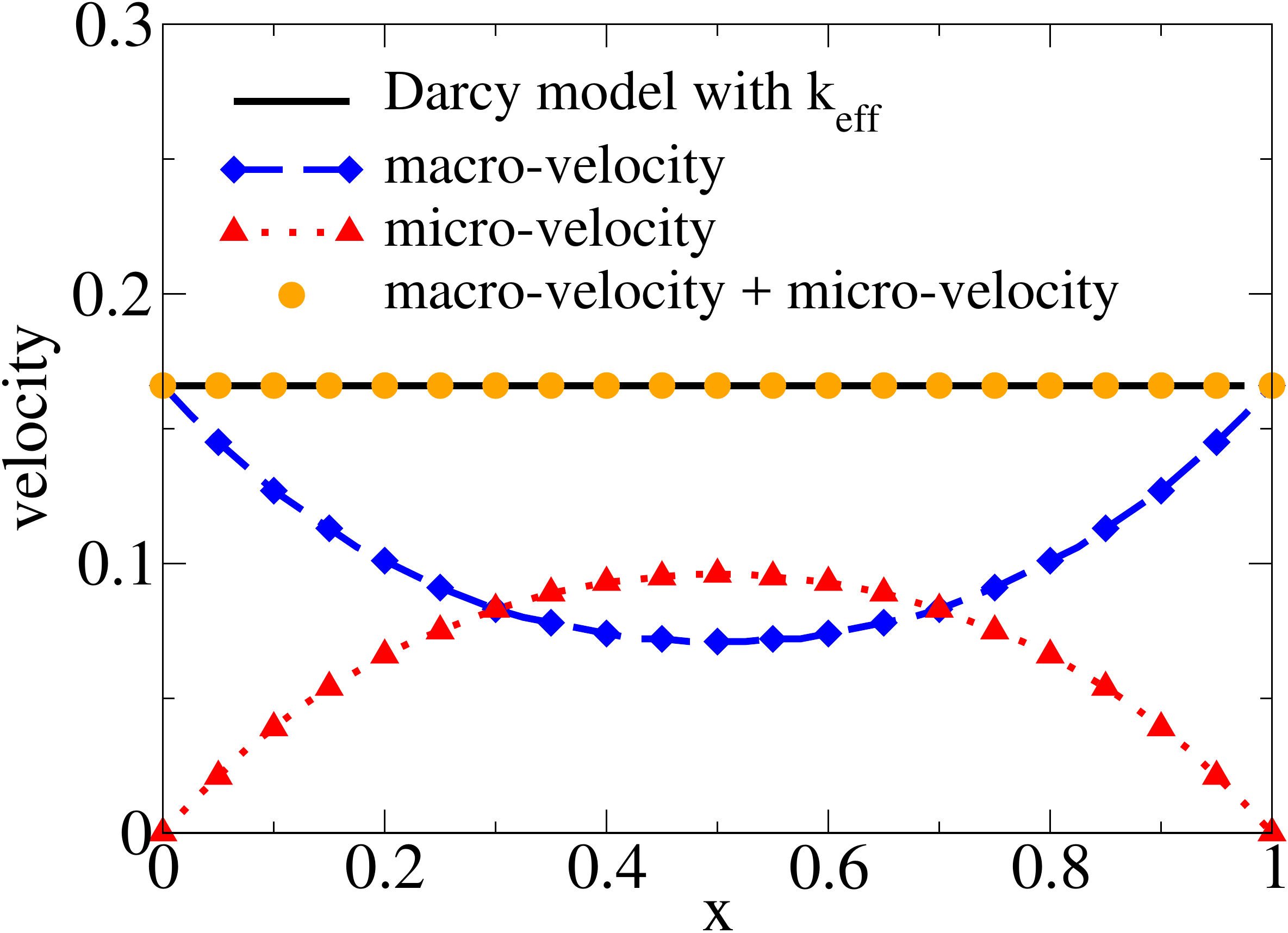}}
	\caption{\textsf{\small One-dimensional problem \#2:}~This
		figure compares the velocities under double
		porosity/permeability model and Darcy model
		for the cases $k_1 > k_2$ and $k_1 < k_2$.
		Macro- and micro-velocities and their summation
		under the double porosity/permeability model as
		well as the velocity under the Darcy model with
		$k=k_{\mathrm{eff}}$ are displayed. As it can be 
		seen, $k_{\mathrm{eff}}$ obtained by the classical 
		Darcy experiment cannot capture the complex internal 
		pore-structure.
		\label{Fig:Comparison_Velocities_Darcy_Dual}}
\end{figure}

\begin{figure}[!h]
	\centering
	\subfigure{
		\includegraphics[scale=0.42]{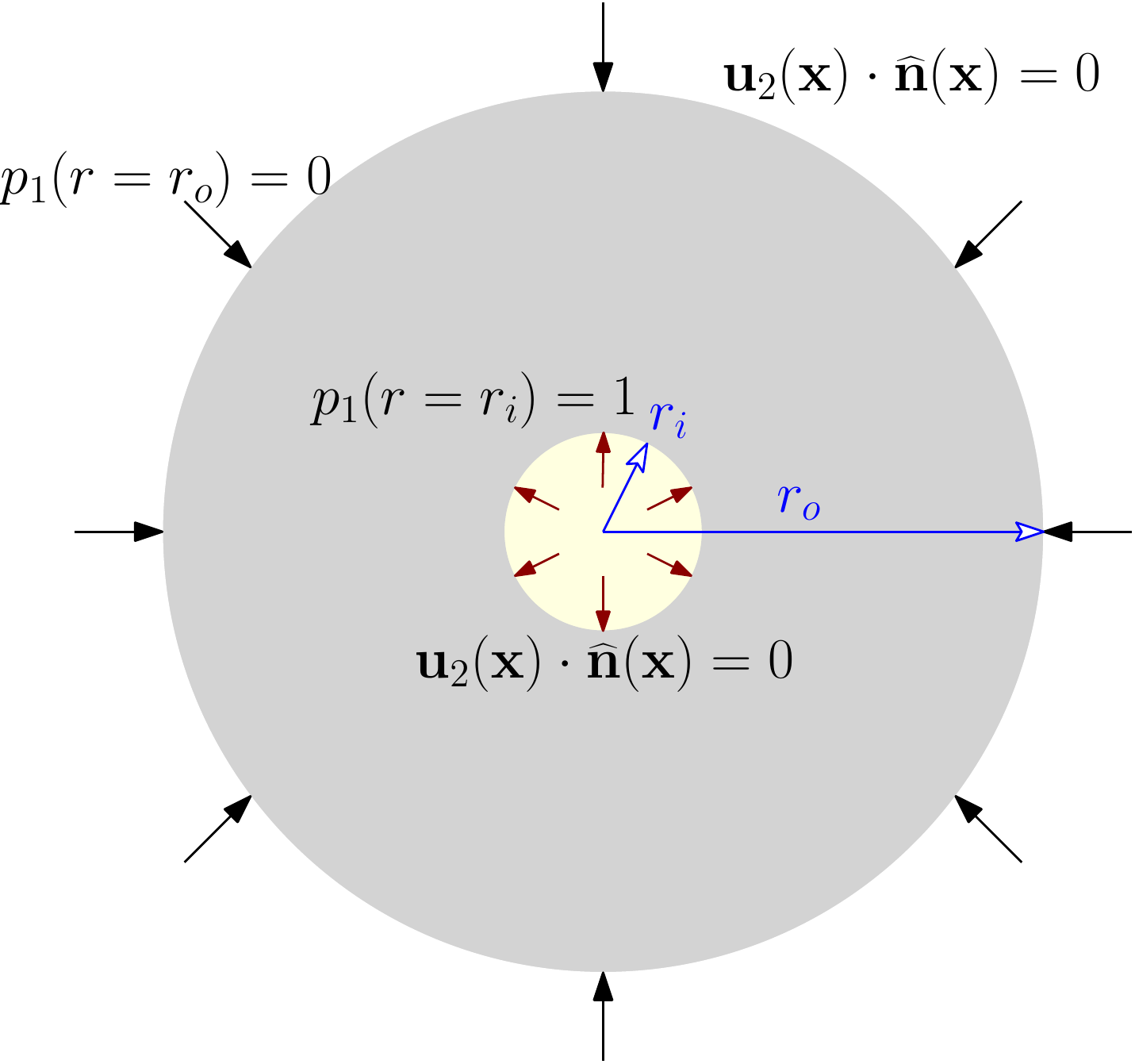}
		\label{Fig:Double_description}}
	\subfigure{
		\includegraphics[clip,scale=0.27]{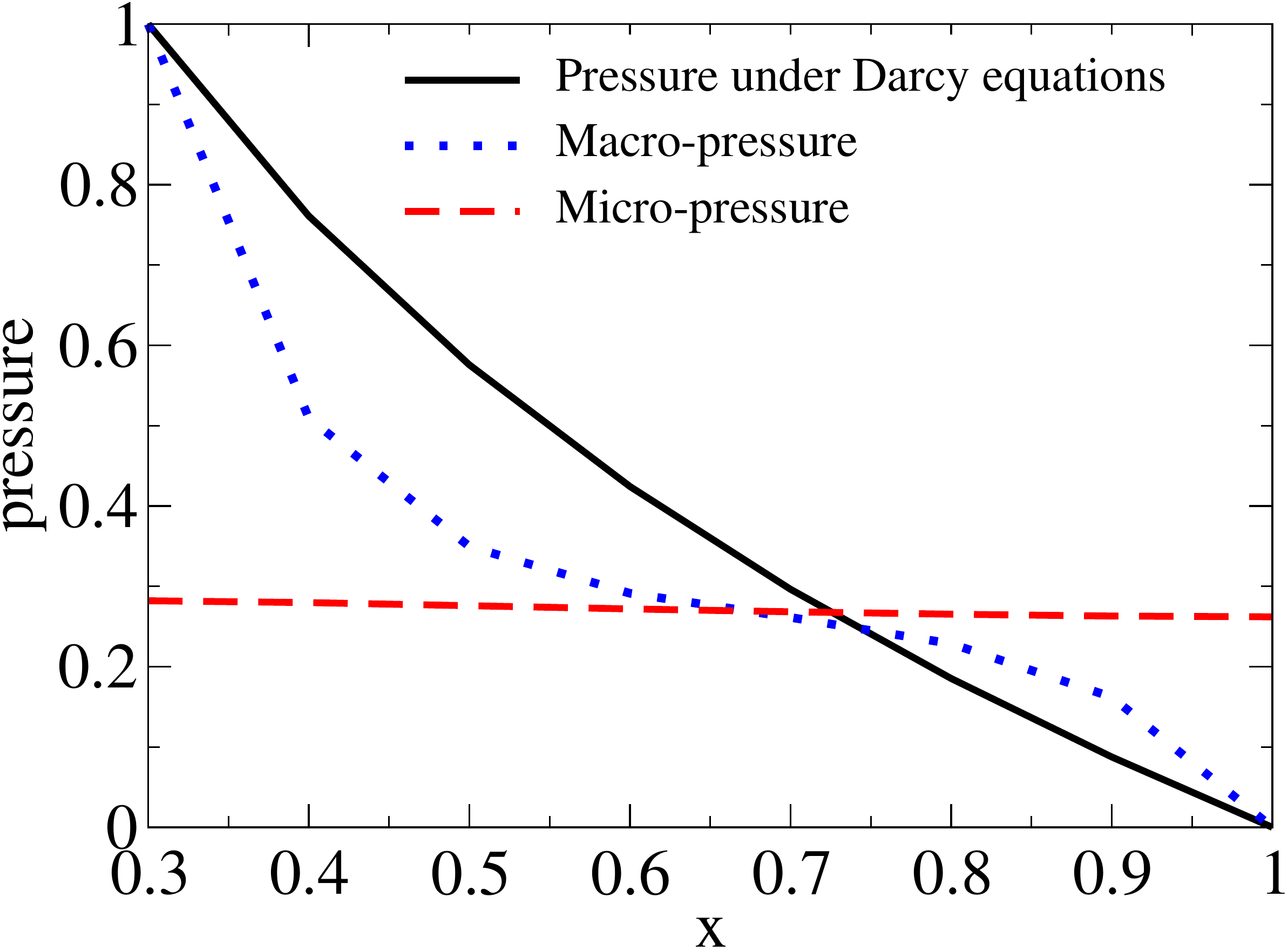}}
	\caption{The left figure provides a pictorial description of the  
		boundary value problem.
		There is no discharge on the inner and outer surfaces of the
		micro-pore network. For the macro-pore network, the inner
		surface is subjected to a pressure of unity, and the outer
		surface is subjected to a pressure of zero. The right figure
		illustrates that the macro-pressure under the double
		porosity/permeability model is qualitatively different from
		the pressure under Darcy equations.
		\label{Fig:Double_description_and_pressure}}
\end{figure}

\end{document}